\documentclass[]{aastex62}
\usepackage{bm}

\graphicspath{{./}{figures/}}

\newcommand{\funits}[1]{erg cm$^{-2}$ s$^{-1}$ \AA{}$^{-1}$}
\newcommand\xmm{\emph{XMM-Newton}}
\newcommand\swift{\emph{Swift}}

\received{November 28, 2022}
\revised{March 11, 2023}
\accepted{March 29, 2023}
\submitjournal{ApJ}

\shorttitle{AU Mic \textsc{i}:  High-Time Resolution Light Curves}
\shortauthors{Tristan et al.}

\begin{document}

\title{A Seven-Day Multi-Wavelength Flare Campaign on AU Mic \textsc{i}: \\
High-Time Resolution Light Curves and the Thermal Empirical Neupert Effect}

\correspondingauthor{Isaiah I. Tristan}
\email{Isaiah.Tristan@colorado.edu}

\author[0000-0001-5974-4758]{Isaiah I. Tristan}
\affiliation{Department of Astrophysical and Planetary Sciences, University of Colorado, Boulder, 2000 Colorado Ave, CO 80305, USA}
\affiliation{Laboratory for Atmospheric and Space Physics, University of Colorado Boulder, 3665 Discovery Drive, Boulder, CO 80303, USA}
\affiliation{National Solar Observatory, University of Colorado Boulder, 3665 Discovery Drive, Boulder, CO 80303, USA}

\author[0000-0002-0412-0849]{Yuta Notsu}
\affiliation{Department of Astrophysical and Planetary Sciences, University of Colorado, Boulder, 2000 Colorado Ave, CO 80305, USA}
\affiliation{Laboratory for Atmospheric and Space Physics, University of Colorado Boulder, 3665 Discovery Drive, Boulder, CO 80303, USA}
\affiliation{National Solar Observatory, University of Colorado Boulder, 3665 Discovery Drive, Boulder, CO 80303, USA}
\affil{Department of Earth and Planetary Sciences, Tokyo Institute of Technology, 2-12-1 Ookayama, Meguro-ku, Tokyo 152-8551, Japan}

\author[0000-0001-7458-1176]{Adam F. Kowalski}
\affiliation{Department of Astrophysical and Planetary Sciences, University of Colorado, Boulder, 2000 Colorado Ave, CO 80305, USA}
\affiliation{Laboratory for Atmospheric and Space Physics, University of Colorado Boulder, 3665 Discovery Drive, Boulder, CO 80303, USA}
\affiliation{National Solar Observatory, University of Colorado Boulder, 3665 Discovery Drive, Boulder, CO 80303, USA}

\author[0000-0003-2631-3905]{Alexander Brown}
\affiliation{Center for Astrophysics and Space Astronomy,
University of Colorado, 389 UCB,
Boulder, CO 80309, USA}

\author[0000-0001-9209-1808]{John P. Wisniewski}
\affil{Department of Physics and Astronomy, George Mason University, 4400 University Drive, MS 3F3, Fairfax, VA 22030, USA}

\author[0000-0001-5643-8421]{Rachel A. Osten}
\affil{Space Telescope Science Institute, Baltimore, MD 21218, USA}

\author[0000-0002-1864-6120]{Eliot H. Vrijmoet}
\affil{Department of Physics and Astronomy, Georgia State University, Atlanta, GA 30303, USA}
\affil{RECONS Institute, Chambersburg, PA 17201, USA}

\author[0000-0002-4914-6292]{Graeme L. White}
\affil{Computational Engineering and Science Research Centre, University of Southern Queensland, Toowoomba 4350, Australia}

\author{Brad D. Carter}
\affil{Computational Engineering and Science Research Centre, University of Southern Queensland, Toowoomba 4350, Australia}

\author[0000-0001-5440-1879]{Carol A. Grady}
\affil{Eureka Scientific, 2452 Delmer, Suite 100, Oakland, CA 94602-3017, USA}

\author{Todd J. Henry}
\affil{RECONS Institute, Chambersburg, PA 17201, USA}

\author{Rodrigo H. Hinojosa}
\affil{Cerro Tololo Inter-American Observatory, CTIO/AURA Inc., La Serena, Chile}

\author[0000-0001-8470-0853]{Jamie R. Lomax}
\affil{Department of Physics, United States Naval Academy, 572c Holloway RD, Annapolis, MD 21402, USA}

\author{James E. Neff}
\affil{Division of Astronomical Sciences, National Science Foundation, Alexandria, VA 22314, USA}

\author[0000-0003-1324-0495]{Leonardo A. Paredes}
\affil{RECONS Institute, Chambersburg, PA 17201, USA}
\affil{Department of Physics and Astronomy, Georgia State University, Atlanta, GA 30302, USA}

\author{Jack Soutter}
\affil{Computational Engineering and Science Research Centre, University of Southern Queensland, Toowoomba 4350, Australia}

\begin{abstract}
We present light curves and flares from a seven day, multi-wavelength observational campaign of AU Mic, a young and active dM1e star with exoplanets and a debris disk.
We report on 73 unique flares between the X-ray to optical data.
We use high-time resolution NUV photometry and soft X-ray (SXR) data from \xmm{} to study the empirical Neupert effect, which correlates the gradual and impulsive phase flaring emissions.
We find that 65\% (30 of 46) flares do not follow the Neupert effect, which is three times more excursions than seen in solar flares, and propose a four part Neupert effect classification (Neupert, Quasi-Neupert, Non-Neupert I \& II) to explain the multi-wavelength responses.
While the SXR emission generally lags behind the NUV as expected from the chromospheric evaporation flare models, the Neupert effect is more prevalent in larger, more impulsive flares. 
Preliminary flaring rate analysis with X-ray and \emph{U}-band data suggests that
previously estimated energy ratios hold for a collection of flares observed over the same time period, but not necessarily for an individual, multi-wavelength flare.
These results imply that one model cannot explain all stellar flares and care should be taken when extrapolating between wavelength regimes.
Future work will expand wavelength coverage using radio data to constrain the nonthermal empirical and theoretical Neupert effects to better refine models and bridge the gap between stellar and solar flare physics.
\end{abstract}

\section{Introduction}  \label{sec:intro} 

M dwarfs will be high-priority targets in the search for habitable exoplanet systems, as they are the most common stars in the solar neighborhood \citep{Henry2006} and their higher planet-to-star contrast relative to more luminous stars allow for easier transit detection.
Stellar activity like flares and coronal mass ejections may affect the habitability of exoplanets through increased high-energy photon and particle fluxes \citep{Linsky2019, Airapetian2020}.
As examples, stellar ultraviolet (UV) spectra help ascertain the effects on exoplanet atmospheric chemistry and escape rates \citep[Sec.\ E-Q2d of][]{decadal, Loyd2018}, and X-rays can influence protoplanetary disk chemistry \citep{Osten2013, Osten2015, Notsu2021} and lead to enhanced NUV radiation rates at the ground level through fluorescence \citep{Smith2004}.

The response at optical wavelengths has historically been the best-observed phenomenon in stellar flares, thanks to many decades of ground-based monitoring in the UBVR bandpasses \citep{Lacy1976, Pettersen1984, Hilton2011} and recent long-baseline, high-precision ancillary white-light data provided by Kepler, K2, and TESS \citep{Hawley2014, Davenport2016, Notsu2019, Howard2022, Mendoza2022}.  
Large observational campaigns spanning the X-ray, UV, optical, and radio have unfortunately been few and far between \citep{Osten2005, Osten2006}.
Empirical multi-wavelength relationships for the full range of flare amplitudes, energies, and light curve morphology (e.g.\ gradual versus impulsive) are severely lacking, and simple slab model extrapolations (i.e.\ modeling flaring emission from a plasma region with static properties) from the optical are largely unjustified \citep{Kowalski2019}. 

It is widely accepted that similar physical processes of magnetic reconnection, particle acceleration, and plasma heating operate in both solar and stellar flares \citep{Shibata2011, Shibata2016}.  
The multi-wavelength relationship that justifies these similarities is known as the Neupert effect, which was first observed in solar flares as the correspondence between the derivative of thermal, soft X-rays ($E \lesssim 15$ keV) and the instantaneous nonthermal gyrosynchrotron cm-wave flux \citep{Neupert1968}.  
The empirical Neupert effect (ENE) has been widely studied in solar flares using the cumulative time-integral of the impulsive phase, power-law hard X-rays ($E \gtrsim 25$ keV) and the soft X-rays emitted from plasma at $T \gtrsim 15$ MK \citep{Dennis1993, McTiernan1999, Veronig2002, Veronig2005, Qiu2013, Namekata2017} and certain timings between the two responses.
The Neupert effect has also been studied in several stellar flares in M dwarfs and RS CVns
\citep{Hawley1995, Gudel1996, Gudel2002, Osten2015, Stelzer2022}.
Note that while gyrosynchrotron emission is best detected through radio observations \citep[see][]{Osten2005, MacGregor2018}, the broadband optical or near-UV (NUV) continuum response is often used as a proxy in stellar studies, as radio or hard X-ray responses can be too faint to easily detect \citep{Osten2007}.
For reference, the thermal empirical Neupert effect uses a proxy, while the nonthermal empirical Neupert effect uses radio/microwave data, and they both measure the same quantities.

Likewise, the theoretical Neupert effect (TNE) explains the multi-wavelength temporal behaviors in terms of chromospheric evaporation and condensation in response to impulsive nonthermal electron beam heating \citep{Antonucci1982, Antonucci1984}. 
The pioneering radiative-hydrodynamic models of \citet{Allred2005} \citep[built upon][]{Livshits1981, McClymont1984, Fisher1985, Fisher1985b, Lee1995, Cargill1995, Abbett1999} established this standard flare paradigm.  
After nonthermal particles gyrate in the magnetic fields, producing gyrosynchrotron radiation, they precipitate into the chromosphere, producing hard X-rays.  
The beam heating is accompanied by increased chromospheric radiation in the optical and NUV just below the upper chromospheric layers that explosively heat to tens of million K, ablate into the corona, and eventually fill the magnetic loops and shine luminously in soft X-rays.
These ideas have been extended to stellar flares using modern radiative-hydrodynamic simulations with larger beam fluxes than inferred on the Sun \citep{Allred2006, Kowalski2015}.

About 20\% of solar flares \citep{Dennis1993} and numerous stellar flare observations clearly do not exhibit the Neupert effect, or the ENE only approximately explains the wavelength-dependent time delays.  
Notable examples are the late impulsive peak in hard X-rays during a solar flare \citep{Warmuth2009} and a giant radio flare from EV Lac with a very impulsive U-band light curve but no temporally consistent response in the soft X-rays \citep{Osten2005}.  
In other event comparisons, the energy partition in the impulsive and gradual phase radiation has been found to vary by nearly an order of magnitude \citep{Osten2015, Osten2016}.  
X-ray and UV flux correlations have been found in stellar flares, with power-law relationships between their energies with exponents between 1 and 2 \citep{Mitra2005}.
\citet{Veronig2002} suggests that flares with relatively more  X-ray luminosity are powered by direct heating of the corona with relatively weaker nonthermal particle heating present; this has also been discussed in relation to energy budget comparisons in solar flares 
\citep{WarmuthMann2016a, WarmuthMann2016b}.
At the high-energy regime however, large stellar flares are shown to be X-ray--weak compared to their solar counterparts \citep{Gudel1996}, supporting higher electron beam flux models.
\citet{Veronig2005} discusses several simplifications in standard models, including assumptions of a static beam low cutoff energy, loop geometry, and heterogeneity of the flare source, all of which may lead to variations in the TNE when critically analyzed at high-time resolution. 
There are also alternative heating sources in the flaring environment, like thermal conduction where heat is transmitted through the loop after magnetic re-connection \citep[see][and references within]{Yokoyama2001}, which could affect the observed ENE and TNE responses.

To better understand the energy partition and multi-wavelength timing in stellar flares, we have executed a large observational campaign on a young flare star, AU Mic, over seven days with a combination of ground- and space-based observatories.  
These observations were designed to challenge our understanding of chromospheric evaporation and condensation in novel ways by constraining the empirical differences in the impulsive and gradual phase radiation from flare to flare.  
Known for its large average flare energies, AU Mic is an ideal target since white-light is not produced in any great amount by direct heating of the corona and subsequent thermal conductive fluxes into the upper chromosphere \citep{Kowalski2017A, Namekata2020}.  
Thus this dataset is intended to push our knowledge of the ENE and TNE into regimes that reveal new physical processes in stellar atmospheric heating by nonthermal particles, which in turn would allow far more realistic inputs into models of atmospheric photochemistry around M dwarfs.

This paper is the first in a series.  Here, we present the data reduction, light curves, and analysis of the ENE from the flare campaign on AU Mic in Oct 2018.  This paper is organized as follows.  In Section \ref{sec:obs}, we describe our target, the observations of AU Mic, and the data reduction and absolute flux calibration.  In Section  \ref{sec:lc}, we present light curves and showcase the flares identified. In Section, \ref{sec:methods}, we outline the methods used and share results. In Section \ref{sec:discussion}, we discuss our findings in the context of the ENE and provide insight into necessary future work. In Section \ref{sec:conclusions}, we summarize and conclude this work.

\section{The 7-Day AU Mic Flare Monitoring Campaign} \label{sec:obs}

\subsection{The Target: AU Mic}
The target of our flare monitoring observations is the M1 star AU Mic, a nearby pre-main sequence star located at a distance of 9.72 parsecs \citep{Gaia, GaiaDR2}.  It is part of the $\beta$ Pictoris moving group, with an age of $23 \pm 3$ Myr \citep{Mamajek2014}. In an H-R diagram constructed from Gaia DR2 data \citep{Gaia, GaiaDR2}, AU Mic is above the main sequence and is consistent with a PARSEC \citep{Bressan2012} color-magnitude 23 Myr isochrone, suggesting that it is still contracting\footnote{Assuming a typical radius for an M1 main sequence star of $0.49R_{\rm{Sun}}$ \citep{Reid2004}, we calculate the radius of AU Mic to be $\sim0.7R_{\rm{Sun}}$ from the Gaia DR2 G-band magnitude.  The AAS abstract from \citet{radius} reports an interferometric radius of  $0.75R_{\rm{Sun}}$, which \citet{Plavchan2020} corroborates.}.
Selected stellar properties of AU Mic are summarized in Table \ref{tbl:aumic}.

AU Mic has been a frequent target of Hubble Space Telescope observations  due to its IR excess \citep[$L_{\text{IR}}/L_{\star} = 0.44$$;$][]{Schneider2014}, lack of mid-IR thermal excess \citep{Chen2005}, and resolved edge-on debris disk with outward-moving features \citep{Kalas2004, Krist2005, Boccaletti2015, Wisniewski2019, Grady2020}. 
Flares from AU Mic may promote stellar wind pressure, contributing to the extended debris disk \citep{Augereau2006}.
AU Mic also has a confirmed planetary system consisting of one Neptune-sized planet, which has an orbital period of 8.46 days and is situated at a distance of 0.07 AU \citep{Plavchan2020}, and one slightly smaller planet with a period of 18.86 days \citep{Martioli2021, Gilbert2022}.
The space weather conditions surrounding AU Mic may make the possibility of planetary atmospheres unlikely, however \citep{Alvarado2022, Cohen2022, Klein2022}.

AU Mic is a member of a relatively rare sub-class of early-type M stars \citep{West2008} that are magnetically active with H$\alpha$ in emission and a variable $V$-band magnitude \citep{Torres1973, Rodono1986,Hebb2007} outside of obvious flare events.  
It is the most X-ray luminous star within 10 parsecs with a quiescent X-ray (0.05 -- 3.5 keV) luminosity of $10^{29.7}$ ergs s$^{-1}$ \citep{Pallavicini1990, Pagano2000}.
Recent studies have also shown that AU Mic's surface-averaged magnetic field is much larger than the Sun's, and its dipole fields exhibit asymmetric components \citep{Kochukhov2020} that may be misaligned with the rotation axis \citep{Wisniewski2019}, thus having an influence on the disk morphology \citep{Wisniewski2019}.

AU Mic has been a source of energetic flares in the X-ray, ultraviolet, optical, and radio wavelength regimes \citep{Robinson1993, Cully1994, Robinson2001, Redfield2002, Mitra2005, Hebb2007, MacGregor2020, Feinstein2022}.  
In the optical and NUV, which directly probe the enigmatic white-light continuum radiation, the flares from early-type (M0 -- M1) M dwarf  stars have rarely been studied with any time resolution or complementary multi-wavelength information, unlike for flares from mid- and late-type M dwarfs \citep[e.g.][]{Hawley1995, Osten2005, Fuhrmeister2008, Fuhrmeister2011, Kowalski2019, MacGregor2021}.  
This is due to their lower contrast against the photospheric background \citep[e.g.\ ][]{Kowalski2009}, allowing only the largest events to rise above the typical noise floor.  
Since the largest events are also the rarest \citep{Lacy1976, Gilbert2022}, long monitoring times are required to guarantee a sample of flares that can be characterized in detail.  
AU Mic is one of the best early-type M-dwarf targets for such a next-generation flare campaign:  it is a single star with a known age,  it is the stellar source at the heart of an intriguing exoplanetary and debris disk system, and its dynamo mechanism falls in the partially convective regime, facilitating interesting comparisons to high-energy solar flares.

\subsection{Observations and Initial Data Reduction}
We observed  AU Mic over seven days spanning 2018 Oct 10 to 2018 Oct 17 with the X-ray Multi-Mirror Mission \citep[\xmm{},][]{xmm_paper}, the Jansky Very Large Array (JVLA), the Neil Gehrels Swift Observatory (\swift{}), the Las Cumbres Observatory Global Telescope network \citep[LCOGT,][]{Brown2013}, the Astrophysical Research Consortium’s 3.5m telescope at the Apache Point Observatory (APO), the Australia Telescope Compact Array (ATCA), and the Small and Moderate Aperture Research Telescope System \citep[SMARTS,][]{Subasavage2010} 0.9m and 1.5m telescopes at the Cerro Tololo Inter-American Observatory (CTIO). 
The observations used in this paper are summarized in Table \ref{tbl:obs_summary}. 
JVLA, APO, and ATCA data will be reduced and analyzed in future studies.
Together, these form the most wavelength-comprehensive, high-time resolution simultaneous observation of AU Mic to date.

\subsubsection{XMM-Newton EPIC-pn X-ray Photometry} \label{sec:xmm_epic}
The \xmm{} observations were conducted in four visits (observations; Obs-IDs: 0822740301, 0822740401, 0822740501, 0822740601), which span 
2018 Oct 10 13:13:59.760 to 2018 Oct 12 01:42:19.177 UTC (131.29 ks),
2018 Oct 12 13:06:33.808 to 2018 Oct 14 01:36:22.232 UTC (131.40 ks),
2018 Oct 14 12:21:18.787 to 2018 Oct 16 00:04:00.273 UTC (71.75 ks), and
2018 Oct 16 23:39:21.362 to 2018 Oct 17 18:17:33.004 UTC (67.10 ks).
X-ray photometry was obtained over a bandpass of $E \approx 0.2 - 12$ keV using the European Photon Imaging Camera \citep[EPIC-pn,][]{epic_camera} Medium filter and 10 second time binning. 
During the least active times (00:00:00.0 -- 07:12:00.0 UTC, 2018 Oct 11), the X-ray flux averages $15.3 \pm 1.6$ counts per second (cps). Data from the MOS 1 and 2 detectors were simultaneously captured and will be used for future analysis.

\subsubsection{Swift XRT X-ray Photometry} \label{sec:swift_xray}
\swift{} provided complementary X-ray and NUV coverage during gaps between the \xmm{} visits.
\swift{} XRT X-ray \citep[$E\approx 0.2-10$ keV,][]{Burrows2005} photon-counting data were taken in two large windows, from
03:26:20.0 to 10:04:50.0 UTC (8.28 ks) on 2018 Oct 12 and 
03:03:16.0 to 09:53:51.0 UTC (7.38 ks) on 2018 Oct 14.
We used AstroImageJ \citep{AIJ} to determine a circular aperture and background annulus. On 2018 Oct 12, an aperture of $37\farcs71$ and annulus between $66\farcs00$ and $99\farcs00$ were used (reported pixel:arcsec conversion of 6.89:16.24), while on 2018 Oct 14, an aperture of $32\farcs98$ and annulus between $58\farcs88$ and $89\farcs50$ were used (reported pixel:arcsec of 5.29:12.46). A light curve was then extracted using XSELECT\footnote{\url{http://heasarc.gsfc.nasa.gov/ftools}} \citep{heasarc} and binned into 5-second intervals. No obvious flares were found in these data, and detailed analysis showed that no time intervals exceeded $2\sigma$ above the daily, average quiescent value.

\subsubsection{XMM-Newton RGS X-ray Spectroscopy} \label{sec:xmm_rgs}
X-ray spectra were obtained using the Reflection Grating Spectrometer (RGS), which consists of two identical spectrographs, RGS1 and RGS2, in order to increase the signal-to-noise ratios. 
Data are collected over the energy range $0.33-2.5$ keV (spectral resolving power ($E/\Delta E$) between 150 and 800) in the ``Spectroscopy'' mode with submode ``HighEventRateWithSES'' and observation mode ``Pointed''.
Pipeline Processing System (PPS; version 17.56\_20190403\_1200) data is supplied, which uses standard Science Analysis System (SAS; version xmmsas\_20190401\_1820-18.0.0) tasks to produce data products like light curves and average spectra \citep{sas_paper}. 
We use the PPS light curves for the four \xmm{} observations, which have time bins of 30, 32, 36, and 27 seconds, per Observation Window.

\subsubsection{Swift UVOT W2 Photometry} \label{sec:swift_uvw2}
The Ultraviolet Optical Telescope \citep[UVOT,][]{Roming2005} was employed on \swift{} using the W2 filter.  Compared to the \xmm{} UVW2 filter (see Section \ref{sec:xmm_om}), the \swift{} W2 is wider but much bluer, with an effective wavelength that is shorter by about 200 \AA{} (see Figure \ref{fig:filters}). AstroImageJ was also used to select the source aperture and background annulus. On the first day, the aperture radius was $5\farcs52$ and the annulus radii were $11\farcs54$ and $35\farcs13$ (reported pixel:arcsec of 8.07:4.05). On the second day, $6\farcs02$ and $10\farcs53-35\farcs11$ (reported pixel:arcsec of 6.36:3.19) were used, respectively. Light curves with time bins of 5, 10, and 30 seconds were then extracted using XSELECT. Again, no obvious flares were found in the \swift{} data, and no data points exceeded $2\sigma$ above the mean.  Two observation windows start with a decreasing rate trend, but there are no corresponding enhancements in the \swift{} XRT X-rays to suggest that these correspond to the tail ends of flares.

\subsubsection{XMM-Newton OM UVW2 Photometry} \label{sec:xmm_om}
We obtained UVW2 photometry with the \xmm{} Optical Monitor (OM) in FAST mode using on-board binning of 10~s.   
Count rates from the PPS TIMESR files were extracted and corrected from Terrestrial Time (TT) to the UTC at the middle of the exposure.    
The UVW2 filter is the bluest of the OM filters and was chosen to build upon the findings from a superflare in \citet{Osten2016} that shows unexpected results from model comparisons to the V band (see their Fig.\ 8).
A more detailed discussion will be presented in Paper II of this study (Kowalski et al. 2023, \emph{in prep}.).

The UVW2 band has an effective wavelength of 2120 \AA{} (for white dwarf stars) and is very broad with a low-level tail which extends to redder wavelengths at $\lambda  \approx 3000 - 7000$ \AA{}. This tail accounts for less than 1\% of the total effective area (see Figure \ref{fig:filters}).  
In quiescence, the count rate of AU Mic is 1.8 cps, and $\sim$50\% is expected to be due to this red-tail emission from wavelengths $\lambda > 4000$ \AA{}.  
For hot stars, the contribution from $\lambda>$ 3,000 \AA{} is on the order of 1\% \citep{Talavera2011}; this red emission is not a concerning source of systematic uncertainty in our analyses because flares are very blue.  
Robust model comparisons are readily achieved by synthesizing flux densities from the full wavelength coverage of the UVW2 effective area curve.

Large coincidence losses occur when the photon arrival rate is greater than the frame rate of the OM \citep{Fordham2000, Page2013}.  Coincidence loss corrections are applied to data with count rates less than 0.97 counts per frametime and are accurate to better than 2\%. 
The largest count rate that occurs in flaring intervals is 105 counts s$^{-1}$, which corresponds to 0.53 counts per frametime of 5 ms\footnote{See \url{https://xmm-tools.cosmos.esa.int/external/xmm_user_support/documentation/uhb/omlimits.html};  we confirmed with OM calibration scientist R. Riestra (priv.\ communication, 2019) that the largest count rate is below the extreme coincidence-loss regime and constitute a clean observation. }.   
FAST mode requires that a very small ($10\times10$\arcsec) window around the star is employed for onboard windowing.  
The pointing variation of \xmm{} among observations caused the star to fall on the edge of this window during the third observation, so there is no UVW2 photometry from 2018 Oct 14 18:48:06.449 to 2018 Oct 15 10:35:09.418 UTC.

\subsubsection{Las Cumbres Observatory Global Telescope \emph{U}- and \emph{V}-band Photometry} \label{sec:lco}
AU Mic was observed in the $U$ and $V$ bands by the Las Cumbres Observatory Global Telescope (LCOGT) network between 2018 Oct 13 and 2018 Oct 29, thus extending the observational campaign past the end of the \xmm{} observations.  
For both bands (see Figure \ref{fig:filters}), we used the reduced images from the automatic \texttt{BANZAI} pipeline, which masks bad pixels, applies an astrometric solution, and performs bias \& dark subtraction.  
Light cuves were extracted using the supplied Kron apertures, which were verified by comparing to standard IRAF procedures and AstroImageJ. 
Among the aperture photometry methods, the Kron aperture was found to exhibit the best signal-to-noise ratio during flares and quiescence.

Bessel $U$-band observations were obtained with the 1m telescopes, with an exposure time of 4 seconds and a cadence of 46 seconds.  
Bessel $V$-band observations were obtained with the 0.4m telescopes within the LCOGT network, with an exposure time and cadence of 2 seconds and 25 seconds, respectively. 
In the $U$-band, the A9V star HD 197673 was used as a comparison star for the relative photometry. 
In the $V$ band, both HD 197673 and an inactive field star, Gaia DR2 6793990654220996352, were used for the relative photometry.

\subsubsection{SMARTS \emph{V}-band Photometry} \label{sec:smarts_v}
AU Mic was observed by the 0.9m telescope at the Cerro Tololo Inter-American Observatory, operated by the SMARTS Consortium (Small and Moderate Aperture Telescope Research System), between 2018 Oct 10 and 2018 Oct 17. 
SMARTS is operated by the REsearch Consortium On Nearby Stars (RECONS), which aims to discover and characterize stars in the solar neighborhood \citep[][]{Henry2018, Vrijmoet2020}.
The exposure time was 15 seconds, and a cadence of 47 seconds is achieved using the V Tek \#2 filter (CTIO/CTIO.5438-1026 on the SVO Filter Database) for these observations.  
The LCOGT $V$ bandpass is slightly bluer than the CTIO/SMARTS 0.9m $V$ band, which is closer to the standard Bessel \emph{V} filter transmission \citep{Bessel2013}. 
Aperture photometry of AU Mic and the background stars Gaia DR2 6794048928337372800 and Gaia DR2 6794044358492150912 is extracted using AstroImageJ. 
A large, 20 pixel aperture is chosen to collect all the flux from the defocused point-spread functions of AU Mic and a sum of nearby comparison stars. 
Defocusing results in $\approx3$ million counts per exposure from AU Mic and a standard deviation of 5 mmag during quiescent times.  
Notably, the $V$-band photometry from the CTIO/SMARTS 0.9m is much higher precision in quiescence than the LCOGT $V$-band.

\subsubsection{SMARTS CHIRON Optical Spectra} \label{sec:smarts_halpha}
We obtained optical spectra from the cross-dispersed, fiber-fed echelle CTIO High ResolutiON (CHIRON) spectrograph \citep{2013PASP..125.1336T} at the CTIO/SMARTS 1.5m with 60 second integration times.  
The usable wavelength range and mean spectral resolving power of our CHIRON data are $\lambda = 4500-8900$ \AA{} and $R=\lambda / \Delta \lambda \approx$ 25,000,
respectively. 
The data were wavelength calibrated using a ThAr lamp, resulting in a resolving power of $R \sim$ 28,000 around H$\alpha$.
In particular, we look at the fluctuations of the equivalent width of the H$\alpha$ line during flaring events.
These observations also extend past the seven-day duration of the \xmm{} observations and will be analyzed in detail in Notsu et al. 2023, \emph{in prep}.

\subsection{Absolute Flux and Energy Calibration of UVW2, U, V, and X-ray Photometry}\label{sec:fluxcalc}

\begin{figure}[!ht]
\centering
\includegraphics[width=0.9\linewidth]{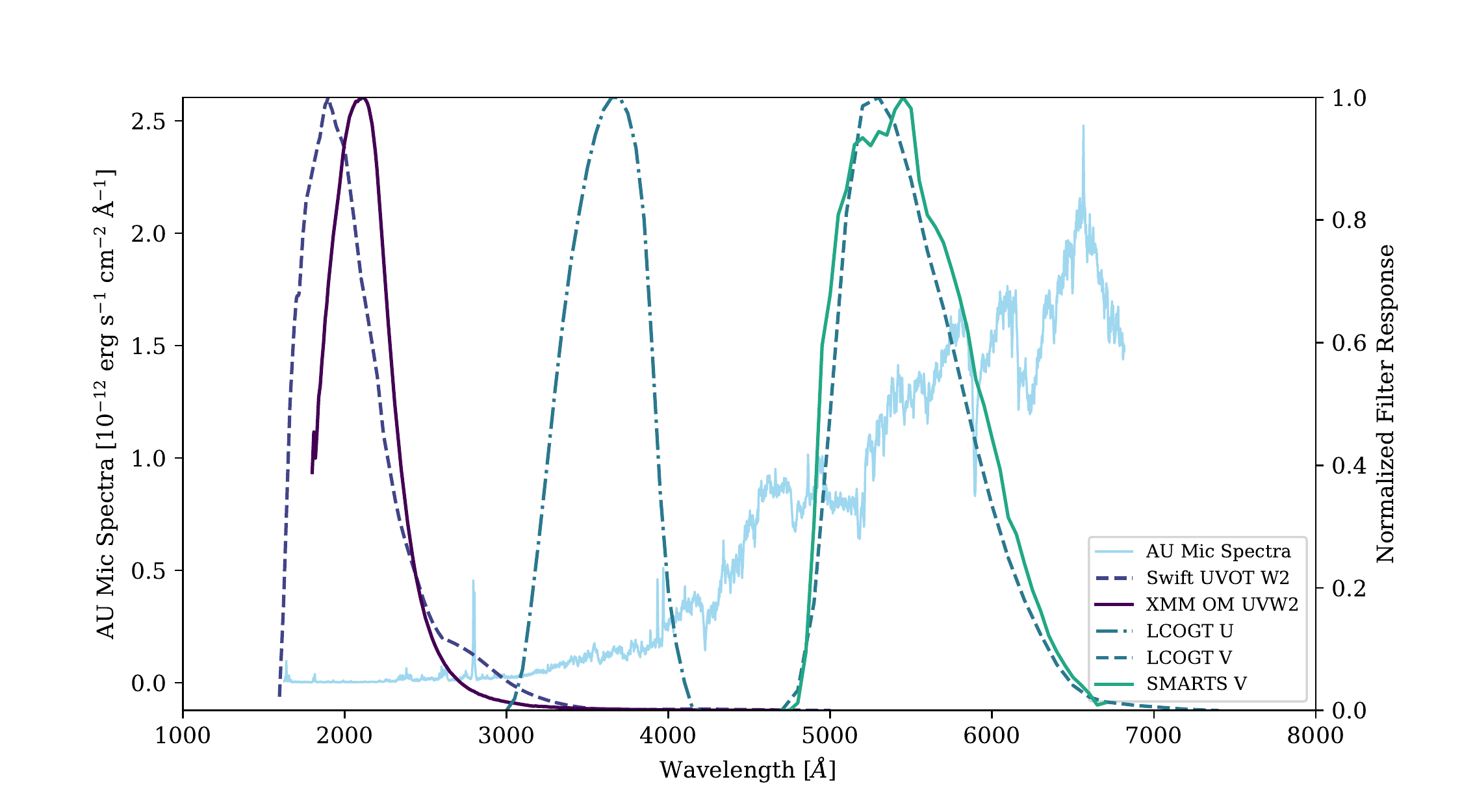}
\caption{Effective area or filter transmission curves (photon weighting) for the broadband filters used in this study are plotted against the quiescent HST/FOS spectrum of AU Mic. All filter responses are normalized from 0 to 1.
\label{fig:filters}
}
\end{figure}

Absolute flux calibration of the photometry is critical for addressing several of the goals of our observational campaign. 
In this section, we summarize our methods, which are described in detail in Appendix \ref{sec:appendix_fluxcalc}.
Note that all of the broadband filters used are shown against the AU Mic spectrum from the HST/FOS \citep[FOS/RD gratings G190H/2300\AA{}, G270H/2650\AA{}, G400H/3600\AA{}, and G570H/4600\AA{} observed on 1991 Sep 11, as used in][available on the MAST archive\footnote{\dataset[DOI: 10.17909/6pe3-yp69]{https://doi.org/10.17909/6pe3-yp69}}]{Augereau2006} in Figure \ref{fig:filters}.

We calculated energies (fluences) using the equivalent duration \citep{Gershberg1972}, defined as
\begin{equation} \label{eq:eqdur}
ED = \int{\frac{I - I_q}{I_q}} \,dt
\end{equation}
\noindent where $I$ is the count rate (calibrated or relative) and $q$ indicates quiescence. The ED is then multiplied by the bandpass ($T$) quiescent luminosity ($L_{q,T}$) to find the bandpass-integrated flare energy, $E_{T} = ED \times L_{q,T}$.

Flux-calibrated light curves are thus obtained by knowledge of the quiescent fluxes.  For known and well-characterized bandpasses, these can be determined using zeropoints \citep[as summarized in][]{Willmer2018} and published apparent magnitudes at low-levels of flare activity ($U$, $V$) or count-rate conversions ($UVW2$).   
We compare the zeropoint method to numerical integration of the observed spectrum of AU Mic from HST/FOS over the bandpass, according to the equation 

\begin{equation} \label{eq:fluxconv}
L_{q,T} = \langle f_{q,\lambda} \rangle_T \, \Delta \lambda_{\text{FWHM}}\, 4 \pi d^2  = \frac{\int{T(\lambda) f_{q,\lambda} (\lambda) \lambda} \,d\lambda}{\int{T (\lambda) \lambda} \,d\lambda} \Delta \lambda_{\text{FWHM}}\, 4 \pi d^2,
\end{equation} 

\noindent where $\langle {f_{q,\lambda}} \rangle_T $ is the system-weighted flux \citep{Sirianni2005}, T($\lambda$) is the total system response or effective area including the bandpass and atmosphere (if applicable), $f_{q,\lambda}(\lambda$) is the quiescent HST/FOS spectrum at Earth in units of \funits{}, $\Delta \lambda_{\text{FWHM}}$ is the Full-Width Half-Max (FWHM) of the bandpass, and $d$ is the distance to AU Mic.  For the $U$, $V$, and $UVW2$ bands, the systematic uncertainties of the flux calibration of the quiescent fluxes are $\approx 10$\%.  The adopted quiescent fluxes, $\langle f_{q,\lambda} \rangle_T$, are thus $1.0 \pm 0.1 \times 10^{-14}$ \funits{} ($UVW2$),  $1.3 \pm 0.1 \times 10^{-13}$ \funits{} ($U$), and $1.2 \pm0.1 \times 10^{-12}$ \funits{} ($V$), also summarized in Table \ref{tbl:flux_num}.

Flux calibration for \swift{} data is not essential, because the lack of any flares only adds monitoring time for better statistics on the flare rates (as in Section \ref{sec:flare_rate}). 
Note that between the \xmm{} and \swift{} instruments, the EPIC-pn X-ray sensitivity is higher than the XRT\footnote{\url{https://xmm-tools.cosmos.esa.int/external/xmm_user_support/documentation/uhb/xmmcomp.html}}, while the UVOT UVW2 sensitivity is higher than that of the OM\footnote{\url{https://swift.gsfc.nasa.gov/about_swift/uvot_desc.html}}.

The energies in the EPIC-pn bandpass were calculated by converting the data from cps to erg cm$^{-2}$ s$^{-1}$ using the online WebPIMMS tool\footnote{\url{https://heasarc.gsfc.nasa.gov/cgi-bin/Tools/w3pimms/w3pimms.pl}} \citep{1993Legac...3...21M}, integrating over the duration of the flare, and multiplying by $4 \pi d^2$.    
Specifically, the data were converted from XMM/PN Med Count Rate 5' region to FLUX, with a default input energy range and $0.4 - 10$ keV energy output range, N$_{\text{H}}$ of $2.29 \times 10^{18}$ cm$^{-2}$ \citep{Wood2005}, and temperatures ranging from $10^7 - 3\times10^7$ K ($k_B T = 0.86 - 2.59$ keV). 
The adopted quiescent X-ray bandpass flux is thus $2.6 \pm 0.3 \times 10^{-11}$ erg cm$^{-2}$ s$^{-1}$. The peak X-ray fluxes of flares at a distance of AU Mic's habitable zone were calculated by normalizing to a distance of 0.3 AU \citep[based on Fig.\ 7 of][]{Kopparapu2013}. X-ray time derivatives were also calculated using the 30-second bin EPIC-pn light curve. As noise was high at many times, we calculated X-ray derivative peak timings for flares by first convolving the light curve with a 15-point box function, taking the derivative of this new curve, and then applying a Scipy \citep{SciPy} UnivariateSpline fit to smooth sharp variations.

For comparison, we use the average X-ray luminosity from \citet{Mitra2005} derived from fitting the EPIC-pn spectrum ranging 0.2 to 12 keV with a 3-temperature collisional ionization equilibrium model. 
Their modeling returned a value of $\bar{L}_x = 2.99 \times 10^{29}$ ergs s$^{-1}$, and they estimate that this luminosity scales linear with the XMM EPIC-pn X-ray light curve count rate.
The non-averaged luminosity is then $L_x = \bar{L}_x/\bar{c}_x \times c_x$, where $c_x$ is count rate and $\bar{c}_x$ is the average count rate of the entire observation. 
This gives a quiescent luminosity of $2.7 \pm 0.3 \times 10^{29}$ erg s$^{-1}$. 
This is about 6\% different than our estimated value from WebPIMMS and is within the uncertainty. 
Using this luminosity for XMM EPIC-pn X-ray Flare 23 (see Section \ref{sec:lc}), we calculate an energy 6.5\% lower than the WebPIMMS estimation. 
We note that these energies are still preliminary as actual X-ray energies will need to account for changing temperatures throughout the flare, and future work will focus on the spectral analysis necessary for this.

\clearpage
\section{The Flare Sample} \label{sec:lc}

    All light curves produced from the observations in Section \ref{sec:obs} are shown in Figure \ref{fig:lc_all}.
    Of interest, we determine the quiescent flux in various bandpasses to be 
    $1.8\pm0.5$ cps ($4.9\pm1.4 \times 10^{-12}$ erg cm$^{-2}$ s$^{-1}$; XMM OM UVW2), 
    $15.3\pm1.6$ cps ($2.6\pm0.3 \times 10^{-11}$ erg cm$^{-2}$ s$^{-1}$; XMM EPIC-pn X-ray), 
    and $9.1\pm0.7 \times 10^{-11}$ erg cm$^{-2}$ s$^{-1}$ (LCOGT U). 
    For context outside of this campaign, \cite{Iwakiri2020} measured a bright AU Mic X-ray flare flux of $2.9^{+0.7}_{-1.2} \times 10^{-9}$ erg cm$^{-2}$ s$^{-1}$ during Apr 2020, and \citet{Kohara2021} measured an X-ray flare flux of $9.0^{+1.7}_{-1.6} \times 10^{-9}$ erg cm$^{-2}$ s$^{-1}$ during Dec 2021 using the MAXI/GSC nova alert system (2 -- 20 keV band).

\subsection{Flare Detection Method \label{sec:flare_detection}}
    
    We determined flare occurrence in each light curve as follows. 
    The start and stop times of each flare were determined by eye from the light curves. 
    The quiet times around each flare were are used to define a standard deviation and estimate the quiescent emission throughout the flare duration using a Scipy Univariate spline interpolation, which was often a linear fit.
    Note that nearby flares are removed before the fit is made, when necessary.
    Flare bounds are then refined based on an initial measure of when the flux is above $2\sigma$ of the quiescent.
    In total, we find 73 flares, with 34 XMM OM UVW2, 38 XMM EPIC-pn X-ray,  15 XMM RGS X-ray, 25 LCOGT \emph{U}-band, 2 LCOGT \emph{V}-band, 4 CTIO/SMARTS 0.9m \emph{V}-band, and 17 CHIRON H$\alpha$ events. Properties of each flare are listed in Appendix \ref{app:flares}, and flares shown in Figures \ref{fig:lc_xmm} through \ref{fig:fl_v}.
    Temporally coincident flares in different bandpasses are labeled with a common flare ID.

    A few notes are warranted.
    Complex flare events with several short-duration sub-peaks are identified as a single flare.  
    Several of the lower-amplitude complex flares are also counted as one when, over short durations, the flux momentarily decreased below 2$\sigma$ above the quiescent level (see XMM OM UVW2 Flares 19 and 31 in Figure \ref{fig:fl_uvw2}). 
    Also, the large X-ray swell (Flare 11, which appears from 37 to 56 hours after the start of 2018 Oct 10 in Figure \ref{fig:lc_xmm}) is not clearly correlated with any other single simultaneous response, so it is given its own ID.
    
\begin{figure}[!ht]
\centering
\includegraphics[width=1.0\linewidth]{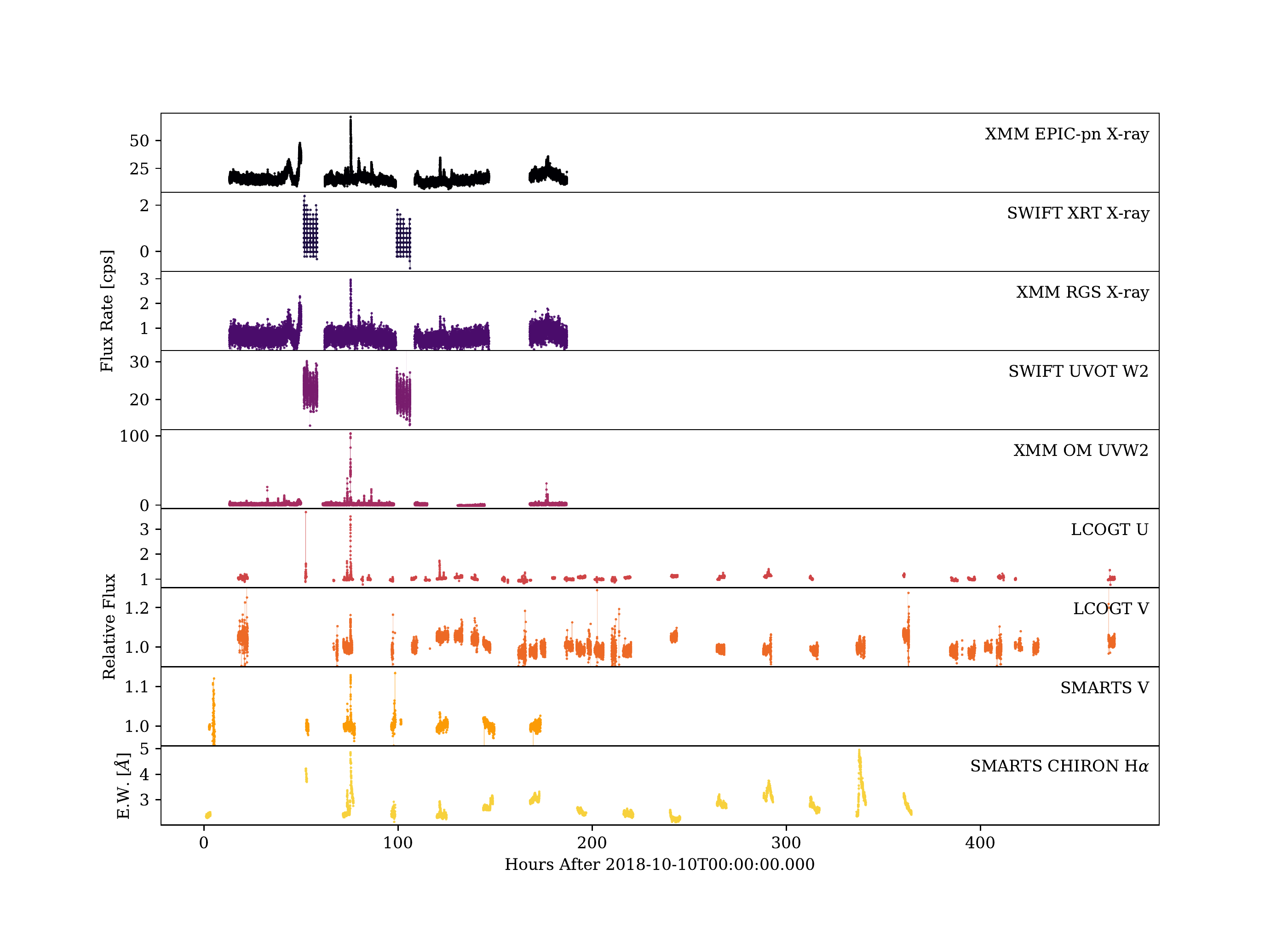}
\caption{Light curve summary of AU Mic during the flare campaign of October 2018. Data from space-based observatories (X-ray and UVW2) are in units of counts per second, while ground-based data (U and V) are in relative flux units. H$\alpha$ equivalent width (E.W.) data are in units of Angstroms.
\label{fig:lc_all}}
\end{figure}

    Flares with significant gaps in observation are left out of the sample. 
    Exceptions were made for XMM OM UVW2 Flares 23, 24, \& 47 (see Figure \ref{fig:fl_uvw2}) and LCOGT \emph{U}-band Flares 72 \& 73 (see Figure \ref{fig:fl_ulco}).
    Flare 24 is missing times early in the impulsive phase.
    It is unclear if there is an initial peak here, as the timing of the measured one lies within expectations.
    Flare 47 is missing times before the peak of an accompanying noisy, complex X-ray flare (see Figure \ref{fig:lc_all} for comparison with the XMM EPIC-pn light curve). 
    Thus, a linear extrapolation is added for each.
    For Flare 23, both linear and exponential decays were modeled for the missing times. 
    The linear extrapolation is used for consistency, with the small difference factored into the error.
    There is no surrounding quiescent area for Flares 72 and 73, and the decay of Flare 73 is incomplete.
    To compensate for this, a quiescent level of 0.95 with a standard deviation of 0.02 is chosen based on the areas lowest point and the standard deviation of a quiet area at a similar relative flux level.
    These 2 flare IDs are the only ones out of chronological order, as the reduction is significantly different.

    Any potential flares that could not pass through our pipeline (see Section \ref{sec:calc_quan}) are considered too small for analysis and left out.
    All flares have at least a few data points above $2\sigma$. 
    
    CHIRON H$\alpha$ flares are identified by eye based upon sudden variations against the gradually varying background flux, which may represent non-flare related changes (e.g.\ rotational modulations) in the stellar active regions \citep{Maehara2020}.
    Rather than spline fitting, we choose the equivalent width quiescent value to be the minimum of the immediate surrounding area (see Figure \ref{fig:fl_halp}).
    Further analysis of H$\alpha$ variations will be discussed in a future paper (Notsu et al., \emph{in prep.}).

\begin{figure}[!ht]
\centering
\includegraphics[width=1.0\linewidth]{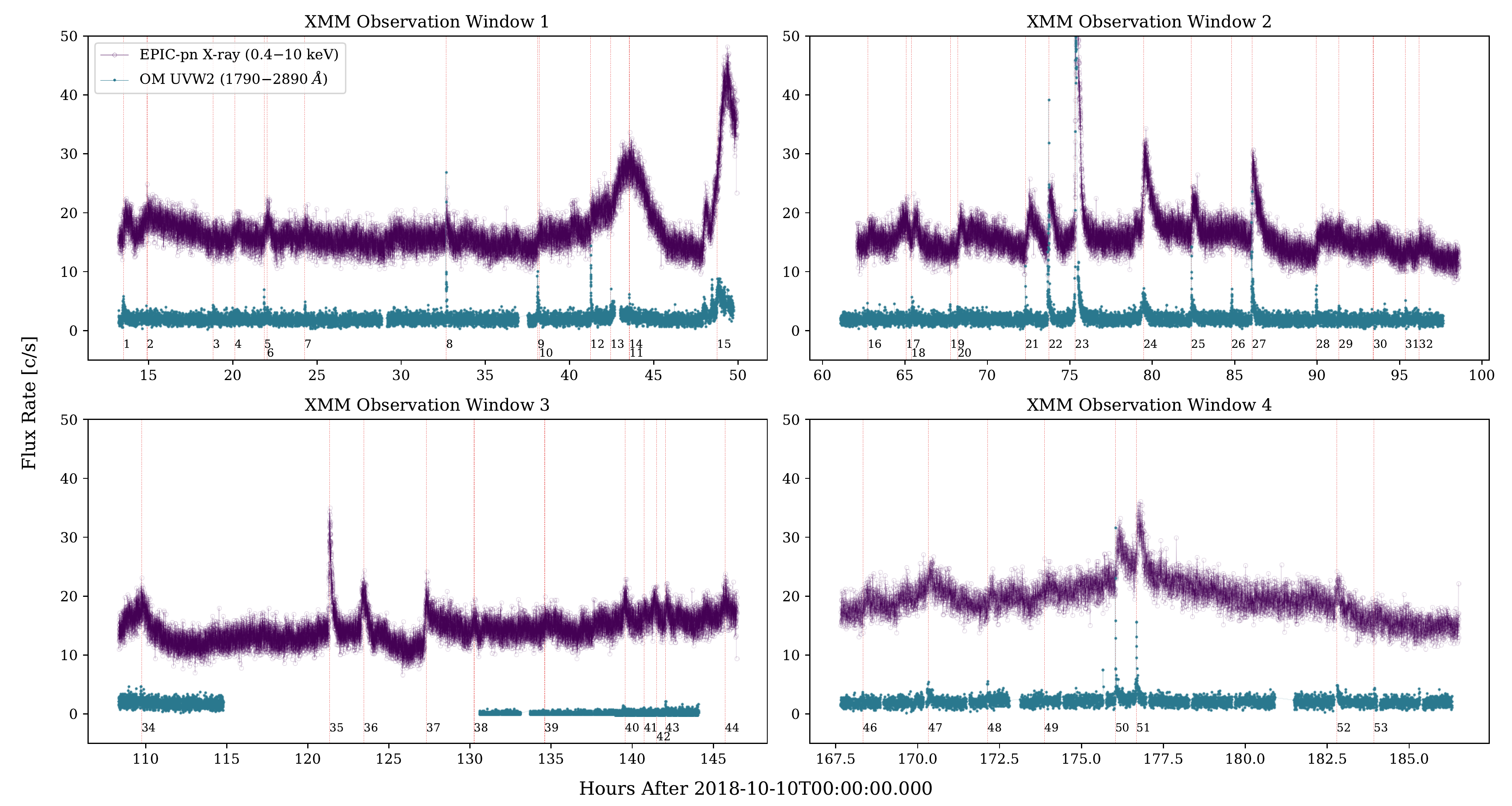}
\caption{Ten-second binned \xmm{} light curves are shown. Vertical red lines correspond to a flare peak, with the Flare ID labeled at the bottom.
\label{fig:lc_xmm}}
\end{figure}

\begin{figure}[!ht]
\centering
\includegraphics[width=1.0\linewidth]{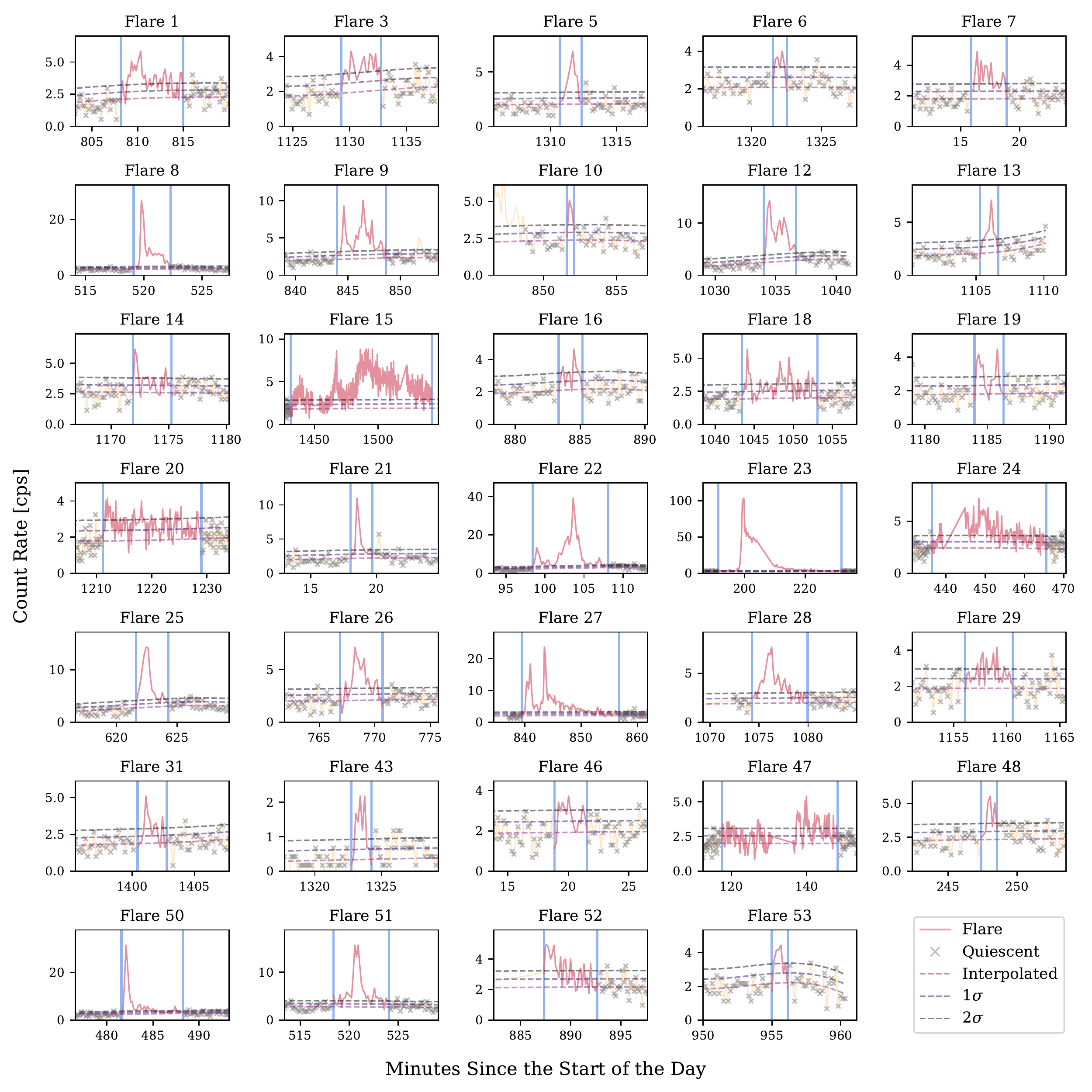}
\caption{Individual flares in the \xmm{} OM UVW2 light curve with ten-second binning.
The solid yellow line outlines the light curve, the solid red line shows flaring times, and gray crosses show times used to calculate the local quiescent level. 
The dashed lines show the estimated quiescent level, as well as the $1\sigma$ and $2\sigma$ local uncertainty levels. 
The vertical solid blue lines mark the chosen beginning and end times of each flare. 
The ID for each flare is listed above its plot.
\label{fig:fl_uvw2} }
\end{figure}

\begin{figure}[!ht]
\centering
\includegraphics[width=0.8\linewidth]{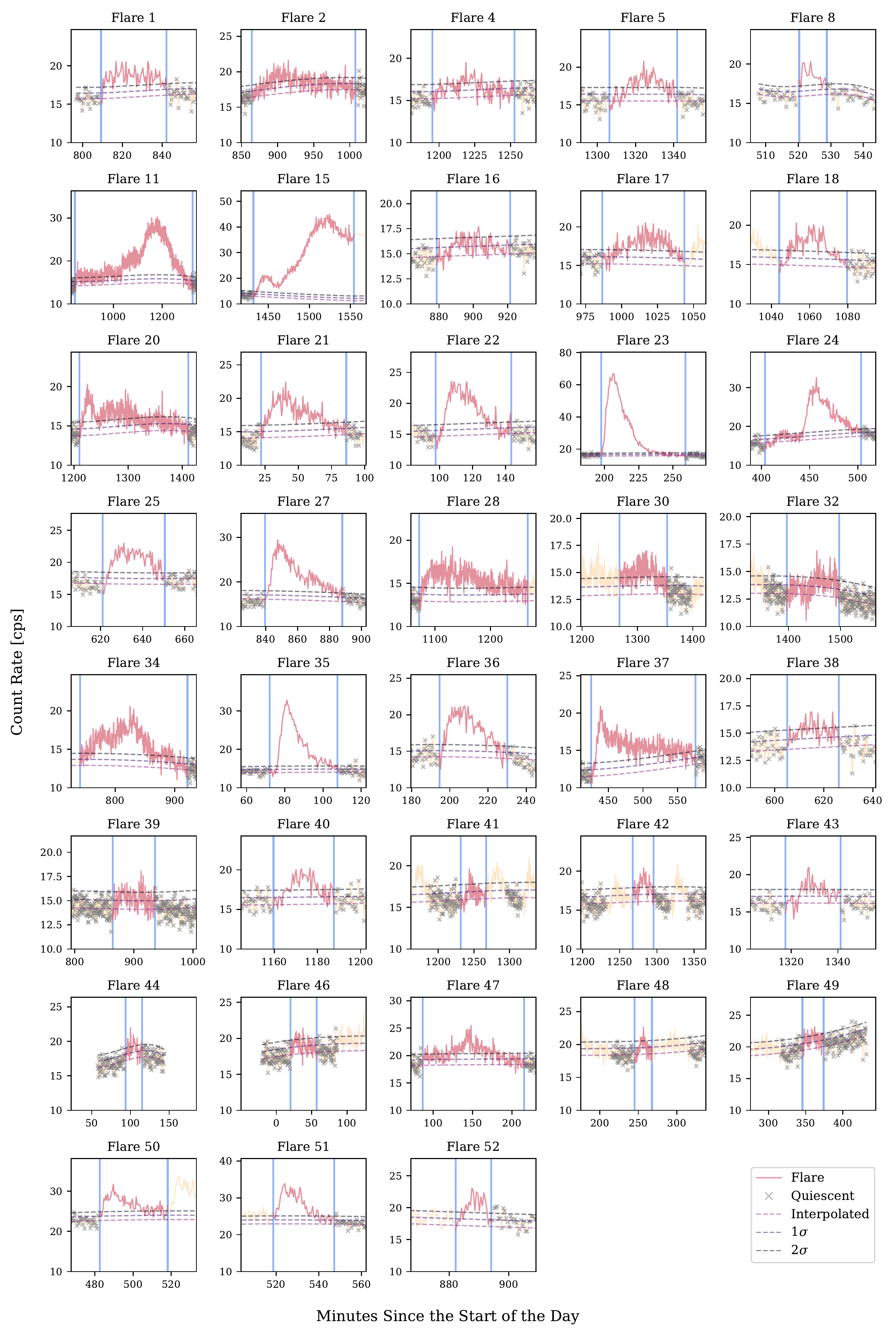}
\caption{Individual flares in the \xmm{} EPIC-pn X-ray light curve with 30-second binning for clarity. Symbols are in the same format as Figure \ref{fig:fl_uvw2}. See Figure \ref{fig:lc_xmm} for ten-second binned data, which were used for the analysis.
\label{fig:fl_xray}}
\end{figure}

\begin{figure}[!ht]
\centering
\includegraphics[width=1.0\linewidth]{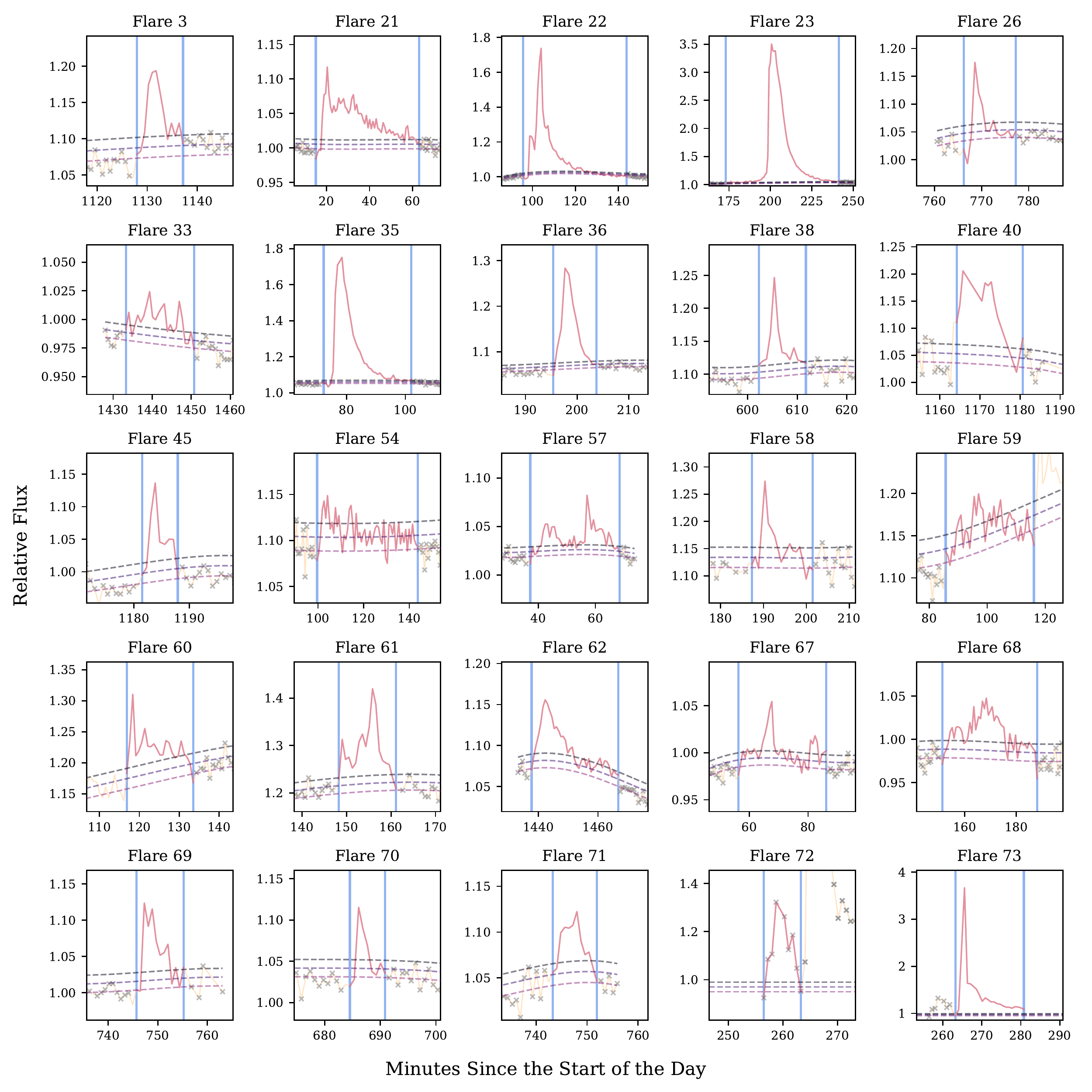}
\caption{Individual flares in the LCOGT \emph{U} band light curve.  
Symbols are in the same format as Figure \ref{fig:fl_uvw2}. 
\label{fig:fl_ulco}}
\end{figure}

\begin{figure}[!ht]
\centering
\includegraphics[width=1.0\linewidth]{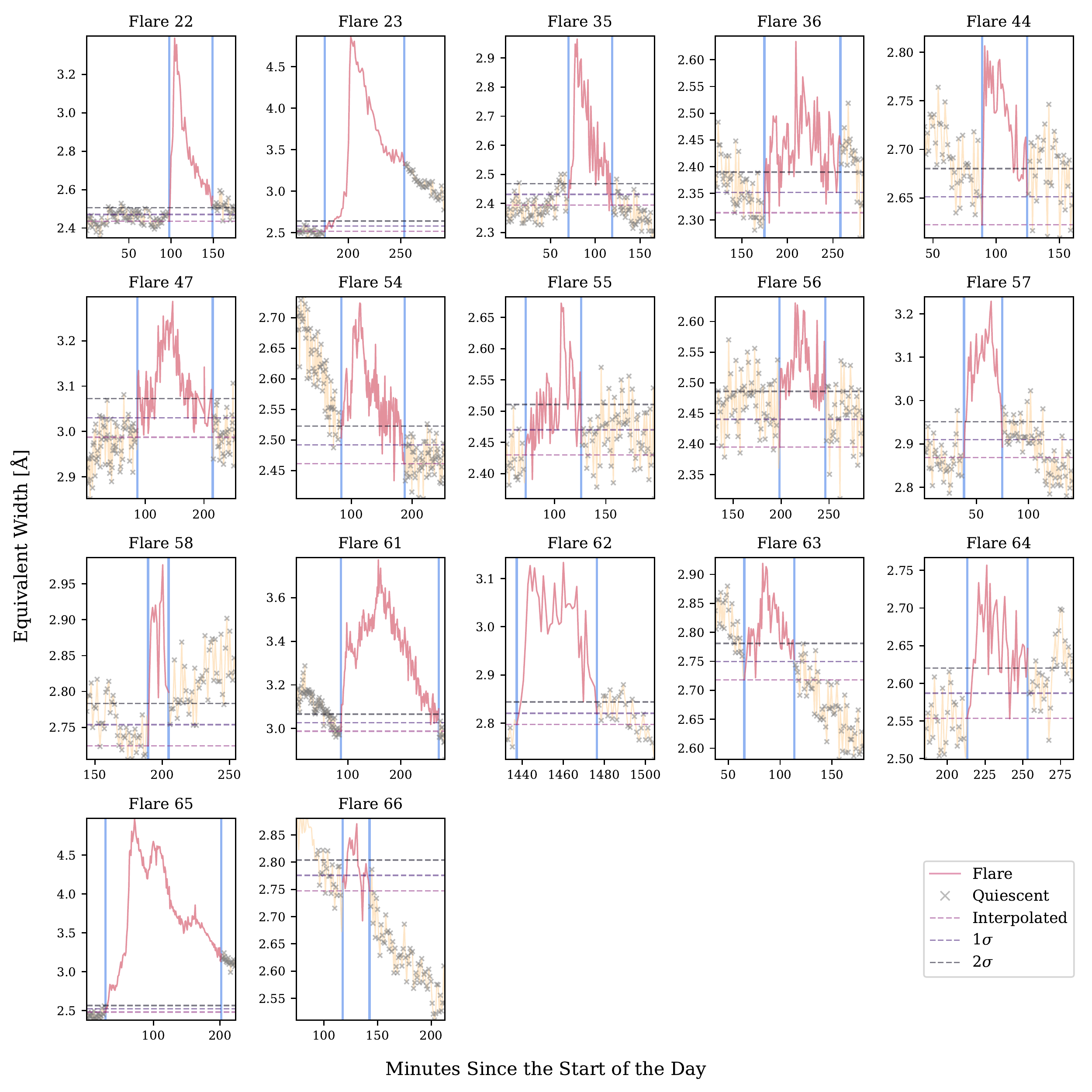}
\caption{Individual flares for the CHIRON H$\alpha$ equivalent widths.  Diagrams are in the same format as Figure \ref{fig:fl_uvw2}. 
\label{fig:fl_halp}}
\end{figure}

\begin{figure}[!ht]
\centering
\includegraphics[width=1.0\linewidth]{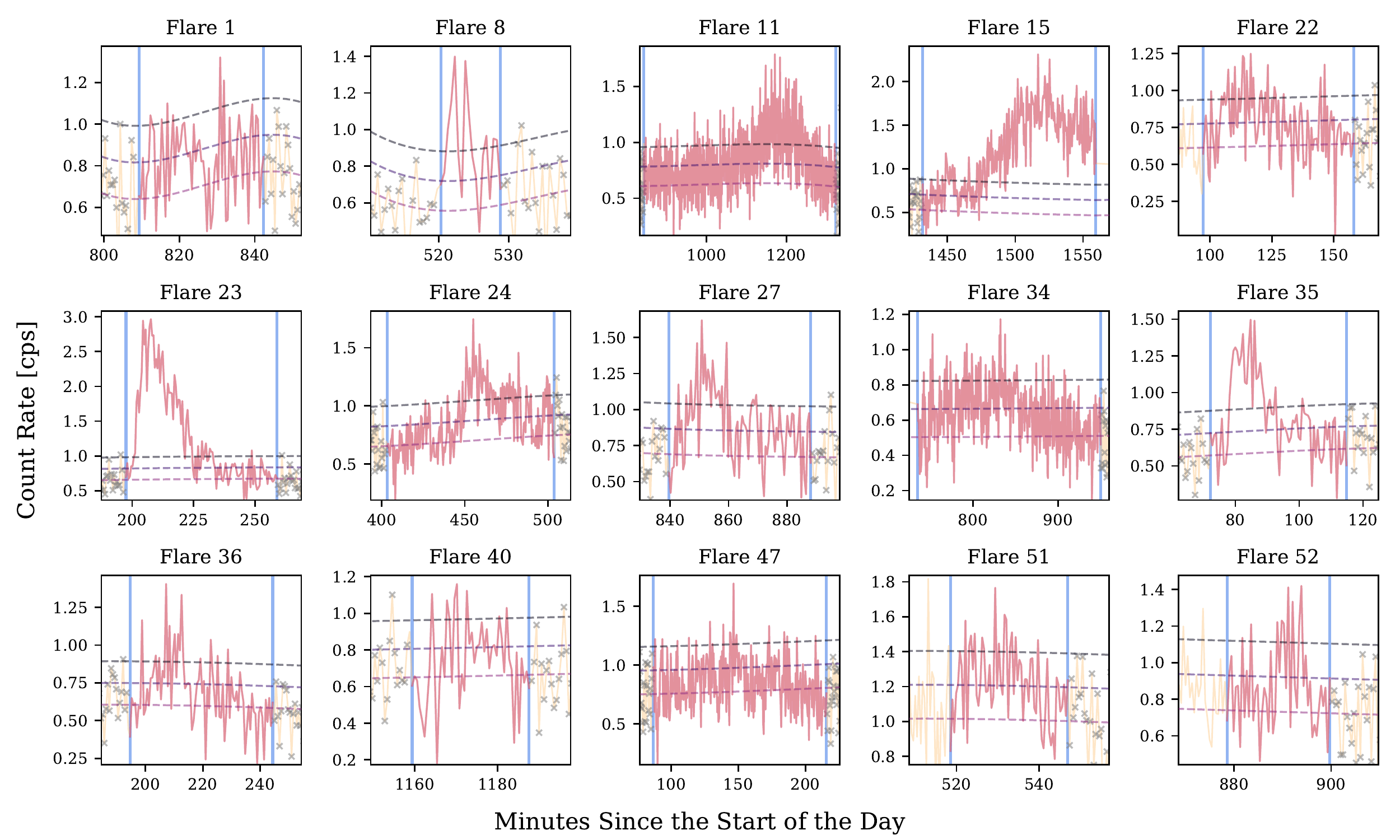}
\caption{Individual flares in the \xmm{} RGS X-ray light curve.  Timing for these flares are adopted from their EPIC-pn counterparts. Symbols and colors are in the same format as Figure \ref{fig:fl_uvw2}. 
\label{fig:fl_rgs}}
\end{figure}

\begin{figure}[!ht]
\centering
\plottwo{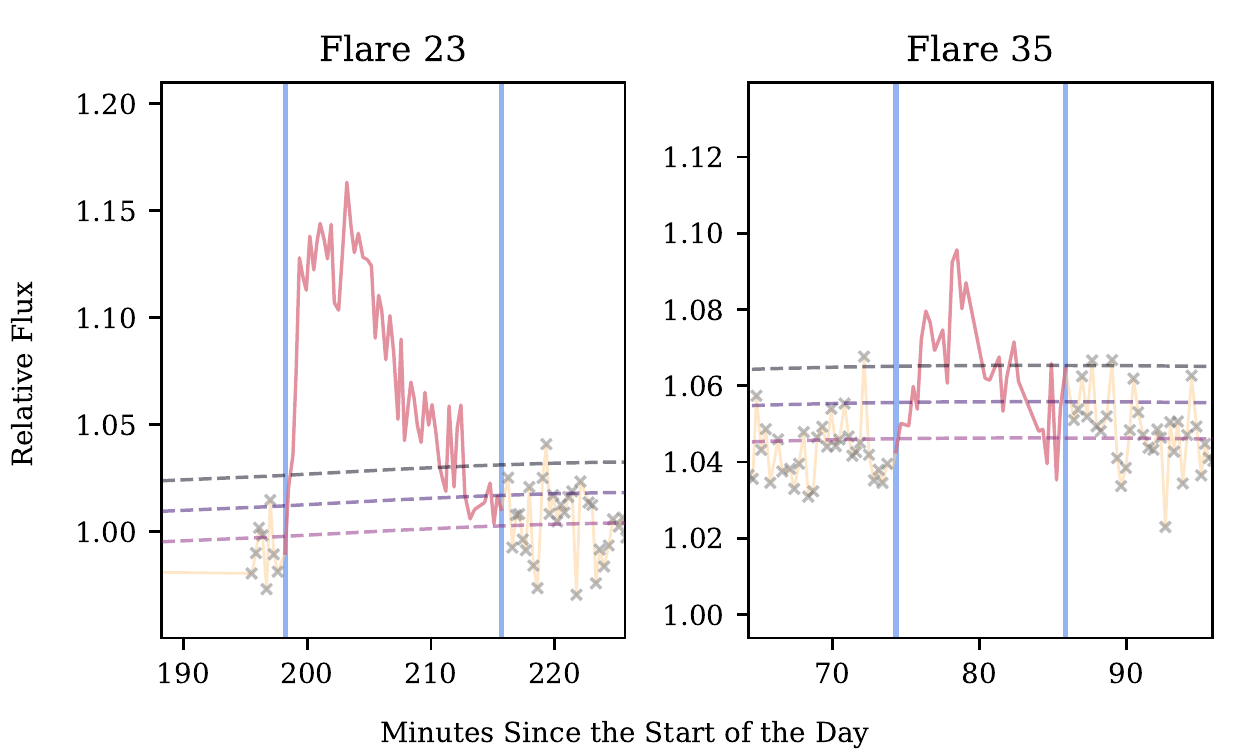}{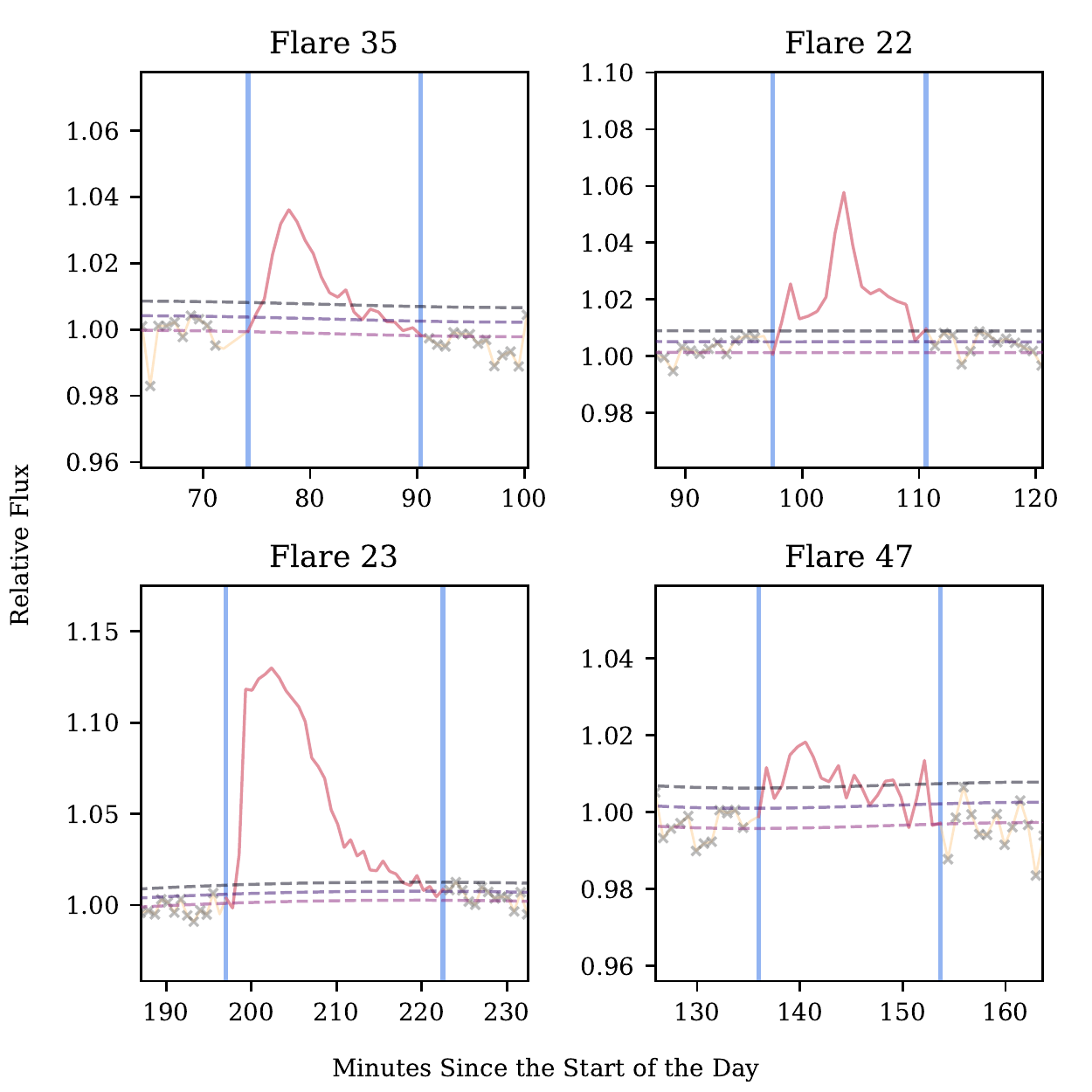}
\caption{Individual flares in the LCOGT \emph{V} band (\emph{left}) and CTIO/SMARTS 0.9m \emph{V} band (\emph{right}) light curves.  Diagrams are in the same format as Figure \ref{fig:fl_uvw2}. 
\label{fig:fl_v}}
\end{figure}

\clearpage

\section{Analysis and Results \label{sec:methods}}

\subsection{Calculated Quantities} \label{sec:calc_quan}

    From the quiescent-subtracted light curves (see Section \ref{sec:flare_detection}), the rise times, peak amplitudes, total durations, $t_{1/2}$ values (FWHM durations), impulsiveness indices, and equivalent durations are calculated.
    Note that we follow \citet{Kowalski2013} and define the impulsiveness index as $\mathcal{I} = I_{f, \text{peak}}/t_{1/2}$, where $I_{f}(t)$ is the intensity contrast, counts$_{\text{target}}$/counts$_{q}-1$, and $t_{1/2}$ is in minutes.
    Equivalent durations are converted to fluences (energies) following Section \ref{sec:fluxcalc}.  
    These quantities are listed between Table \ref{tbl:neupert_quantities} and Appendix \ref{app:flares}.  
    
    For energy calculations, we use a Monte Carlo simulation that varies the flare count rate by $\pm1\sigma$ for every datapoint. 
    We use 1,000 iterations to produce consistent results between different runs.
    Reported values and errors are the mean and standard deviation of these trials. 
    Reported errors for the XMM EPIC-pn X-ray are based on the difference between temperature values chosen in WebPIMMS (see Section \ref{sec:fluxcalc}), as these dominate over the statistical rate uncertainty.

\subsection{Flare Rate Analysis} \label{sec:flare_rate}
\begin{figure}[!ht]
\centering
\includegraphics[width=0.6\linewidth]{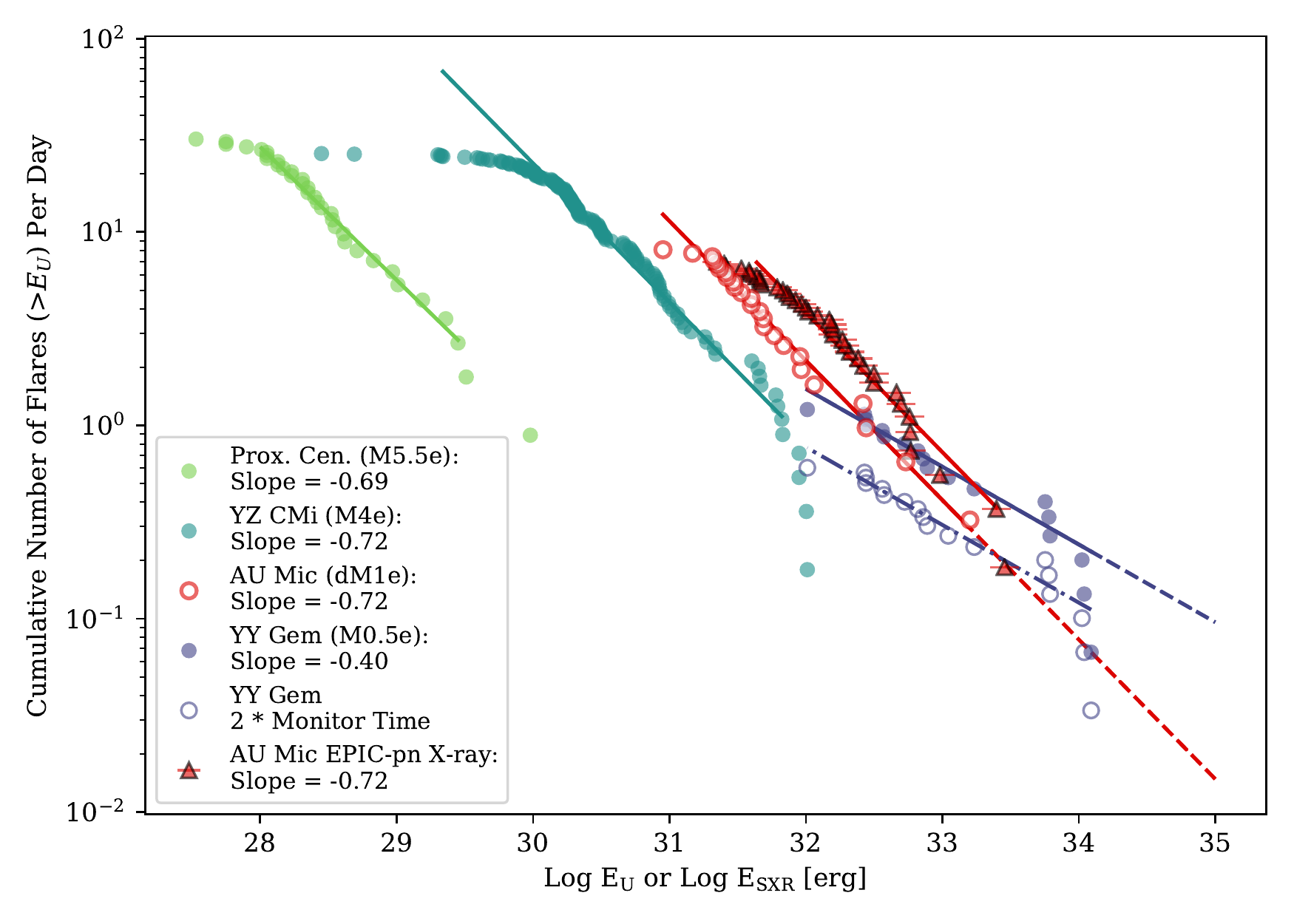}
\caption{AU Mic's LCOGT \emph{U}-band and XMM EPIC-pn X-ray cumulative flare frequency distribution compared to other U-band FFDs from active dMe's reported in the literature (Section \ref{sec:flare_rate}). 
We did not record any flares in the extreme high-energy regime where the slope steepens sharply, and only one in the lower regime where the slope stays roughly constant. 
We note the vertical shift if YY Gem's monitoring time is doubled, accounting for its status as a binary system \citep{Lacy1976}. 
\label{fig:cffd}}
\end{figure}

The $U$-band energies are organized into a cumulative flare frequency distribution (FFD) by calculating the number of flares with equivalent duration greater than $E$, $N = \int_E^\infty n(E)\,dE$ and dividing by the total monitoring time (74.15 hr).  
This FFD is shown in Figure \ref{fig:cffd}.  
To our knowledge this is the first systematic flare rate analysis of AU Mic in the $U$-band and there are few similarly detailed studies for other M0 -- M1 stars.

\begin{figure}[!ht]
\centering
\includegraphics[width=0.6\linewidth]{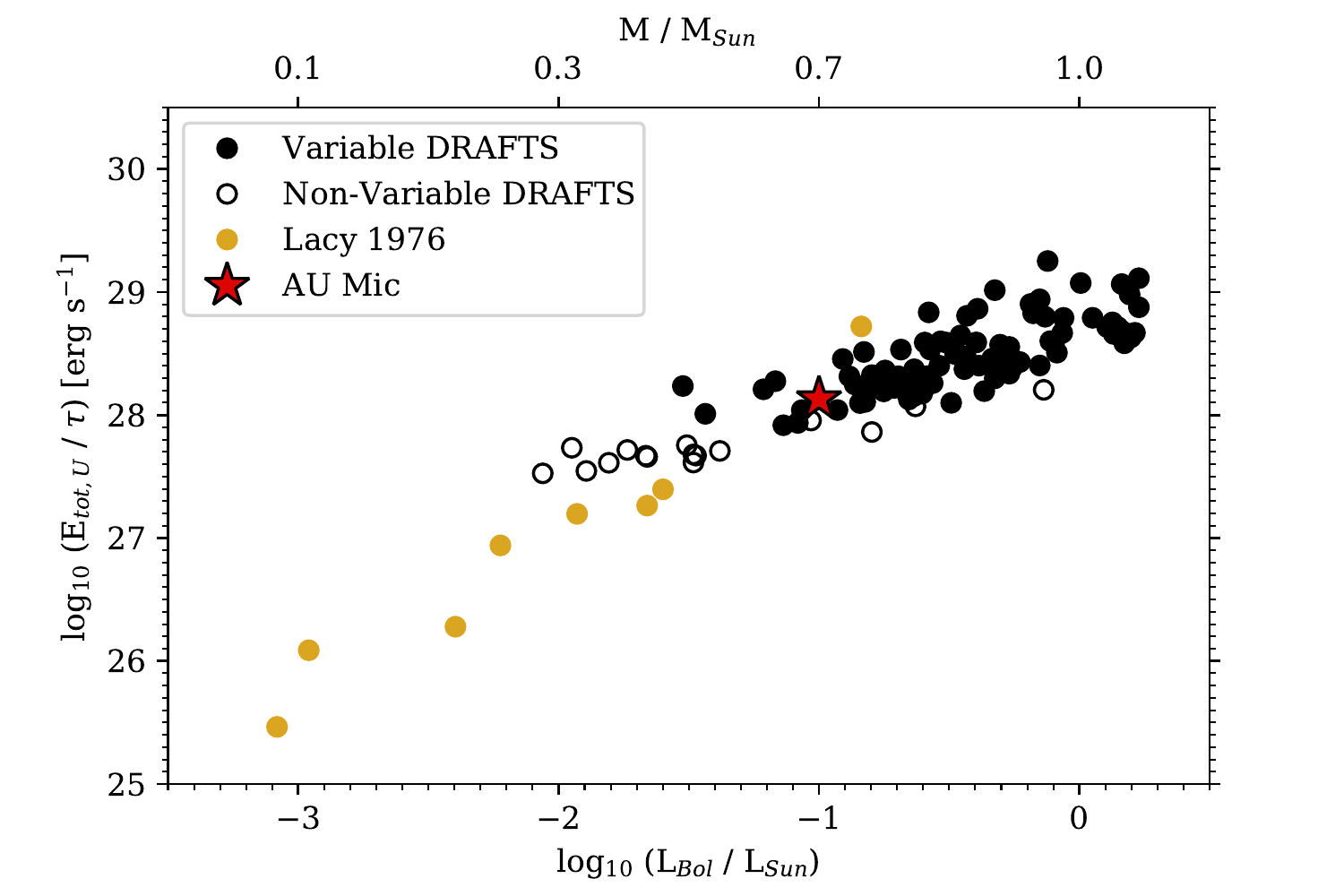}
\caption{We re-create Fig.\ 8 of \citet{Osten2012} showing the \emph{U}-band average energy loss due to flares compared to bolometric luminosity, with AU Mic. DRAFTS is the Deep Rapid Archival Flare Transient Search aimed at the older ($\sim10$ Gyr) flaring stellar population within the Galactic bulge. AU Mic is consistent with the slope extended from the other single M dwarfs and the apparent turnover with old solar-type stars from the Galactic bulge.
\label{fig:ubandeloss}
}
\end{figure}

From the 25 LCOGT $U$-band flares, we calculate an average energy of $\langle \log E_U \rangle = 31.74$ erg and an average energy loss rate of $\log \mathcal{L}_U = 28.13$ erg s$^{-1}$. 
Taking into account losses from 
The only other early-type M dwarf flare star system with similar quantities determined is the eclipsing binary YY Gem \citep{Lacy1976}.  
Compared to YY Gem, AU Mic has around a 18 times smaller average flare energy and a 4 times smaller energy loss rate due to flaring.  
We extrapolate flaring rates for AU Mic and YY Gem to estimate the occurrence of extreme flares. 
For $10^{34}$ erg flares, the expected number of flares greater than this energy (``flares $>E_U$'') is 0.078 and 0.242 flares $>E_U$ per day for AU Mic and YY Gem, respectively. 
For $10^{35}$ erg flares, we expect 0.015 and 0.096 flares $>E_U$ per day for AU Mic and YY Gem.
The FFD is fit to a power-law of form $ \log{\nu} = \alpha + \beta \log{E_U}$ (see Figure \ref{fig:cffd}), and the resulting values are $\alpha = 23.42 \pm 4.78$ and $\beta = -0.72 \pm 0.15$.  This value of $\beta$ is consistent with other M dwarfs from \citet{Lacy1976}, who reports values ranging from $-0.4$ to $-1.1$.  
We compare the FFD of AU Mic to that of a late-type M dwarf star Proxima Centauri \citep{Walker1981} which has a much lower average flare energy, as expected from the other later-type M dwarfs in \citet{Lacy1976}.   
The energy loss rate of AU Mic is compared to a wider range of M dwarfs in Figure \ref{fig:ubandeloss}, which compiles results from the much older 10 Gyr Galactic bulge population of flare stars reported in \citet{Osten2012}. 
We can further estimate the expected mean energy loss rate using $\alpha$ and $\beta$, along with estimated minimum and maximum flare energies, $E_{min}$ and $E_{max}$, to calculate 
$\mathcal{L}'$ following \citet{Lacy1976}.
$\mathcal{L}'$ accounts for numerous flares much below the observational detection limit.
However, given the calculated $\beta$ value and that most of the total flaring energy is contributed by the larger flares, lower $E_{min}$ values will not change the final value of $\mathcal{L}'$ significantly.
Assuming $E_{min} = 10^{30}$ erg and $E_{max} = 10^{35}$ erg, we find $\log \mathcal{L}' = 28.18$ erg s$^{-1}$.
This value is very similar to the value of $\log \mathcal{L}_U$ in Figure \ref{fig:ubandeloss}, as expected for early-type M stars \citep{Lacy1976}.

A preliminary X-ray FFD using XMM EPIC-pn flares is shown for comparison in Figure \ref{fig:cffd}.  
The slopes of the X-ray and $U$-band FFDs are remarkably similar, but the X-ray flares are shifted to larger energies. 
In fitting the slope, several X-ray flares in the lower energy regime and one in the high energy regime have been excluded. 
The slope is $-0.66$ if all flares are used, for reference.
\citet{Paudel2021} finds an FFD slope of $-0.65\pm0.19$ for NICER flares on EV Lac, which is consistent with our preliminary values of $\beta$ regardless of whether all flares or only flares from the middle energy regime are used for the FFD fitting.
However, other studies have found a generally wide range of X-ray FFD slopes for M dwarfs.
\citet{Audard2000} found slopes ranging from $-0.5$ to $-1.2$ and averaging around $-0.8$.
For reference, their value for EV Lac was $-0.76 \pm 0.33$. 
\citet{Kashyap2002} finds higher slopes around $-1.6$ for FK Aqr and V1054 Oph. 
In a larger study, \citet{Caramazza2007} found a cumulative slope of $-1.2 \pm 0.2$ for 165 low mass ($0.1 - 0.3 M_\sun$) stars from the Orion Nebula Cluster (ONC). 
Similarly, \citet{Stelzer2007} found slopes of $-1.4 \pm 0.5$ and $1.9 \pm 0.2$ for stars in Taurus and the ONC, respectively. 
For comparison, stars with masses above $1 M_\sun$ in the ONC and Cygnus OB2 have slopes around $-1.1 \pm 0.1$ \citep{Albacete2007}. 
Compared to these results, the slope derived from AU Mic is not as steep as many other M dwarfs.

Based on the energy budget from \citet{Osten2015}, we expect the $E_{\text{SXR}}/E_{U}$ ratio to be about 2.72. 
For the 6 flares with both X-ray and U-band responses (see Appendix \ref{app:flares}), this ratio is 1.5 on average. 
However, the ratio between the FFD power-laws at a given cumulative flare rate is 2.68, which is consistent with the previous budget.
This implies that while the energy budget varies from flare to flare, the average emission during a given time period (at a given cumulative rate) may lead to the previously established energy budget obtained from non-contemporaneous data or FFD comparisons.
This may be due in part to detection thresholds and the resulting flare durations between the $U$-band and X-rays, but warrants further investigation.  Specifically, we intend to explore the relations among coronal thermal energy and XEUV energies from temperature analysis of RGS spectra following \citet{Audard2000} and \citet{Pillitteri2022} in future work.  

In summary, the $U$-band flaring rate and FFD slope of AU Mic is consistent with other M dwarfs. However, the empirical energy relationships between the X-ray and U-band suggest that simultaneous observations of each flare give a different picture than energy conversions obtained from non-contemporaneous flare frequency distributions at different wavelength regimes.

\subsection{Light Curve and Flare Cross-Correlations \label{sec:crosscorrelation} }

We perform cross-correlations for XMM Observation Windows 2 \& 4 (refer to Figure \ref{fig:lc_xmm}) by following a modified version of the cross-correlation method in \citet{Mitra2005}. 
First, a time grid with 10 second spacing is defined on the XMM OM UVW2 light curve along times where observations overlap with the XMM EPIC-pn X-ray. Count rates are interpolated along this common time grid. Then, the XMM EPIC-pn X-ray light curve is shifted $\pm$3,000 seconds off from the original starting point, in steps of 10 seconds. 
For each shift, a sample Pearson correlation coefficient ($r_{\text{Lag}}$) is calculated between the overlapping times of the light curves. 
The time shift at $r^{\text{max}}_{\text{Lag}}$ represents the time when the light curves are most correlated, and the lags at 90\% of this value represents the $1\sigma$ uncertainty.
Window 2 gives $r_{\text{max}} = 0.70$ at a time lag of $-250_{-200}^{+150}$ seconds, where the X-ray lags behind the XMM OM UVW2 response. 
Window 4 gives $r_{\text{max}} = 0.32$ at $-200_{-370}^{+80}$ seconds. 
As the entirety of the $1\sigma$ region lies below 0 seconds for Windows 2 and 4, flares peak earlier in UVW2 band than in X-rays on average. 
This is consistent with findings of \citet{Mitra2005} and showcases that this relation holds in the presence of the X-ray's large scale variations in Window 4. 
For completeness, Window 1 reports $r_{\text{max}} = 0.575$ at a lag of $-780_{-1270}^{+780}$ seconds, but these are skewed due to the large complex X-ray flares. 
Window 3 is left out due to missing UVW2 data.

We perform the same cross-correlation for each individual flare with both \xmm{} OM UVW2 and EPIC-pn data.
The $\pm$3,000 second time shift used for the Observational Windows sometimes preferentially selects the largest nearby flare (e.g.\ Figure \ref{fig:neuperttests}).
Therefore, we explore a range of time shifts around each flare.

Shifts of 1 ks are preferred in almost all cases, as $r_{\text{Lag}}$ tends to decease with increasing range and large flares in surrounding areas can dominate widescale calculations.
However, some time ranges are widened as peak differences were greater than 1 ks away or specific shapes in the light curves affect the outcome.
The coefficients for all flares are positively correlated, with most having medium ($r>0.3$), if not high ($r>0.5$), associations with the exception of the small Flares 5 and 16.
The X-ray lags behind the UVW2 in all cases, with an average of $-$392 seconds.

Coefficients, lags, and time shift adjustments are listed in Table \ref{tbl:neupert_quantities}. 
Values are also compared with the original study of \citet{Mitra2005} in Figure \ref{fig:crosscorrelationscatter}.
Of note, this analysis implies that X-ray flaring emission lags behind the UVW2 when both responses are observed, as expected from the chromospheric evaporation model and the theoretical Neupert effect.

\subsection{Thermal Empirical Neupert Effect Criteria \label{sec:neupertanalysis}}

    The time lags in Figure \ref{fig:crosscorrelationscatter} are qualitatively consistent with the empirical Neupert Effect. We leverage the high-time resolution properties of our data set to examine the stellar empirical Neupert effect in more detail.

	Though there are multiple metrics for quantifying the empirical Neupert effect, we choose the timing between the peaks of the XMM OM UVW2 (as a proxy for HXR) and XMM EPIC-pn X-ray (here, SXR) time derivative, $\Delta t_{\text{peaks}}^{\text{der}}$, \citep{Neupert1968, Dennis1993}. 
	Other methods are then tested against this. We favor this timing metric over others (e.g.\ UVW2 end and SXR peak) as it does not rely on flare start or end definitions. We combine this with the two part Neupert criteria of \citet{Veronig2002}, which included the timing difference normalized by the HXR (here, UVW2) duration, $\Delta t_{\text{norm}}$ , to mitigate bias towards intense flares or against long-duration flares. The original conditions of this metric are ($|\Delta t_{\text{peaks}}^{\text{der}}| < 1$ min) or ($|\Delta t_{\text{norm}}| < 0.5$ units). However, based on the uncertainties due to differences in the UVW2 and SXR time grid, as well as using a 30-second binning for the SXR derivative, we increase the window slightly to ($|\Delta t_{\text{peaks}}^{\text{der}}| < 1.5$ min) or ($|\Delta t_{\text{norm}}| < 0.6$ units). 
	Results are shown in Figure \ref{fig:neupertnumbers}, compared to the UVW2 impulsiveness and peak luminosity ratio. 
	Notably, the most impulsive flares also show the highest peak luminosity ratios and pass the timing criteria.
	Results are also tabulated in Table \ref{tbl:neupert_quantities}. 
	Figures and classification of each flare are shown in Appendix \ref{appendix:neupert_figures}.
	
	We calculate the Neupert effect score (NES) of \citet{McTiernan1999}. Namely, for each data point along the UVW2 flare, we check if the signs of the UVW2 flare-only flux and the X-ray time derivative are the same. If they are, +1 is added to the NES, and -1 otherwise. The score is then normalized by the total number of data points. The Monte Carlo simulation from Section \ref{sec:calc_quan} is used to report average values and standard deviations. 
	
	The NES is intended to give a straightforward and quantifiable indication that a flare follows the Neupert effect. Unfortunately, this method returns near-zero values for all flares, with highly variable standard deviations. This is likely due to the noise of the SXR derivative, which fluctuates between positive and negative between most points. We attempt to improve this score by using a 3-point running median smoothing before calculating the derivative, but changes are minimal. If the noise floor is taken into account rather than relying only on the sign of the data points, the NES is often close to 1, as the relatively slow decay of the SXR puts the SXR derivative within the uncertainty range. This score is heavily improved with intense smoothing, such as the box function convolution used to estimate the derivative peak (see Section \ref{sec:fluxcalc}), but this is not applied in the final analysis as the resulting curves away from the peaks can be unrepresentative of the exact derivative. 
	
\begin{figure}[!ht]
\centering
\gridline{\fig{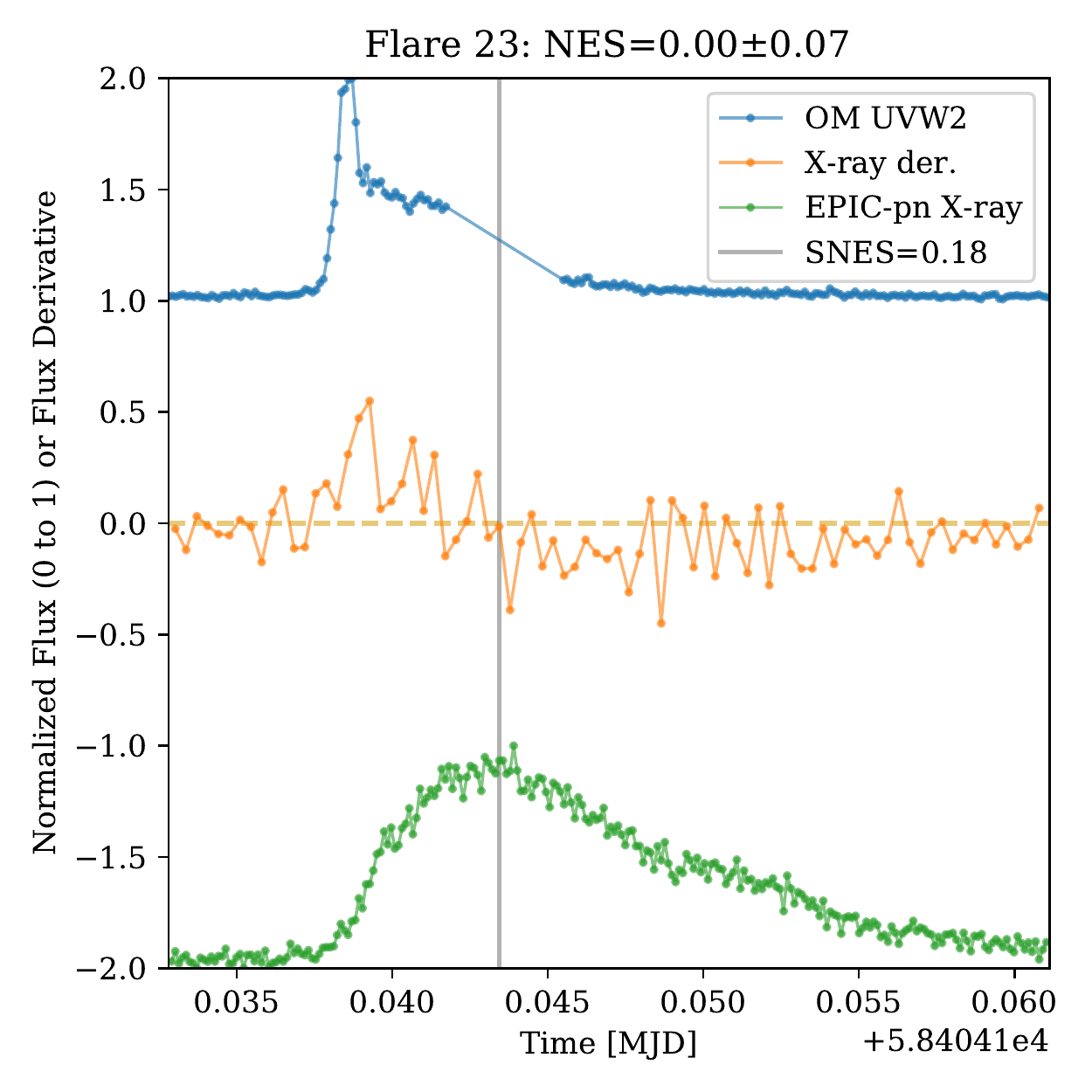}{0.3\textwidth}{(a)}
          \fig{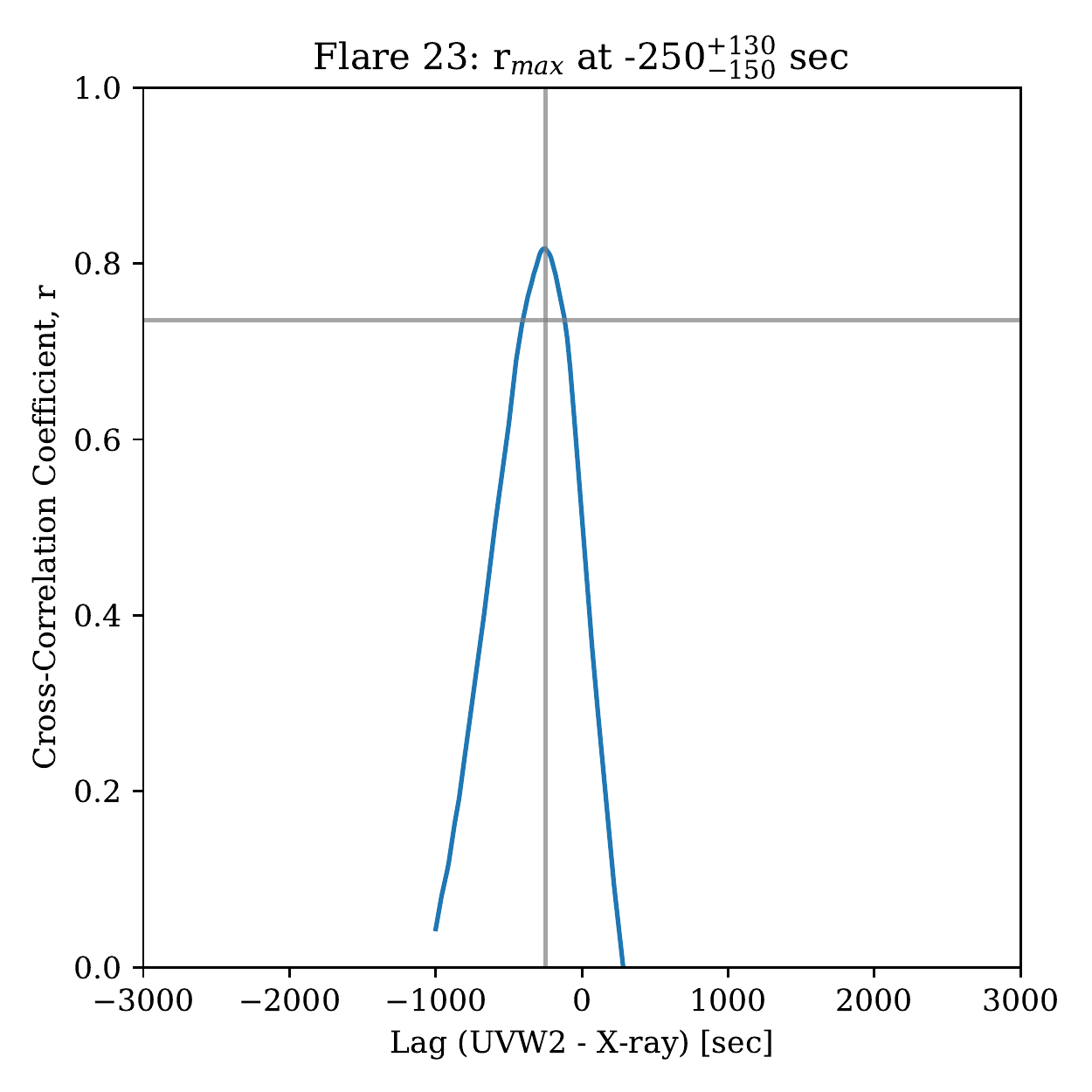}{0.3\textwidth}{(b)}
          \fig{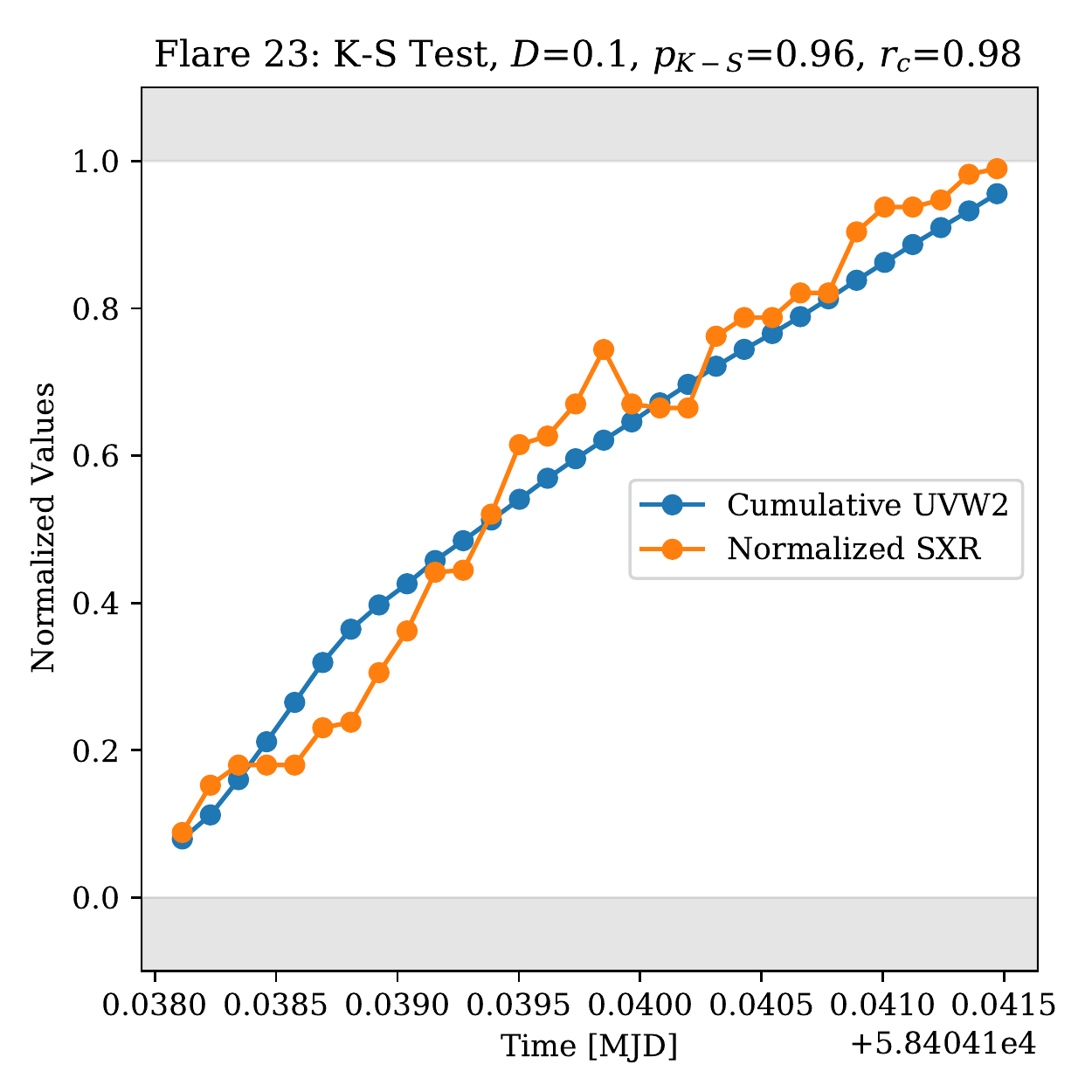}{0.3\textwidth}{(c)}}
\caption{Summarizing figures for the methods used in Section \ref{sec:neupertanalysis}. (a) The light curves for the UVW2 and X-ray response, as well as the X-ray time derivative, are shown for Flare 23. The vertical gray line shows the time at highest SNES. (b) The cross-correlation coefficients for different lag times. The vertical gray line indicates the lag time at $r_{\text{max}}$. The horizontal gray line intersects the 90\% percentile, which gives the lag time uncertainty. (c) The cumulative distributions used for the K-S test. Gray bars show the regions where points were discounted to create a normalized distribution for the X-ray response.
\label{fig:neuperttests}
}
\end{figure}

\begin{figure}[!ht]
\centering
\plottwo{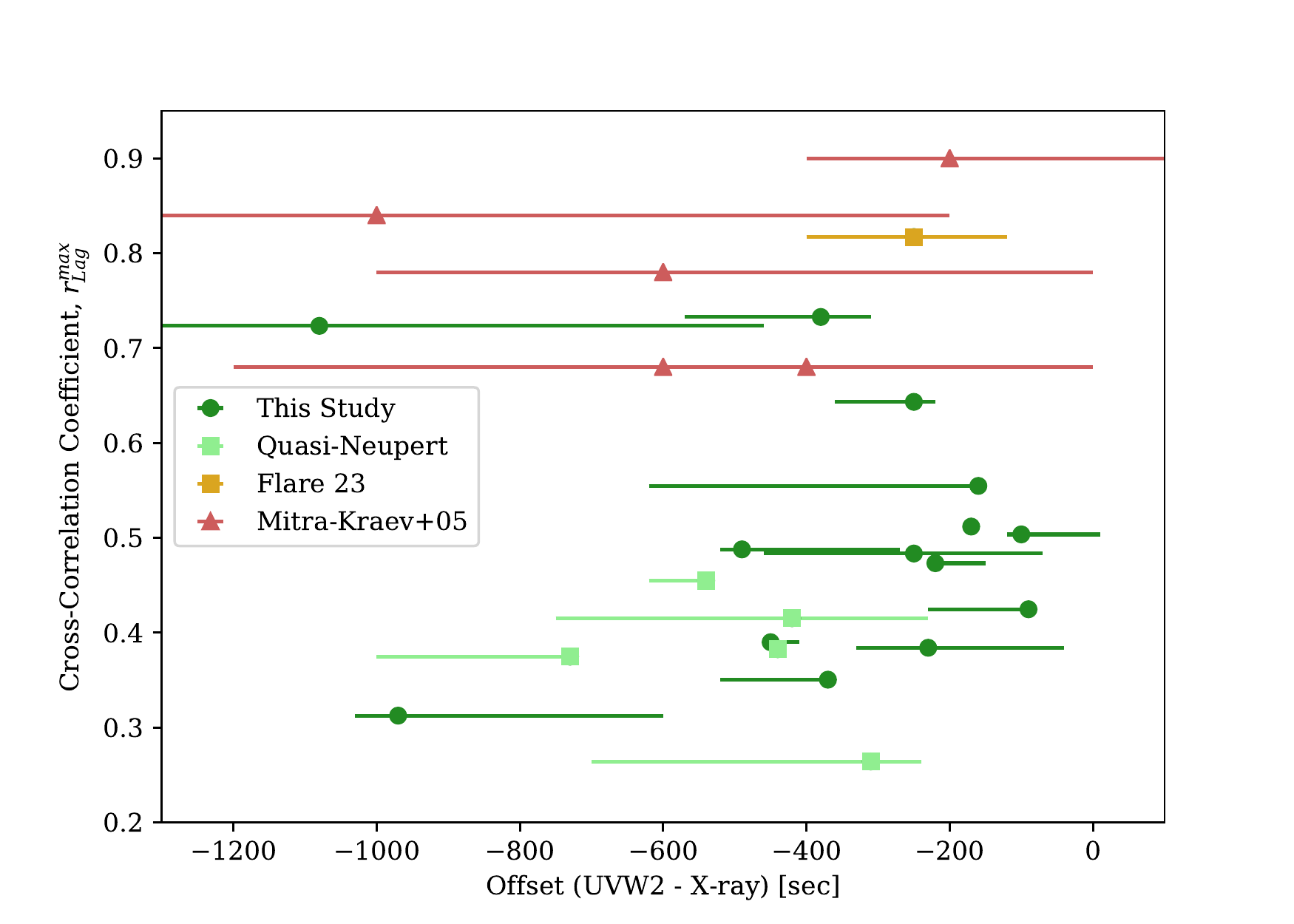}{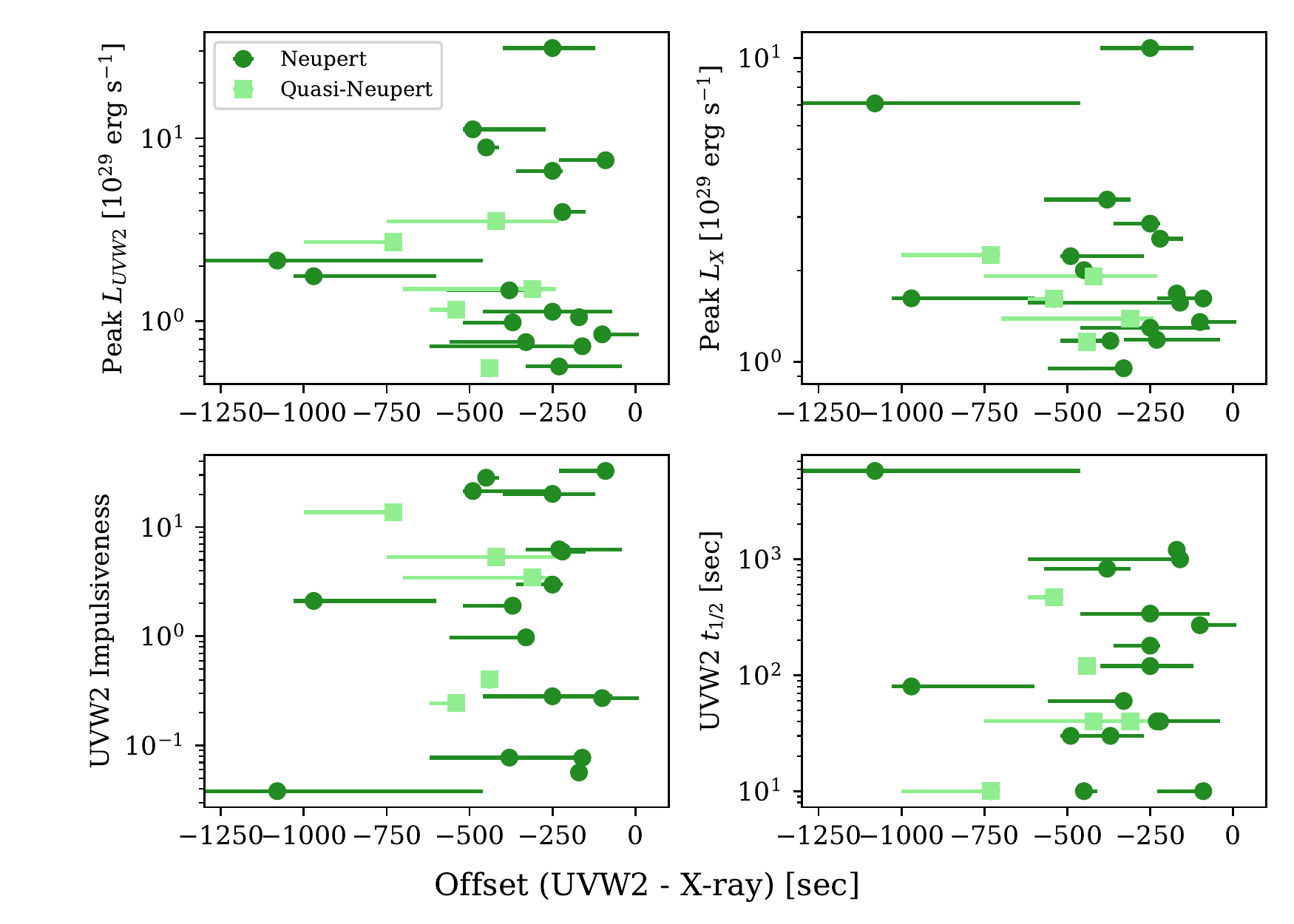}
\caption{\textit{Left}: X-ray response lags for the 21 \xmm{} flares with both components.
On average, these lag uncertainties are improved by a factor of five over the previous study of \citet{Mitra2005} with lower time resolution.
We see an intrinsic spread in lags from 0 to $-$1,000 seconds and an average of $-$392 seconds. Flare 23 and Quasi-Neupert flares are indicated. \\
\textit{Right}: The lag is compared to peak luminosities, impulsiveness, and $t_{1/2}$ from Table \ref{tbl:neupert_quantities}. No significant correlations are found.
\label{fig:crosscorrelationscatter}
}
\end{figure}

\begin{figure}[!ht]
\centering
\plottwo{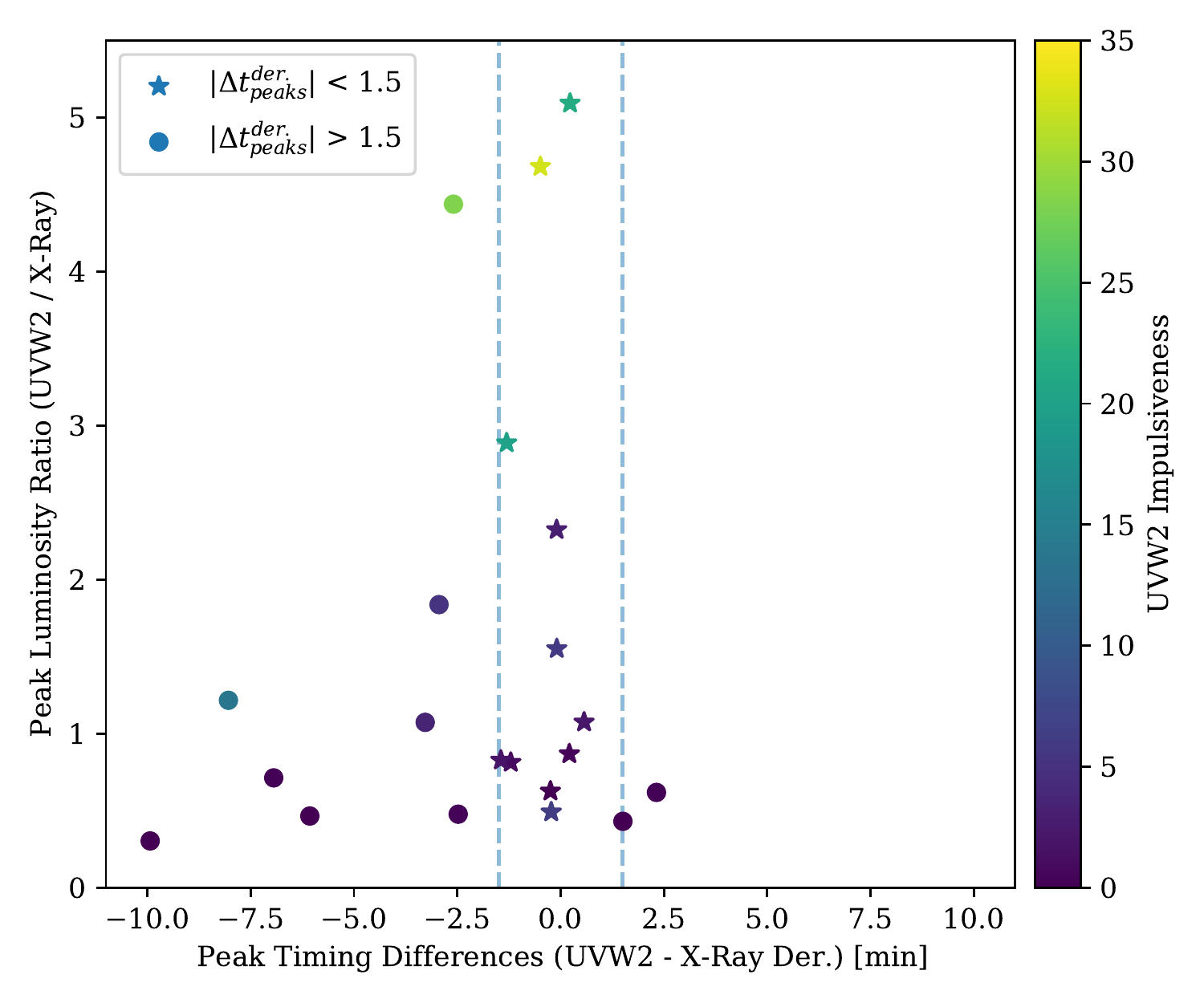}{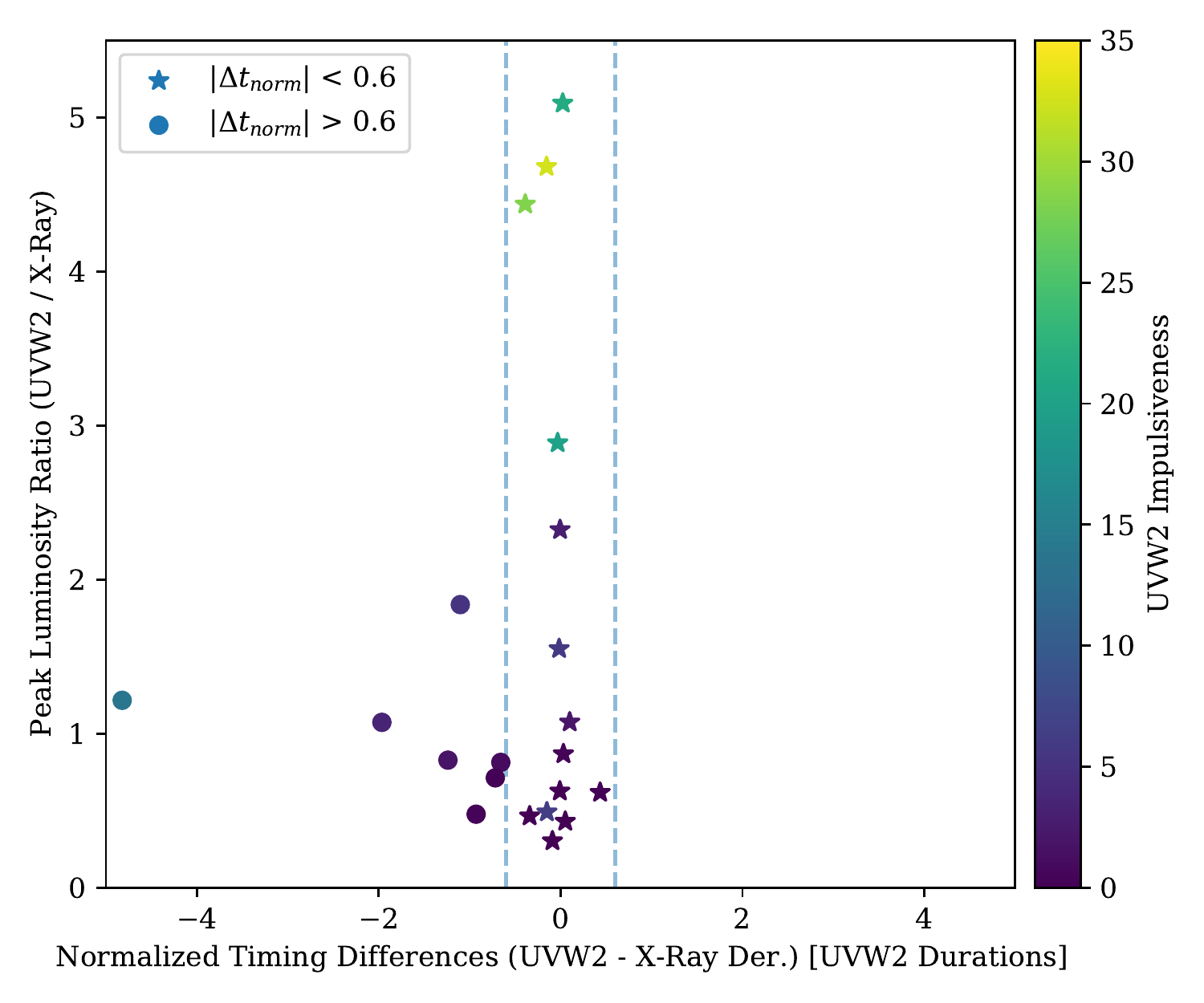}
\caption{The Neupert effect timing differences are plotted against the peak luminosity ratios for the 21 \xmm{} flares that have both UVW2 and X-ray components, color-coded by impulsiveness. 
We see a suggestive trend that flares with higher impulsiveness tend to have higher UVW2/X-ray peak luminosity ratios and exhibit the Neupert effect as well. \\
\emph{Left}: $\Delta t_{\text{peaks}}^{\text{der}}$ is shown, with vertical dashed lines showing the $\pm1.5$ minute mark, which 11 flares lie between.\\
\emph{Right}: $\Delta t_{\text{norm}}$ ($\Delta t_{\text{peaks}}^{\text{der}}$ normalized by UVW2 duration) is shown, with dashed lines showing the $\pm0.6$ units mark. 14 flares pass this metric.
\label{fig:neupertnumbers}
}
\end{figure}

	We define a second Neupert effect score (SNES), which is the maximum value of the time-cumulated NES. 
	This is done in an effort to minimize the effects of long decay phases and to focus on the differences around peak values of the UVW2 and X-ray derivative. 
	While these scores generally show higher and varied values, they do not show a correlation with the Neupert timing criteria and are rather dependent on flare shape. 
	For example, the SNES of the largest flare is smaller than expected due to a long rise phase, shown in Figure \ref{fig:neuperttests}.
	
	Another facet of the Neupert effect is a similarity in the shape between the UVW2 cumulative and the SXR response of the flare.
    This similarity implies a common origin between responses, as the total energy deposited by the electrons would contribute to X-ray emission generation \citep[similar to the microwave emission in][]{Neupert1968}.
	A sample Pearson correlation coefficient ($r_c$, see Section \ref{sec:crosscorrelation}) is calculated to test the similarity in the simultaneous rise of these responses.
	Values of this coefficient are extremely high ($>0.9$) for some Neupert flares and generally high for all.
	We also perform a Kolmogorov–Smirnov test, which gives a test statistic, $D$, the farthest vertical distance between two cumulative distribution functions (CDFs), along with a p-score, $p_{\text{K-S}}$.
	To simulate CDFs, the UVW2 cumulative is first normalized between 0--1. 
	Then, a SXR light curve normalization is created by using a 3-point running median to remove outliers, dividing by an average of the last 5 data points, and applying a 0--1 cutoff.

	There are a few flares which show higher $p_{\text{K-S}}$ only in certain phases.
	For Flare 22, testing only the rise phase reports a high $p_{\text{K-S}}$, while the peak and decay do not.
	For Flare 23, the slow rise phase preceding the sharp increase in emission is removed to improve the score. 
	Flare 15, however, reports a low $p_{\text{K-S}}$ despite having an extremely high $r_c$, as the simulated X-ray CDF values are larger than the UVW2 CDF, especially during the first peak.
	Overall, while there are a few high $p_{\text{K-S}}$ values, there is not a clear trend that separates groups of flares with respect to the Neupert effect timing criteria.
	Examples are shown in Figure \ref{fig:neuperttests}.

	The relation between SXR peak flux and UVW2 energy is expected to be $F_{\text{P,SXR}} = k \cdot \text{E}_{\text{UVW2}}$ (see \citet{Lee1995} and \citet{Veronig2002}), where $F_{\text{P,SXR}}$ is the peak SXR flux and $\text{E}_{\text{UVW2}}$ is the total UVW2 flare energy.
	Note $F_{\text{P,SXR}}$ and E$_{\text{UVW2}}$ are connected linearly if $k$ is a constant, implying a singular heating mechanism that produces a reliable relationship between responses \citep[see][]{Veronig2002}.
	Using all available flares, we calculate $k = 3.9 \pm 0.2 \times 10^{-33}$ W m$^{-2}$ erg$^{-1}$, where the SXR flux is scaled to AU Mic’s habitable zone (see Section \ref{sec:methods}). 
	Values are plotted in Figure \ref{fig:neupertline}.
	Linear regression slopes in log-log space ($b$) are calculated for all flares and large (E$_{\text{UVW2}} > 10^{32}$ erg) flares only, as larger flares show a clear regime change.
	Neither group approaches the expected linear value of 1, but the slope from the larger flares is greater by a factor of two. 
	This is discussed further in Section \ref{sec:discussion}.

    In summary, we use two timing criteria to determine if flares are consistent with the thermal empirical Neupert effect. 
    The first is the time difference between the UVW2 response peak and the X-ray time derivative peak. 
    The second is the time difference normalized by the UVW2 flare duration.
    This normalized criteria is meant to mitigate the bias towards larger flares, which more readily show the Neupert effect. 
    Of the 46 flares with both UVW2 and SXR times available, 35\% pass at least one timing criteria.
    We test the Neupert effect score against the timing results, but found that all scores were near-zero without significant X-ray time derivative smoothing.
    We also measure the similarity in shapes between the UVW2 cumulative and SXR responses. We find expected agreements with some of the largest flares, but ambiguous results otherwise. 
    Finally, we look at the SXR flux and UVW2 energy relationship expected if all flares were Neupert.
    We find that larger flares follow the expected trend much more than smaller flares, but neither are as close as anticipated.

\subsection{Classification of Stellar Flares According to the Thermal Empirical Neupert Effect} \label{sec:neupertclassification}

The high-time resolution of this data set has allowed us to explore several analysis metrics in stellar flares for the first time. 
We use the timing between the UVW2 response peak and the X-ray derivative peak to categorize flares into four categories under the thermal empirical Neupert effect.
Representative flares in each category are shown in Figure \ref{fig:neuperttypes} (all flares are shown in Appendix \ref{appendix:neupert_figures}), and the results for the 51 \xmm{} flares are summarized in Table \ref{tbl:neupert_classifications}. 

The first category is ``Neupert'' (N), in which the UVW2 flux and SXR time derivative peaks nearly coincide and both timing criteria are passed.  This behavior is most similar to the traditional empirical Neupert effect in solar flare HXR/radio and high-temperature SXR emission.
About 20\% (9 flares) of the sample belongs to this group. 
The second category is ``Quasi-Neupert'' (Q).  These flares exhibit a response in both UVW2 and SXR emission, but the light curve peaks are well-separated in time.
About half as many events fall into this group.
We also categorize flares that pass a single timing criterion only as being ``N/Q''; most of these events only pass the normalized criteria.
However, this designation is mainly used to identify marginal cases, and we intend to revise classification in future work as complementary data sets are brought into the analysis. 
For this study, ``N/Q'' flares should be considered Neupert.

Of particular interest are the 25 events that exhibit a response in only one spectral region between the X-ray and UVW2. These may be due to weak responses that are within the noise, but their prevalence may also suggest alternative explanations (see Section \ref{sec:discussion}). 
We classify these as Non-Neupert flares.  
The 12 ``Non-Neupert Type I'' flares are those where there is a SXR response but no UVW2 response.  
In 13 ``Non-Neupert Type II'' events, there is a UVW2 response but no detected SXR emission.  
Note that there are also 5 ``Undetermined'' (Un) \xmm{} flares, where no UVW2 observations were taken.

At high-time cadence, we find a large variety of multi-wavelength behavior of thermal emissions among the flares of AU Mic.  
In future papers, we plan to connect to the nonthermal radio data for a comprehensive empirical Neupert effect relationship among the flares from this star.  
Radiative-hydrodynamic modeling will also be pursued to explain each of the four types of empirical behaviors in terms of the theoretical Neupert effect (chromospheric evaporation and condensation) and possible physical origins.

\section{Discussion} \label{sec:discussion}

According to the standard electron beam model of solar and stellar flares, nonthermal electrons rapidly heat the mid chromosphere, which causes mass and heat advection into the corona (see references in Section \ref{sec:intro}).  The impulsive energy deposition of the chromosphere produces the NUV and optical continuum and emission lines, while the SXR radiation is emitted as the coronal loops fill with hot plasma to several tens of MK.  The loops  then cool down over tens of minutes and shine brightly in subsequently lower temperature SXR and Extreme-UV (EUV) radiation signatures \citep{Aschwanden2001}.  All the while, this process takes place over tens of minutes to hours through sequential heating and cooling of many loops and chromospheric footpoints \citep{Warren2006}, with the heating rates early on being much larger and the reconnected loops much smaller than in the gradual phase.

We interpret the four categories of flares (see Section \ref{sec:neupertclassification}) in terms of this fundamental process, drawing from several, similar observations that have been reported in the literature.

Neupert flares follow the timing of the chromospheric evaporation model as expected.
Of the 46 flares with both \xmm{} UVW2 and SXR times available, 35\% (9 ``N'' and 7 ``N/Q'' flares in Table \ref{tbl:neupert_classifications}) pass at least one timing criteria, where the timing between the UVW2 response peak and the X-ray time derivative peak are within either 1.5 minutes or 0.6 UVW2 durations.
In the 21 flares with both responses (``N'', ``N/Q'', and ``Q'' flares in Table \ref{tbl:neupert_classifications}), 76\% pass at least one of these timing criteria. 
In addition, we find that the X-ray response significantly lags behind the UVW2, which is expected under this model.
We calculate the NES and SNES of the flares, but find the method to be inconclusive without heavy light curve smoothing. 
These metrics may be more useful for brighter stars or solar studies and should be kept in consideration in future follow up analyses.
We find that the shapes of the cumulative UVW2 response and the SXR flux are similar, indicating a common origin.

It should be emphasized that ``N/Q'' flares are considered to be consistent with the Neupert effect. 
Timing differences across the soft X-rays occur due to the flare temperature evolution \citep{Gudel1996, McTiernan1999}, which depends on both conductive and radiative plasma cooling timescales.
The ratio between the two changes over time, with radiation dominating the decay (i.e.\ lower temperature) phase \citep[see Fig.\ 9 of][]{Osten2016}. 
SXR emission from hotter plasma peaks earlier than from cooler plasma, and high-energy, hotter flares tend to have smaller timing delays \citep[see Sec.\ 4.2 and Fig.\ 5 of][]{Gudel1996}.
Generally, high-temperature SXR derivatives peak around the same time as hard X-rays, while low-temperature SXR derivatives show little resemblance to the hard X-rays \citep[see Figs.\ $7-9$ of][]{McTiernan1999}.
Flares that would be labeled ``N/Q'' in our classification scheme from \citet{Gudel1996, Gudel2002} may be fully consistent with another generalized Neupert effect that considers the energy balance between radiative cooling and impulsive heating, as constrained by nonthermal gyrosynchrotron emission.
Thus, the choice of X-ray band and plasma temperature can bias the classification of flares and reveal non-correlations, especially after the peak UVW2 phase.
Note that the normalized timing criteria attempts to mitigate this bias towards intense flares, but it is not expected to cover all possibilities.

The timing differences of the Quasi-Neupert flares may also be caused by significant contributions from additional heating mechanisms. 
A strong candidate for this is thermal conduction \citep[see][and references within, for a comprehensive overview]{Longcope2014}, where reconnection energy is transferred through heat transfer in the plasma along the flaring loop top-down to the chromosphere \citep[see also][]{Ashfield2022}.
Contributions from this could explain delayed X-ray time derivative peaks, and relatively small contributions from this may explain near-late timings like that of Flare 23.
Likewise, cooling rates are also of interest.
As noted before, \citet{McTiernan1999} finds flares from high-temperature plasma ($T>16.5$ MK) are more likely to show the empirical Neupert effect than low-temperature plasma. 
The XMM EPIC-pn instrument measures flare plasma at 10--30 MK, encompassing both regimes.
Thus, contributions from both regimes may skew Neupert effect results as the high-temperature plasma cools.
This cooling may contribute to the poor K-S test correlations, even in Neupert flares.
Alternatively, \citet{Li1993} finds long-durations flares can display a SXR peak before the HXR end, which contrasts some formulations of the empirical Neupert effect, due to evaporation-driven density enhancements failing to overcome hot plasma cooling.
\citet{Veronig2002} further suggests that the power-law break in the SXR peak to NUV fluence (see Figure \ref{fig:neupertline}) is due to these flares that are dominated by thermal conduction transport into the chromosphere.
Quasi-Neupert flares introduce a wide array of possibilities for flaring energy transfer, and future X-ray spectroscopic and temperature analysis, along with stellar flare modeling, will help determine individual flare origins.

Non-Neupert Type I flares are the most likely candidate for an alternative heating method, like thermal conduction. 
Following \citet{Petschek1964}, shock heating may occur after magnetic reconnection without a nonthermal electron beam.
The heat energy would then be transferred down into the chromosphere through thermal conduction, driving chromospheric evaporation and producing SXR emission \citep{Forbes1989}.
Consistent models have been designed for these types of flares \citep[see][]{Yokoyama1998, Yokoyama2001}, and future work will determine if Non-Neupert Type I flares fall under this category or require additional physics.

\citet{Osten2005} gives possible explanations for Non-Neupert Type II flares, which have UVW2 responses yet lack X-ray emission. 
They suggest low ambient electron densities within the loop, heating at lower (or higher) temperatures than measured by the instrument, high amounts of magnetic trapping, or particles accelerated to MeV energies that penetrate into the  lower-chromosphere or photosphere and show continuum emission without significant chromospheric evaporation. 
Flares 3 and 26 have both UVW2 and $U$-band responses, similar to the event featured in \citet{Osten2005}.
For the former, there is an arguably faint X-ray response and the UVW2 response is small, which lends credibility to the low-temperature heating scenario.
For the latter, the UVW2 and \emph{U}-band responses are reasonably sized, though short-lived.
This likely rules out low-temperature heating by comparison to Flare 3.
Efficient magnetic trapping may also be unlikely, as a longer decay would be expected. 
We predict that Flare 3 will have a low radio response while Flare 26 will have a bright response.
We plan to analyze these time intervals in detail in future work to test these hypotheses and constrain the presence of accelerated electrons in the flares.

There are a few Non-Neupert peculiarities that should be mentioned.
Events like Flares 6 and 10 in the UVW2 are likely secondary (sympathetic) flares of a preceding larger one despite being temporally separated. 
In these cases, their corresponding responses may be hidden within the larger preceding one or too small to observe.
Likewise, there are arguably tiny responses to other bands, like in Flares 34 and 40, which did not pass flare selection. 
These situations all contribute to a potentially diverse flare production environment and are important to keep in mind for future flare modeling.
Whether H$\alpha$ equivalent width flaring variations occur during Non-Neupert flares is also of interest for models. Unfortunately, there are no accompanying CHIRON observations for any of the Non-Neupert flares in this study.

The percentage of Neupert flares among the \xmm{} flare pairs (76\%) is similar to \citet{Dennis1993}, which found 80\% agreement in 66 solar flare pairs.
This percentage is also higher than the 44\% of solar flares from the large (1114 flare pairs) statistical study of \citet{Veronig2002}.
Note these studies discounted complex flares, flares with multiple possible pairings, and flares where the two responses did not start near each other to various degrees.
Therefore, there are no comparable Non-Neupert statistics available.
Despite the high percentage of Neupert flares, the SXR flux to UVW2 energy relationship is lower than expected (Figure \ref{fig:neupertline}). 
\citet{Veronig2002} showed a log-log slope near 1.0 ($b = 0.96 \pm 0.07$) for large flares and $b = 0.83 \pm 0.03$ including mid-sized flares. 
\citet{Namekata2017} similarly found a slope of 0.8 between white light energy and GOES X-ray flux for solar flares above GOES class M2. 
\citet{Stelzer2022} expanded on that work with a scaled superflare from AD Leo and found a close, but steeper, slope of $b = 1.150 \pm 0.005$.
While our results indicate that the chromospheric evaporation model plays a significant role in M dwarf stellar flares, the large deviation from the ideal slope of 1 in Figure \ref{fig:neupertline} implies other heating mechanisms besides nonthermal electron beam heating are necessary to understand many stellar flares.
The existence and prevalence of Non-Neupert flares further indicates that flare energy budgets do not follow one slightly variable relation.

Since ion production in Earth’s thermosphere and ionosphere is sensitive to EUV radiation \citep{Mitra1974}, extrapolations from X-ray flare observations to the EUV regime are considered useful in gauging exoplanet habitability \citep[as in][]{Chadney2017}. \citet{Qian2011} finds that solar flare models showing the Neupert Effect have a larger EUV enhancement during the impulsive phase. This leads to a greater ion production early on, but weaker electron and neutral density enhancements overall due to the short-lived impulsiveness. Broadly based on the findings of \citet{Qian2011}, Neupert and Quasi-Neupert flares in M dwarfs may also have comparable effects on an exoplanet’s ionosphere.

Due to the similar energy ranges of the SXR (0.2 -- 12 keV) and EUV (10 -- 124 eV) regimes, it is expected that the presence of one response implies the other, as they both stem from ablated plasma (see Section \ref{sec:intro}). However, coupling their time evolution too tightly is known to cause low accuracy in solar modeling \citep{Nishimoto2021}, as the EUV emission peaks after the SXR. Some solar EUV flares also have a late phase which is not easily estimated by X-ray comparisons, where a second peak is observed hours after the primary event \citep{Woods2011, Chen2020}.

With this in mind, Neupert, Quasi-Neupert, and Non-Neupert Type I (X-ray only) flares may help determine minimum EUV radiation effects, but they are unlikely to estimate maximum limits without proper multi-wavelength EUV observations.
However, Non-Neupert Type II (UVW2 only) flares are unlikely to correspond to EUV emissions, as significant plasma heating is not expected.
Thus, it may be inadvisable to do any EUV extrapolations based on only NUV observations.

Another possible complication is the large X-ray swell on 2018 Oct 10 (Flare 11).
Notably, most of the enhancements during the swell match up with UVW2 flares (e.g. Flare 9 at the start, Flare 14 at the peak), however there is also at least one clear increase with no UVW2 response about an hour before Flare 12 (see Figure \ref{fig:lc_xmm}).
The maximum enhancement also has a slower decay than other X-ray flares of comparable or greater size, like Flares 23 or 27.
Given this, the X-ray swell seems to be composed of coupled flaring events of different Neupert types, and it is currently unknown if any quiescent fluctuations play a part in sustaining the emission enhancement.
Spectral analysis of the different regimes of Flare 11 may be able to shed light on these issues, but similar events in M dwarfs, if prevalent, may complicate auto-detection or separation of individual X-ray events, Neupert classification studies, and extrapolations between wavelengths.

\section{Summary and Conclusions} \label{sec:conclusions}

\begin{figure}[!ht]
\centering
\includegraphics[width=0.6\linewidth]{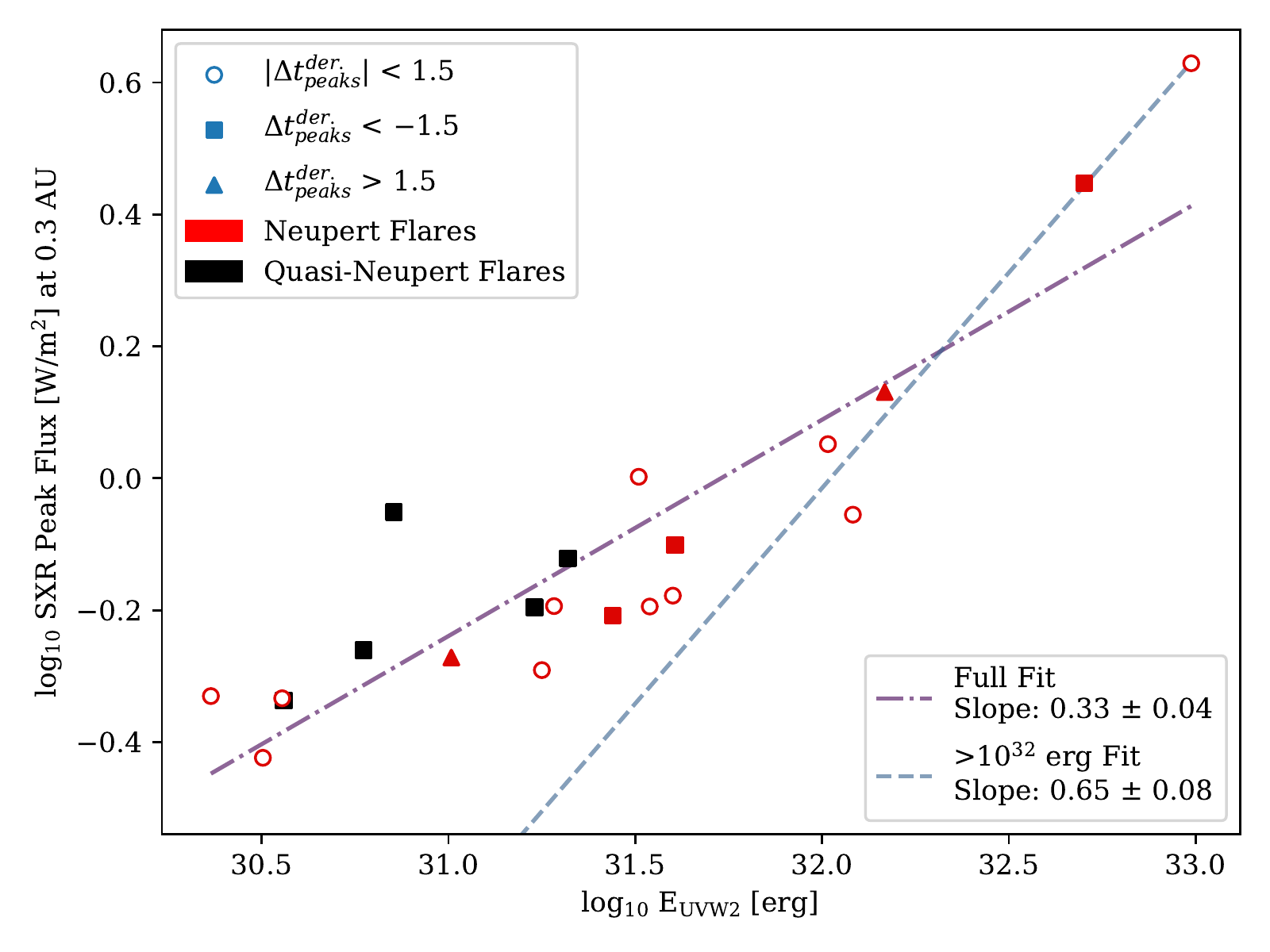}
\caption{SXR peak flux versus UVW2 energy is shown for the 21 \xmm{} flares with both components. Square markers indicate that a flare has a value of $\Delta t_{\text{peaks}}^{\text{der}} < -1.5$ minutes (lag between UVW2 and X-ray derivative peaks), triangle markers indicate $\Delta t_{\text{peaks}}^{\text{der}} > 1.5$, and open circles indicate that $\Delta t_{\text{peaks}}^{\text{der}}$ falls between these. Bright red coloring indicates that the flare shows the Neupert effect, while black points are classified as Quasi-Neupert. We calculate slopes for the entire set of flares, as well as only for the largest ones ($E_{UVW2}>10^{32}$ erg), in log-log space. We find that the expected slope of $b=1$ for a linear relationship \citep{Veronig2002} is neither met nor close in either case.
\label{fig:neupertline}
}
\end{figure}

    The purpose of this campaign is to build upon past studies and explore the physical parameters of M dwarf flares using observations with long-duration temporal coverage and high-time resolution of a single flaring star.
    In this paper, we comprehensively analyze a new, multi-wavelength dataset of AU Mic spanning the X-ray to the optical over 7 days, with the $U$-band, $V$-band, and H$\alpha$ data extending to $\sim$500 hours.
    The data used here includes X-ray and UVW2 bands from \xmm{} and \swift{}, the H$\alpha$ line and \emph{V}-band from CTIO/SMARTS 0.9m, and the \emph{V}- and \emph{U}-bands from LCOGT (summarized in Table \ref{tbl:obs_summary}). 
    This dataset allowed us to analyze the thermal empirical Neupert effect in large sample of stellar flares.
    Quantitative results are summarized as follows:

\begin{enumerate}
    \item We find 73 unique flares, with 51 in the \xmm{} OM UVW2 and EPIC-pn X-ray data, 21 of which overlap in time (see further classification in Table \ref{tbl:neupert_classifications}).
    \item Of the 21 overlapping flares, 16 show the Neupert effect, where the timing between the UVW2 emission peak and X-ray time derivative peak are within 1.5 minutes or 0.6 UVW2 durations. 
    5 flares then show significant timing differences between responses (Quasi-Neupert). 
    Of the others, 12 flares show only X-ray responses (Non-Neupert Type I), while 13 show only UVW2 (Non-Neupert Type II) (see Table \ref{tbl:neupert_classifications}). From this, 65\% of flares from AU Mic do not follow the thermal empirical Neupert effect.
    \item We find that the Neupert effect score (NES) of \citet{McTiernan1999} returns null results for our data without significant smoothing of the X-ray light curve. Defining a second Neupert effect score (SNES) which takes the highest cumulative score along the flare improves the results. However, these remain inconsistent between Neupert and Quasi-Neupert flares.
    \item The sample Pearson coefficients between the cumulative UVW2 response and the X-ray response are generally high, indicating similar curve shapes.
    These are extremely high ($>0.9$) for a few Neupert flares.
    K-S Tests between these can also give favorable results, but careful formatting of the curves is sometimes required.
    \item The SXR peak flux to UVW2 energy relation shows two distinct regimes, whereas a linear relation ($F_{\text{P,SXR}} = b \cdot \text{E}_{\text{UVW2}}$, $b=1$ in log-log space) is expected under the Neupert effect. We calculate $b=0.33\pm0.04$ using all flares with both responses and $b=0.65\pm0.08$ for flares with $E_{\text{UVW2}}>10^{32}$ erg (see Figure \ref{fig:neupertline}).
    Fig.\ 11 of \citet{Veronig2002} also shows this division for solar flares, though both regimes have slopes closer to 1.0.
    \item We find that the X-ray response lags behind the UVW2 for all overlapping flares in this sample, with an average lag of $-$392 seconds. Note that the Neupert timing relation does not follow this trend.
    \item We calculate a \emph{U}-band FFD slope of $\beta=-0.72\pm0.15$, which is similar to other M dwarfs in literature. We also calculate a preliminary X-ray FFD, which has a similar slope with higher flare energies.
    We find that among flares with X-ray and \emph{U}-band responses, the empirical $E_{\text{SXR}}/E_U$ ratio averages to 1.5. However, the ratio between a set cumulative number of flares per day is 2.68, similar to the canonical energy partition (2.72) of \citet{Osten2015}.
\end{enumerate}

    The relationships between \xmm{} OM UVW2 and EPIC-pn X-ray flares and the variety between Neupert, Quasi-Neupert, and Non-Neupert types and shows that while the chromospheric evaporation model plays a large part in stellar flares, it cannot explain all M dwarf flare observations. 
    Other heating mechanisms like thermal conduction are needed to explain the timing differences of Quasi-Neupert flares and the existence of Non-Neupert Type I flares, which do not show the UVW2 response expected from nonthermal beam heating.
    In Non-Neupert Type II flares, we find profiles that suggest both  low-temperature heating and either low ambient densities or highly accelerated particles as origins, but further work using radio data is needed to draw conclusions.
    Last, the \emph{U}-band and X-ray FFDs imply that the canonical energy partition may apply to an average of flares rather than individual ones, so caution should be used when extrapolating flare energies from only one wavelength regime.
    The X-ray energies will be refined with the analysis of the RGS spectra, and the number flares is still low and should be supplemented through future campaigns.

    In future work, we will add in the radio data from the Jansky Very Large Array (JVLA) and the Australia Telescope Compact Array (ATCA) to perform the same type of analysis on the nonthermal empirical Neupert effect, which does not rely on a proxy for the radio/HXR response. 
    This will allow us to better constrain the processes of the flares found in this study.
    X-ray spectral analysis will also assess flare temperature evolution to further studies of the theoretical Neupert effect through chromospheric evaporation and condensation modeling.

\acknowledgements
    We acknowledge and thank Dr.\ Wei-Chun Jao for assistance with scheduling CHIRON observations and for guidance on CTIO 0.9m data reduction, Dr.\ Glenn H.\ Schneider for helpful discussions about young stars and AU Mic, and Dr.\ Joel C.\ Allred for helpful discussions about flare heating models that contributed to the observing proposal.
    I.I.T.\ thanks Dr.\ Meredith A.\ MacGregor and Dr.\ Allison Youngblood for discussion on stellar flare analysis.
    We also thank an anonymous referee for comments that helped to clarify results and improve the manuscript.
    
    This work was supported by NASA ADAP award program Number 80NSSC21K0632, NASA XMM-Newton Guest Observer AO-17 Award 80NSSC19K0665, and the Space Telescope Science Institute's Director's Discretionary Research Fund 52079.
    
    I.I.T.\ acknowledges support from the NSF Graduate Research Fellowship Program (GRFP). 
    J.E.N.\ was supported by the Independent Research/Development program for program officers at the National Science Foundation.
    Any findings and conclusions are those of the author(s) and do not necessarily reflect the views of the National Science Foundation.
    
    Y.N.\ was supported by JSPS (Japan Society for the Promotion of Science) KAKENHI Grant Number 21J00106, JSPS Overseas Research Fellowship Program, and JSPS Postdoctoral Research Fellowship Program. 
    
    We also acknowledge the International Space Science Institute and the supported International Teams 464: “The Role Of Solar And Stellar Energetic Particles On (Exo)Planetary Habitability (ETERNAL, \url{http://www.issibern.ch/teams/exoeternal/})” and 510: “Solar Extreme Events: Setting Up a Paradigm (SEESUP, \url{https://www.issibern.ch/teams/solextremevent/})’.
    
    This work is based on observations obtained with XMM-Newton, an ESA science mission with instruments and contributions directly funded by ESA Member States and NASA.
    
    This work makes use of observations from the Las Cumbres Observatory global telescope network.
    
    This research has used data from the SMARTS 1.5m and 0.9m telescopes, which are operated as part of the SMARTS Consortium.

\clearpage

\begin{deluxetable}{ccccccc}
\tablewidth{0pt}
\tablecaption{Stellar Properties of AU Mic \label{tbl:aumic}}
\tablehead{
\colhead{Spectral Type} & \colhead{R$^{\text{(1,2)}}$} & \colhead{M$^{\text{(2)}}$} & 
\colhead{Dist.$^{\text{(3)}}$} & \colhead{Age$^{\text{(4)}}$} & \colhead{Rot. Period$^{\text{(2)}}$} & \colhead{Projected Radial Vel.$^{\text{(1,2)}}$} \\
\colhead{} & \colhead{R$_\sun$} & \colhead{M$_\sun$} & 
\colhead{pc} & \colhead{Myr} & \colhead{day} & \colhead{km s$^{-1}$}
}
\startdata 
dM1e & $0.75\pm0.03$ & $0.5\pm0.03$ & 9.72 & $23\pm3$ & $4.863\pm0.01$ & $8.7\pm0.2$\\
\enddata
\tablerefs{(1) \cite{radius}, (2) \cite{Plavchan2020}, (3) \cite{GaiaDR2}, (4) \cite{Mamajek2014}}
\end{deluxetable}

\begin{deluxetable}{llllccc}[!ht]
\tablewidth{0pt}
\tablecaption{Observation Log - Summary \label{tbl:obs_summary}}
\tablehead{
\colhead{Instrument} & \colhead{Band} & \colhead{Start Date} & \colhead{End Date} & \colhead{Monitoring Time} & \colhead{Cadence} & \colhead{Exposure} \\
\colhead{} & \colhead{\AA{}} & \colhead{} & \colhead{} & \colhead{hrs} & \colhead{sec} & \colhead{sec}
}
\startdata 
\xmm/EPIC-pn/X-ray & 1.03 - 62 & 2018-10-10 & 2018-10-17 & 130.04 & 10 & 10 \\
\swift/XRT/X-ray & 1.24 - 62  &  2018-10-12  &  2018-10-14 &  4.35 & 5 & 5 \\
\xmm/RGS/X-ray & 5 - 35 &  2018-10-10 & 2018-10-17  &  130.83 & $\sim$35 & $\sim$35 \\
\swift/UVOT/W2 & 1600 - 3480  &  2018-10-12  &  2018-10-14  & 4.35 & 5 & 5 \\
\xmm/OM/UVW2 & 1790 - 2890   & 2018-10-10 & 2018-10-17 & 108.74 & 10 & 10 \\
LCOGT/1m/U band & 3030 - 4170  & 2018-10-10 & 2018-10-29 &  74.15 & 46 & 4 \\
LCOGT/0.4m/V band & 4780 - 6350  & 2018-10-10 & 2018-10-29 & 79.84 & 25 & 2 \\
SMARTS/2KCCD/V band & 4780 - 6350   & 2018-10-10 & 2018-10-17 &  26.69 & 47 & 15 \\
SMARTS/CHIRON/H$\alpha$  & 4500 - 8900  & 2018-10-10 & 2018-10-25 &  61.21 & 65 & 60 \\
\enddata
\tablecomments{Summary of all observations used in this analysis. A detailed log of all observation windows is given in Appendix \ref{app:obs}. Note that while CHIRON observes over a broad wavelength range, the analysis in this paper is focused on the H$\alpha$ line, 6562.8 \AA{}.
}
\end{deluxetable}

\begin{deluxetable}{lccccc}
\tablewidth{0pt}
\tablecaption{Adopted and Synthetic Broadband Quiescent Fluxes and Luminosities of AU Mic \label{tbl:flux_num}}
\tablehead{
\colhead{Bandpass $T$} & \colhead{$\langle f_{q,\lambda} \rangle_T$ Zero-Point Method$^{\dagger}$} & \colhead{$\langle f_{q,\lambda} \rangle_T$ HST Spectrum} & \colhead{$\Delta \lambda_{\rm{FWHM}}$} & \colhead{$\lambda_{\rm{mean}}$} & \colhead{$L_{q,T}$} \\ 
\colhead{} & \colhead{$10^{-14}$  \funits{}} & \colhead{$10^{-14}$  \funits{}} & \colhead{\AA{}} & \colhead{ \AA{} }  & \colhead{$10^{28}$ erg s$^{-1}$} }
\startdata 
UVW2 (XMM OM) &$ 1.03$ & $1.14- 1.18$ & $475$ & $2150$ & $5.5$ \\
Bessel $U$ (LCOGT) & $12.2-14.0 $ & $13.5 - 14.1$ & $700$ & 3610 & 100  \\
Bessel $V$ (CTIO/SMARTS 0.9m) & $110-133$ & $122$ & 850 & 5510 & 1200  \\
\enddata
\tablecomments{  Bandpass zeropoints are obtained from \citet{Willmer2018} and the SAS.  The values of $\Delta \lambda_{\rm{FWHM}}$ are adopted from \citet{Moffett1974} to compare more directly to the study of \citet{Lacy1976} with similar $U$ and $V$ bandpasses. $^{\dagger}$Adopted quiescent fluxes of AU Mic; see text and Appendix \ref{sec:appendix_fluxcalc}.}
\end{deluxetable}

\clearpage

\begin{deluxetable}{cccccccccccccc}[!ht]
\tabletypesize{\scriptsize}
\tablewidth{0pt}
\tablecaption{Calculated Quantities and Neupert Classification Criteria \label{tbl:neupert_quantities} }
\tablehead{
\colhead{ID} & \colhead{$t_{1/2}^{(1)}$} & \colhead{$\mathcal{I}$$^{(1)}$} & \colhead{$L^{\text{peak}}_{\text{UVW2}}$} & \colhead{$L^{\text{peak}}_{\text{X}}$} & \colhead{$\Delta t_{\text{peaks}}^{\text{der}}$} & \colhead{$\Delta t_{\text{norm}}$} & \colhead{SNES} & \colhead{$D$} & \colhead{$p_{\text{K-S}}$} & \colhead{$r_{c}$} & \colhead{$r^{\text{max(2)}}_{\text{Lag}}$} & \colhead{Lag} & \colhead{Classification$^{(3)}$}\\
\colhead{} & \colhead{min} & \colhead{} & \colhead{$10^{29}\frac{\text{erg}}{\text{s}}$} & \colhead{$10^{29}\frac{\text{erg}}{\text{s}}$} & \colhead{min} & \colhead{} & \colhead{} & \colhead{} & \colhead{} & \colhead{} & \colhead{} & \colhead{sec}  & \colhead{}
}
\startdata 
1 & 5.98 & 0.28 & 1.13 & 1.30 & 0.21 & 0.03 & 0.21 & 0.5 & 0.00 & 0.73 & 0.48 & -250$^{+180}_{-210}$ & N \\ 
5 & 0.69 & 3.46 & 1.50 & 1.39 & -3.28 & -1.97 & -0.09 & 0.7 & 0.00 & 0.58 & 0.26 & -310$^{+70}_{-390}$ & Q   \\ 
8 & 0.35 & 32.66 & 7.57 & 1.62 & -0.49 & -0.15 & 0.20 & 0.8 & 0.04 & 0.64 & 0.42 & -90$^{+0}_{-140}$ & N \\ 
15 & 99.37 & 0.04 & 2.14 & 7.09 & -9.93 & -0.09 & 0.14 & 0.2 & 0.00 & 0.98 & 0.72 & -1080$^{+620}_{-820}$ & N/Q \\ 
16 & 1.20 & 0.98 & 0.77 & 0.95 & -1.21 & -0.66 & 0.08 & 0.7 & 0.00 & 0.33 & 0.17 & -330$^{+10}_{-230}$ & N/Q \\ 
18 & 8.00 & 0.24 & 1.16 & 1.61 & -6.94 & -0.72 & 0.19 & 0.5 & 0.12 & 0.85 & 0.45 & -540$^{+10}_{-80}$ & Q   \\ 
20 & 17.22 & 0.08 & 0.73 & 1.57 & -6.07 & -0.34 & 0.24 & 0.5 & 0.94 & 0.83 & 0.55 & -160$^{+10}_{-460}$ & N/Q \\ 
21 & 0.30 & 13.68 & 2.71 & 2.25 & -8.04 & -4.82 & 0.64 & 0.8 & 0.00 & 0.52 & 0.37 & -730$^{+10}_{-270}$ & Q   \\ 
22 & 0.63 & 21.37 & 11.18 & 2.23 & 0.23 & 0.02 & 0.51 & 0.2 & 0.82 & 0.94 & 0.49 & -490$^{+220}_{-30}$ & N \\ 
23 & 2.21 & 20.13 & 31.09 & 10.78 & -1.30 & -0.03 & 0.18 & 0.1 & 0.96 & 0.99 & 0.82 & -250$^{+130}_{-150}$ & N \\ 
24 & 25.51 & 0.08 & 1.47 & 3.42 & 1.51 & 0.05 & 0.21 & 0.3 & 0.69 & 0.97 & 0.73 & -380$^{+70}_{-190}$ & N/Q \\ 
25 & 0.76 & 5.32 & 3.52 & 1.91 & -2.94 & -1.10 & 0.53 & 0.8 & 0.00 & 0.73 & 0.42 & -420$^{+190}_{-330}$ & Q   \\ 
27 & 3.54 & 2.97 & 6.63 & 2.85 & -0.10 & -0.01 & 0.24 & 0.2 & 0.44 & 0.98 & 0.64 & -250$^{+30}_{-110}$ & N \\ 
28 & 1.40 & 2.11 & 1.76 & 1.62 & 0.57 & 0.10 & 0.29 & 0.8 & 0.00 & 0.62 & 0.31 & -970$^{+370}_{-60}$ & N \\ 
43 & 0.87 & 6.25 & 0.57 & 1.18 & -0.23 & -0.15 & 0.70 & 0.5 & 0.39 & 0.65 & 0.38 & -230$^{+190}_{-100}$ & N \\ 
46 & 2.32 & 0.40 & 0.55 & 1.16 & -2.48 & -0.93 & 0.53 & 0.7 & 0.02 & 0.32 & 0.38 & -440$^{+0}_{-0}$ & Q   \\ 
47 & 30.35 & 0.06 & 1.05 & 1.68 & -0.25 & -0.01 & 0.11 & 0.9 & 0.15 & 0.33 & 0.51 & -170$^{+0}_{-0}$ & N \\ 
48 & 0.72 & 1.91 & 0.98 & 1.17 & -1.45 & -1.24 & 0.00 & 0.9 & 0.00 & 0.32 & 0.35 & -370$^{+10}_{-150}$ & N/Q \\ 
50 & 0.41 & 28.31 & 8.90 & 2.01 & -2.59 & -0.39 & 0.41 & 0.6 & 0.00 & 0.77 & 0.39 & -450$^{+40}_{-10}$ & N/Q \\ 
51 & 0.78 & 5.97 & 3.95 & 2.54 & -0.09 & -0.02 & 0.43 & 0.4 & 0.12 & 0.85 & 0.47 & -220$^{+70}_{-10}$ & N \\ 
52 & 4.72 & 0.27 & 0.85 & 1.35 & 2.32 & 0.44 & 0.09 & 0.8 & 0.77 & 0.64 & 0.50 & -100$^{+110}_{-20}$ & N/Q \\ 
\enddata
\tablecomments{
Terms: $t_{1/2}$ is the Full-Width Half-Max timing, $\mathcal{I}$ is impulsiveness, L$^{\text{peak}}$ is the luminosity at the peak of the flare, SNES is the maximum cumulative Neupert Effect Score, D is the maximum difference in the K-S Test, $p_{\text{K-S}}$ is the p-value of the K-S Test, $r_c$ is the sample Pearson coefficient between the cumulative UVW2 response and X-ray response, and $r^{\text{max}}_{\text{Lag}}$ is the maximum sample Pearson coefficient calculated during the cross-correlation of the light curves around a flare. \\
(1) XMM OM UVW2 values, specifically. 
(2) Time shift ranges are changed to 500 seconds for Flares 46 \& 52, 1,500 for Flare 16, 2,000 for Flares 5, 21, \& 23, and 3,000 for Flares 15, 25, \& 28.
(3) N -- Neupert flare (pass both timing criteria), Q -- Quasi-Neupert (pass neither). N/Q pass only one timing criteria. These should generally be considered Neupert, but continuing studies will further refine placement.
}
\end{deluxetable}

\begin{deluxetable}{llrrr}
\tablewidth{0pt}
\tablecaption{Multi-Wavelength Classification Statistics
\label{tbl:neupert_classifications}}
\tablehead{
\colhead{Classification} & \colhead{Label} & \colhead{Count} & \colhead{Percent Total} & \colhead{Percent Adjusted} 
}
\startdata 
Neupert                          & N     & 9   & $20\%$  & $43\%$ \\
Neupert (passes one criteria)    & N/Q   & 7   & $15\%$  & $33\%$ \\
Quasi-Neupert                    & Q     & 5   & $11\%$  & $24\%$ \\
Non-Neupert Type I               & NN-I  & 12  & $26\%$  & \nodata \\
Non-Neupert Type II              & NN-II & 13  & $28\%$  & \nodata \\
Undetermined                     & Un    & 5   & \nodata & \nodata \\
\enddata
\tablecomments{ 
Classifications for the 51 flares analyzed in the \xmm{} UVW2 and X-ray data. 
As timing choices can lead to significant changes in the Neupert and Quasi-Neupert categories, flares that passed only one Neupert timing criteria are labeled with N/Q for posterity.
$n_{\text{Un}}$ are undetermined due to missing UVW2 times.
Percent Total uses the 46 flares with data available for both observations, while Percent Adjusted only uses the 21 flares with responses in both. }
\end{deluxetable}

\begin{figure}[!ht]
\centering
\includegraphics[width=0.6\linewidth]{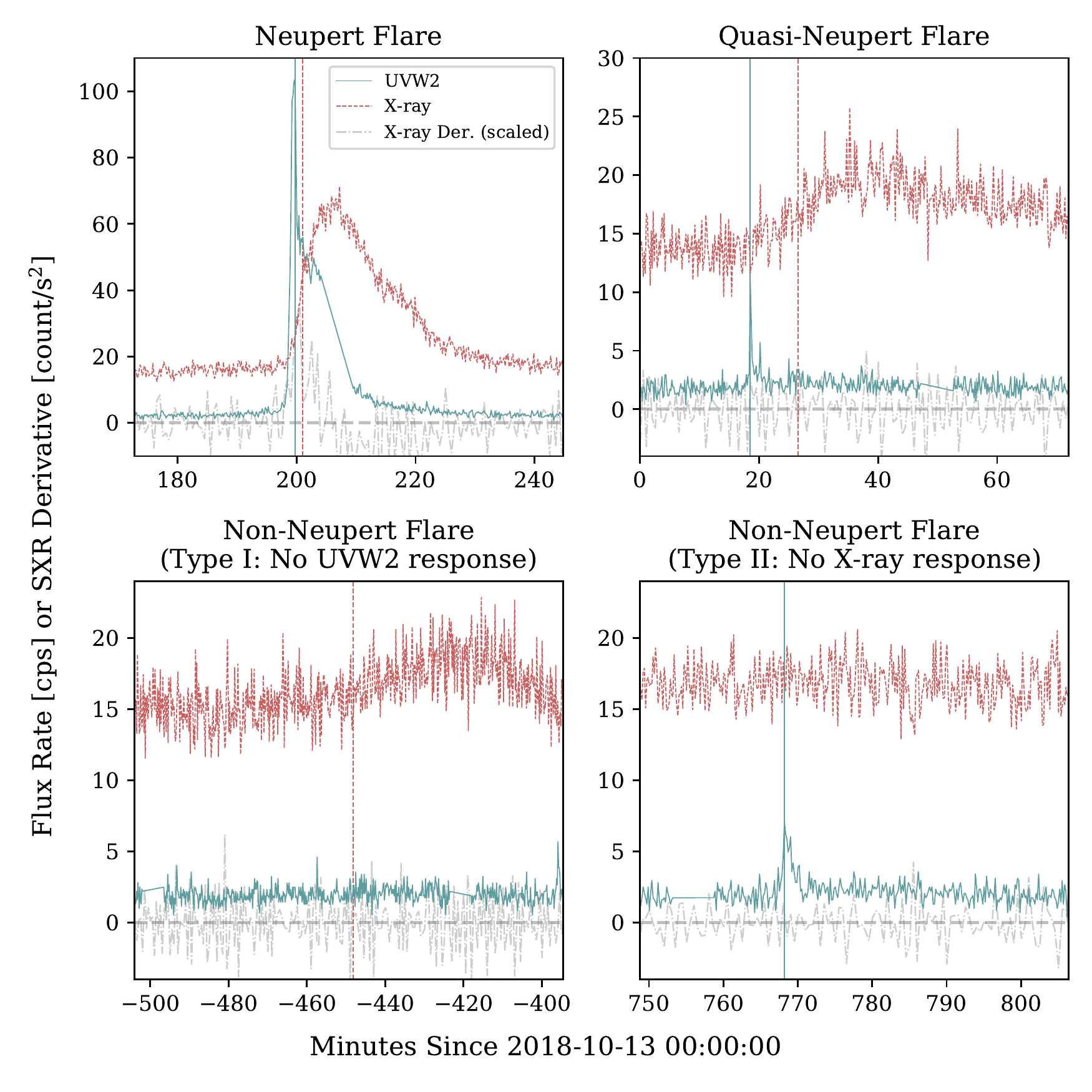}
\caption{We propose a four part Neupert classification system based on this \xmm{} UVW2 and X-ray data. The dashed vertical lines show the times of the UVW2 and X-ray derivative peaks. \\
\emph{Top}: The left panel shows a Neupert Flare, where the peak timings of the UVW2 and X-ray time derivative nearly coincide. The right panel shows a Quasi-Neupert Flare, where a clear flare is present in both, but these peak timings do not match. \\
\emph{Bottom}: These lower panels show Non-Neupert Flares, where either the UVW2 or X-ray response is missing (Type I and Type II on the left and right, respectively).
\label{fig:neuperttypes}
}
\end{figure}

\clearpage

\appendix
\section{Calibration of the Quiescent Flux of AU Mic in $U$, $V$, and UVW2} \label{sec:appendix_fluxcalc}

\subsection{OM/UVW2}

Following the SAS calibration tables\footnote{\url{https://www.cosmos.esa.int/web/xmm-newton/sas-watchout-uvflux}}, count rates from XMM OM UVW2 are multiplied by $5.71\times10^{-15}$ to convert to the flux at Earth (\funits{} ).  This conversion was determined for white dwarf standard stars, which is accurate to 10\%.  We confirm that this calibration is reasonable for AU Mic by integrating the HST/FOS spectrum of AU Mic over the UVW2 effective area curve (Eq.\ \ref{eq:fluxconv}), giving a quiescent flux of $1.14-1.18 \times 10^{-14}$ \funits{},  which is consistent with the adopted value of $1.03 \times 10^{-14}$ \funits{} that is calculated from the SAS count rate conversion.  The mean wavelength of UVW2 is $\approx 2150$ \AA{} and the FWHM is $475$ \AA{} giving a quiescent XMM OM UVW2 luminosity of $5.5 \times 10^{28}$ erg s$^{-1}$.  For reference, the \swift{} UVOT W2 quiescent flux that we calculate from the quiescent AU Mic spectrum is $2.1 \times 10^{-14}$ \funits{} .  Compared to the XMM OM UVW2, \swift{} W2 has a factor of eight larger peak effective area, a shorter mean wavelength, but a more pronounced red tail (``leak'') that results in larger bandpass-weighted fluxes for very red stars during quiescence.

\subsection{$U$ band}

 The apparent $U$-band magnitude of AU Mic is estimated in several ways.  First, we use the (Johnson) $U-B$ color of 1.11 that was reported in \citet{Leggett1992} from a compilation of measurements in \citet{The1984} and \citet{Celis1986}.  This color is consistent with the  colors that have been reported more recently \citep{Cut,Cut2,Cut3}.  Given a $B$-band magnitude of 10.26 from \citet{Leggett1992} and \citet{Reid2004} and the Johnson $U$-band  zeropoint of $4.3\times10^{-9}$ \funits{}
  from \citet{Willmer2018}, an estimated quiescent $U$-band magnitude is  thus $11.37$, which corresponds to a flux of $1.22 \times10^{-13}$ \funits{}.  Second, we compare the quiescent count rate of AU Mic to the comparison A9V star HD 197673, obtaining a count-rate ratio of $0.126$ in quiescence.  We estimate the Johnson $U$-band magnitude of this star to be $9.0$ using representative colors\footnote{Compiled on the website at \url{https://www.stsci.edu/\~inr/intrins.html}.} from \citet{Ducati2001} and the transformations from its measured Tycho-2 $B_T$ and $V_T$ magnitudes\footnote{\url{http://vizier.cfa.harvard.edu/ftp/cats/I/239/version_cd/docs/vol1/sect1_03.pdf}}.  This gives a quiescent $U$-band magnitude of 11.22 and a flux of $1.4 \times10^{-13}$ \funits{} for AU Mic.  Third, we integrate over the HST/FOS spectrum of AU Mic with the LCO $U$-band transmission curve, which is nearly identical to the standard Bessel $U$ filter from \cite{Bessel2013}, obtaining a quiescent flux of $1.35\times10^{-13}$ \funits{} above the atmosphere and $1.41 \times10^{-13}$ \funits{} at a representative airmass of sec $z = 1.35$.  We thus adopt a value of $1.3\pm0.1\times10^{-13}$ \funits{} as the quiescent $U-$band flux of AU Mic.  Using the distance of 9.72 pc from Gaia DR2 \citep{Gaia, GaiaDR2}, this flux corresponds to a quiescent luminosity  of $L_U \sim 10^{30}$ erg s$^{-1}$ for a filter FWHM of 700 \AA{}.

 \subsection{$V$ band} \label{sec:vband}

There is a significant variation of single-epoch, $V$-band magnitudes ranging from $8.6$ to $8.8$ for AU Mic in the literature, which is likely in part due to the intrinsic, rotational flux modulation  \citep{Cut2, Cut3,Cut4, Hebb2007, Ibanez2019};  see also the peak-to-trough white-light variation in TESS data \citep{Wisniewski2019}.  There is also a range of $V-I_C$ colors from 2.00 to 2.09 in the literature \citep{Cut4,Winters2015,Leggett1992}; some of this variation may be due to color changes due to rotational modulation as well \citep{Cut4,Hebb2007}.    Most recently, \citet{Winters2015} find that $V=8.65$, which is consistent with the All Sky Automatic Survey (ASAS) value of 8.63$\pm0.04$ \citep[][and see discussion and Figs 3 and 4 of \citealt{Ibanez2019}]{Poj, ASAS}.  
\citet{Reid2004}\footnote{\url{http://www.stsci.edu/\~inr/phot/allphotpi.sing.2mass}} quote $BVR$ magnitudes of AU Mic that are systematically fainter; their value of $V=8.81$ comes from \citet{Leggett1992}, who compiled Johnson photometry from \citet{Celis1986}.  
A fainter magnitude is also quoted in the Hipparcos catalog, with $V_J = 8.8$ \citep{Hipparcos, Hipparcos2012}, and variation of $H_p$ from 8.71 to 8.82 is notable over the 70 observations in Hipparcos \citep{Hip97a, Hip97Cat}.  
A value of $V=8.65$ \citep{Winters2015}  is consistent with the Gaia DR2 \citep{Gaia, GaiaDR2} magnitude that is transformed\footnote{\url{https://gea.esac.esa.int/archive/documentation/GDR2/Data_processing/chap_cu5pho/sec_cu5pho_calibr/ssec_cu5pho_PhotTransf.html}} to $V$. 
We thus take a conservative range of $V$-band magnitudes for the systematic uncertainty in the quiescent flux of AU Mic.  Using the zeropoints from \citet{Willmer2018}, we calculate the quiescent flux at Earth in the $V$ band of AU Mic to be $1.1-1.33\times10^{-12}$ \funits{} for a magnitude range of 8.6 to 8.8.  Integrating the  HST/FOS spectrum of AU Mic over the SMARTS/CTIO $V$-band transmission curve gives $1.22\times10^{-12}$ \funits{} in the middle of these estimates.  We thus adopt $1.2 \pm 0.1\times10^{-12}$ \funits{} as the quiescent flux\footnote{This is consistent with the range of the fluxes quoted in the Simbad photometry viewer tool: \url{http://vizier.unistra.fr/vizier/sed/}.}  of AU Mic in the Bessel $V$-band, giving $L_V=1.2\times10^{31}$ erg s$^{-1}$ through a filter with FWHM of 850 \AA{}.

A comparison of the LCO $V$ band, SMARTS $V$ band, and the standard \citet{Bessel2013} $V$-band transmission curves is shown in Figure \ref{fig:filters}.  The LCO $V$ band is slightly bluer than the SMARTS $V$ band and the standard Bessel $V$ band.  We integrate the AU Mic HST/FOS spectrum and find that the filter-weighted\footnote{We assume that all filter transmission curves are provided as appropriate for photon-counting CCDs;  many filter curve files in the VSO database are listed as ``energy counting'' for historical reasons only and are intended to be corrected as appropriate in future updates (C.\ Rodrigo, priv.\ communication, 2019). } flux density \citep[][see also \citealt{Bessel2013}]{Sirianni2005} in the LCO $V$ band is lowest by 10\%, while the other two synthesized $V$-band fluxes are within 1\% of each other.  We test this against the dM1e template spectrum from \citet{Bochanski2007} as well.   Note that the (intractable) sensitivity differences between the CCDs and telescope optics are not accounted for in this comparison.

\section{Flare Quantities} \label{app:flares}

\begin{longrotatetable}
\begin{deluxetable}{cllccccccc}
\tabletypesize{\scriptsize}
\tablewidth{0pt}
\tablecaption{Common Flare Properties \label{tbl:flare_properties}}
\tablehead{
\colhead{ID} & \colhead{Obs.} & \colhead{Band} & \colhead{$t_{\text{Total}}$} & \colhead{$t_{\text{Rise}}$} & \colhead{Peak Rate}  & \colhead{Energy} & \colhead{Start Time} & \colhead{End Time} & \colhead{Peak Time} \\
\colhead{} & \colhead{} & \colhead{} & \colhead{min} & \colhead{min} &  \colhead{(1)} & \colhead{erg} & \colhead{UTC} & \colhead{UTC} & \colhead{UTC}
}
\startdata 
1 & OM & UVW2 & 6.83 & 2.17 & 3.67 & $(1.8\pm0.1) \times 10^{31}$ & 2018-10-10T13:28:09.760 & 2018-10-10T13:34:59.760 & 2018-10-10T13:30:19.760 \\
 & EPIC-pn & X-ray & 33.00 & 10.33 & 6.68 & $(10.0\pm2.2) \times 10^{31}$ & 2018-10-10T13:29:17.030 & 2018-10-10T14:02:17.030 & 2018-10-10T13:39:37.030 \\
 & EPIC-pn & X-ray Der. & \nodata & \nodata & \nodata & \nodata & \nodata & 2018-10-10T13:34:57.030 & 2018-10-10T13:30:07.030 \\
 & RGS & X-ray & 33.00 & 21.50 & 0.59 & \nodata & 2018-10-10T13:29:18.868 & 2018-10-10T14:02:18.868 & 2018-10-10T13:50:48.868 \\
2 & EPIC-pn & X-ray & 143.83 & 30.50 & 8.05 & $(2.1\pm0.5) \times 10^{32}$ & 2018-10-10T14:24:07.030 & 2018-10-10T16:47:57.030 & 2018-10-10T14:54:37.030 \\
 & EPIC-pn & X-ray Der. & \nodata & \nodata & \nodata & \nodata & \nodata & 2018-10-10T14:59:57.030 & 2018-10-10T14:43:37.030 \\
3 & OM & UVW2 & 3.50 & 0.83 & 2.38 & $(8.9\pm0.4) \times 10^{30}$ & 2018-10-10T18:49:16.656 & 2018-10-10T18:52:46.656 & 2018-10-10T18:50:06.656 \\
 & LCOGT & U & 5.37 & 3.82 & 0.11 & $(2.1\pm0.1) \times 10^{31}$ & 2018-10-10T18:47:58.000 & 2018-10-10T18:53:20.000 & 2018-10-10T18:51:47.000 \\
4 & EPIC-pn & X-ray & 57.50 & 11.50 & 4.97 & $(7.3\pm1.6) \times 10^{31}$ & 2018-10-10T19:55:17.030 & 2018-10-10T20:52:47.030 & 2018-10-10T20:06:47.030 \\
 & EPIC-pn & X-ray Der. & \nodata & \nodata & \nodata & \nodata & \nodata & 2018-10-10T20:23:57.030 & 2018-10-10T19:57:17.030 \\
5 & OM & UVW2 & 1.67 & 1.00 & 4.89 & $(5.9\pm0.3) \times 10^{30}$ & 2018-10-10T21:50:40.525 & 2018-10-10T21:52:20.525 & 2018-10-10T21:51:40.525 \\
 & EPIC-pn & X-ray & 35.83 & 18.17 & 7.17 & $(7.6\pm1.7) \times 10^{31}$ & 2018-10-10T21:46:07.030 & 2018-10-10T22:21:57.030 & 2018-10-10T22:04:17.030 \\
 & EPIC-pn & X-ray Der. & \nodata & \nodata & \nodata & \nodata & \nodata & 2018-10-10T22:04:47.030 & 2018-10-10T21:54:57.030 \\
6 & OM & UVW2 & 1.00 & 0.67 & 1.94 & $(2.3\pm0.3) \times 10^{30}$ & 2018-10-10T22:01:30.525 & 2018-10-10T22:02:30.525 & 2018-10-10T22:02:10.525 \\
7 & OM & UVW2 & 3.00 & 0.50 & 3.13 & $(7.6\pm0.4) \times 10^{30}$ & 2018-10-11T00:15:54.389 & 2018-10-11T00:18:54.389 & 2018-10-11T00:16:24.389 \\
8 & OM & UVW2 & 3.17 & 0.67 & 24.70 & $(3.5\pm0.0) \times 10^{31}$ & 2018-10-11T08:39:07.720 & 2018-10-11T08:42:17.720 & 2018-10-11T08:39:47.720 \\
 & EPIC-pn & X-ray & 9.17 & 3.67 & 8.35 & $(2.5\pm0.6) \times 10^{31}$ & 2018-10-11T08:39:57.030 & 2018-10-11T08:49:07.030 & 2018-10-11T08:43:37.030 \\
 & EPIC-pn & X-ray Der. & \nodata & \nodata & \nodata & \nodata & \nodata & 2018-10-11T08:51:17.030 & 2018-10-11T08:40:17.030 \\
 & RGS & X-ray & 8.50 & 2.00 & 0.84 & \nodata & 2018-10-11T08:40:18.868 & 2018-10-11T08:48:48.868 & 2018-10-11T08:42:18.868 \\
9 & OM & UVW2 & 4.67 & 2.50 & 7.85 & $(2.3\pm0.0) \times 10^{31}$ & 2018-10-11T14:03:56.240 & 2018-10-11T14:08:36.240 & 2018-10-11T14:06:26.240 \\
10 & OM & UVW2 & 0.50 & 0.17 & 2.72 & $(1.8\pm0.2) \times 10^{30}$ & 2018-10-11T14:11:36.240 & 2018-10-11T14:12:06.240 & 2018-10-11T14:11:46.240 \\
11 & EPIC-pn & X-ray & 482.33 & 331.50 & 18.70 & $(2.8\pm0.6) \times 10^{33}$ & 2018-10-11T14:02:27.030 & 2018-10-11T22:04:47.030 & 2018-10-11T19:33:57.030 \\
 & EPIC-pn & X-ray Der. & \nodata & \nodata & \nodata & \nodata & \nodata & 2018-10-11T19:11:57.030 & 2018-10-11T17:16:27.030 \\
 & RGS & X-ray & 481.50 & 328.50 & 1.15 & \nodata & 2018-10-11T14:02:48.868 & 2018-10-11T22:04:18.868 & 2018-10-11T19:31:18.868 \\
12 & OM & UVW2 & 2.67 & 0.50 & 11.90 & $(2.7\pm0.0) \times 10^{31}$ & 2018-10-11T17:14:00.205 & 2018-10-11T17:16:40.205 & 2018-10-11T17:14:30.205 \\
13 & OM & UVW2 & 1.33 & 0.83 & 4.95 & $(4.8\pm0.3) \times 10^{30}$ & 2018-10-11T18:25:17.162 & 2018-10-11T18:26:37.162 & 2018-10-11T18:26:07.162 \\
14 & OM & UVW2 & 3.33 & 0.17 & 3.55 & $(6.1\pm0.5) \times 10^{30}$ & 2018-10-11T19:31:53.637 & 2018-10-11T19:35:13.637 & 2018-10-11T19:32:03.637 \\
15 & OM & UVW2 & 110.93 & 54.63 & 6.99 & $(5.0\pm0.4) \times 10^{32}$ & 2018-10-11T23:51:23.346 & 2018-10-12T01:42:19.177 & 2018-10-12T00:46:01.280 \\
 & EPIC-pn & X-ray & 123.17 & 90.17 & 36.57 & $(2.5\pm0.5) \times 10^{33}$ & 2018-10-11T23:51:27.030 & 2018-10-12T01:54:37.030 & 2018-10-12T01:21:37.030 \\
 & EPIC-pn & X-ray Der. & \nodata & \nodata & \nodata & \nodata & \nodata & 2018-10-12T01:52:27.030 & 2018-10-12T00:55:57.030 \\
 & RGS & X-ray & 127.39 & 85.00 & 1.83 & \nodata & 2018-10-11T23:51:48.868 & 2018-10-12T01:59:12.093 & 2018-10-12T01:16:48.868 \\
16 & OM & UVW2 & 1.83 & 1.17 & 2.51 & $(3.2\pm0.3) \times 10^{30}$ & 2018-10-12T14:43:20.711 & 2018-10-12T14:45:10.711 & 2018-10-12T14:44:30.711 \\
 & EPIC-pn & X-ray & 43.00 & 24.67 & 4.92 & $(4.4\pm1.0) \times 10^{31}$ & 2018-10-12T14:38:33.020 & 2018-10-12T15:21:33.020 & 2018-10-12T15:03:13.020 \\
 & EPIC-pn & X-ray Der. & \nodata & \nodata & \nodata & \nodata & \nodata & 2018-10-12T15:07:03.020 & 2018-10-12T14:45:43.020 \\
17 & EPIC-pn & X-ray & 57.33 & 38.00 & 7.80 & $(1.6\pm0.3) \times 10^{32}$ & 2018-10-12T16:26:33.020 & 2018-10-12T17:23:53.020 & 2018-10-12T17:04:33.020 \\
 & EPIC-pn & X-ray Der. & \nodata & \nodata & \nodata & \nodata & \nodata & 2018-10-12T16:55:03.020 & 2018-10-12T16:31:53.020 \\
18 & OM & UVW2 & 9.67 & 0.67 & 3.77 & $(1.7\pm0.1) \times 10^{31}$ & 2018-10-12T17:23:26.626 & 2018-10-12T17:33:06.626 & 2018-10-12T17:24:06.626 \\
 & EPIC-pn & X-ray & 35.83 & 9.17 & 8.32 & $(9.3\pm2.0) \times 10^{31}$ & 2018-10-12T17:24:03.020 & 2018-10-12T17:59:53.020 & 2018-10-12T17:33:13.020 \\
 & EPIC-pn & X-ray Der. & \nodata & \nodata & \nodata & \nodata & \nodata & 2018-10-12T17:38:23.020 & 2018-10-12T17:31:03.020 \\
19 & OM & UVW2 & 2.33 & 1.83 & 2.67 & $(4.3\pm0.4) \times 10^{30}$ & 2018-10-12T19:43:59.035 & 2018-10-12T19:46:19.035 & 2018-10-12T19:45:49.035 \\
20 & OM & UVW2 & 17.83 & 0.83 & 2.38 & $(2.8\pm0.1) \times 10^{31}$ & 2018-10-12T20:11:09.035 & 2018-10-12T20:28:59.035 & 2018-10-12T20:11:59.035 \\
 & EPIC-pn & X-ray & 201.33 & 15.33 & 8.09 & $(4.6\pm1.0) \times 10^{32}$ & 2018-10-12T20:09:43.020 & 2018-10-12T23:31:03.020 & 2018-10-12T20:25:03.020 \\
 & EPIC-pn & X-ray Der. & \nodata & \nodata & \nodata & \nodata & \nodata & 2018-10-12T20:23:53.020 & 2018-10-12T20:18:03.020 \\
21 & OM & UVW2 & 1.67 & 0.50 & 8.82 & $(7.1\pm0.3) \times 10^{30}$ & 2018-10-13T00:18:00.903 & 2018-10-13T00:19:40.903 & 2018-10-13T00:18:30.903 \\
 & EPIC-pn & X-ray & 64.67 & 13.50 & 11.60 & $(2.4\pm0.5) \times 10^{32}$ & 2018-10-13T00:21:43.020 & 2018-10-13T01:26:23.020 & 2018-10-13T00:35:13.020 \\
 & EPIC-pn & X-ray Der. & \nodata & \nodata & \nodata & \nodata & \nodata & 2018-10-13T00:35:53.020 & 2018-10-13T00:26:33.020 \\
 & LCOGT & U & 47.98 & 5.33 & 0.12 & $(1.1\pm0.0) \times 10^{32}$ & 2018-10-13T00:15:01.000 & 2018-10-13T01:03:00.000 & 2018-10-13T00:20:21.000 \\
22 & OM & UVW2 & 9.67 & 5.17 & 36.47 & $(1.2\pm0.0) \times 10^{32}$ & 2018-10-13T01:38:26.759 & 2018-10-13T01:48:06.759 & 2018-10-13T01:43:36.759 \\
 & EPIC-pn & X-ray & 46.67 & 19.33 & 11.50 & $(1.9\pm0.4) \times 10^{32}$ & 2018-10-13T01:37:13.020 & 2018-10-13T02:23:53.020 & 2018-10-13T01:56:33.020 \\
 & EPIC-pn & X-ray Der. & \nodata & \nodata & \nodata & \nodata & \nodata & 2018-10-13T02:09:33.020 & 2018-10-13T01:43:23.020 \\
 & RGS & X-ray & 60.80 & 19.20 & 0.63 & \nodata & 2018-10-13T01:37:13.491 & 2018-10-13T02:38:01.491 & 2018-10-13T01:56:25.491 \\
 & LCOGT & U & 48.12 & 8.33 & 0.71 & $(2.6\pm0.0) \times 10^{32}$ & 2018-10-13T01:35:41.000 & 2018-10-13T02:23:48.000 & 2018-10-13T01:44:01.000 \\
 & 2KCCD & V & 13.12 & 6.09 & 0.06 & $(2.0\pm0.1) \times 10^{32}$ & 2018-10-13T01:37:29.285 & 2018-10-13T01:50:36.263 & 2018-10-13T01:43:34.439 \\
 & CHIRON & H$\alpha$ & 51.29 & 6.45 & 0.95 & \nodata & 2018-10-13T01:37:45.300 & 2018-10-13T02:29:02.800 & 2018-10-13T01:44:12.100 \\
23 & OM & UVW2 & 40.78 & 8.50 & 101.39 & $(9.7\pm0.3) \times 10^{32}$ & 2018-10-13T03:11:14.735 & 2018-10-13T03:52:01.749 & 2018-10-13T03:19:44.735 \\
 & EPIC-pn & X-ray & 61.67 & 9.83 & 55.61 & $(9.6\pm2.1) \times 10^{32}$ & 2018-10-13T03:17:23.020 & 2018-10-13T04:19:03.020 & 2018-10-13T03:27:13.020 \\
 & EPIC-pn & X-ray Der. & \nodata & \nodata & \nodata & \nodata & \nodata & 2018-10-13T04:04:43.020 & 2018-10-13T03:21:03.020 \\
 & RGS & X-ray & 61.33 & 10.13 & 2.31 & \nodata & 2018-10-13T03:17:29.491 & 2018-10-13T04:18:49.491 & 2018-10-13T03:27:37.491 \\
 & LCOGT & U & 68.28 & 27.67 & 2.42 & $(1.6\pm0.0) \times 10^{33}$ & 2018-10-13T02:53:02.000 & 2018-10-13T04:01:19.000 & 2018-10-13T03:20:42.000 \\
 & LCOGT & V & 17.46 & 4.98 & 0.16 & $(9.5\pm0.1) \times 10^{32}$ & 2018-10-13T03:18:13.367 & 2018-10-13T03:35:40.717 & 2018-10-13T03:23:12.377 \\
 & 2KCCD & V & 25.44 & 5.33 & 0.13 & $(9.8\pm0.1) \times 10^{32}$ & 2018-10-13T03:17:03.017 & 2018-10-13T03:42:29.470 & 2018-10-13T03:22:22.524 \\
 & CHIRON & H$\alpha$ & 75.86 & 24.89 & 2.34 & \nodata & 2018-10-13T02:57:26.600 & 2018-10-13T04:13:18.500 & 2018-10-13T03:22:19.800 \\
24 & OM & UVW2 & 29.13 & 11.96 & 4.81 & $(1.5\pm0.3) \times 10^{32}$ & 2018-10-13T07:16:25.715 & 2018-10-13T07:45:33.505 & 2018-10-13T07:28:23.505 \\
 & EPIC-pn & X-ray & 100.67 & 53.67 & 17.65 & $(5.0\pm1.1) \times 10^{32}$ & 2018-10-13T06:43:13.020 & 2018-10-13T08:23:53.020 & 2018-10-13T07:36:53.020 \\
 & EPIC-pn & X-ray Der. & \nodata & \nodata & \nodata & \nodata & \nodata & 2018-10-13T08:09:33.020 & 2018-10-13T07:26:53.020 \\
 & RGS & X-ray & 100.27 & 51.73 & 1.05 & \nodata & 2018-10-13T06:43:21.491 & 2018-10-13T08:23:37.491 & 2018-10-13T07:35:05.491 \\
25 & OM & UVW2 & 2.67 & 0.83 & 11.47 & $(2.1\pm0.0) \times 10^{31}$ & 2018-10-13T10:21:36.573 & 2018-10-13T10:24:16.573 & 2018-10-13T10:22:26.573 \\
 & EPIC-pn & X-ray & 30.00 & 23.83 & 9.87 & $(1.2\pm0.3) \times 10^{32}$ & 2018-10-13T10:20:43.020 & 2018-10-13T10:50:43.020 & 2018-10-13T10:44:33.020 \\
 & EPIC-pn & X-ray Der. & \nodata & \nodata & \nodata & \nodata & \nodata & 2018-10-13T10:40:43.020 & 2018-10-13T10:25:23.020 \\
26 & OM & UVW2 & 3.83 & 1.33 & 5.05 & $(1.2\pm0.0) \times 10^{31}$ & 2018-10-13T12:46:52.338 & 2018-10-13T12:50:42.338 & 2018-10-13T12:48:12.338 \\
 & LCOGT & U & 11.06 & 2.38 & 0.13 & $(1.5\pm0.1) \times 10^{31}$ & 2018-10-13T12:46:12.237 & 2018-10-13T12:57:16.096 & 2018-10-13T12:48:35.081 \\
27 & OM & UVW2 & 17.33 & 4.00 & 21.62 & $(1.0\pm0.0) \times 10^{32}$ & 2018-10-13T13:59:27.286 & 2018-10-13T14:16:47.286 & 2018-10-13T14:03:27.286 \\
 & EPIC-pn & X-ray & 48.83 & 9.17 & 14.71 & $(3.1\pm0.7) \times 10^{32}$ & 2018-10-13T13:59:33.020 & 2018-10-13T14:48:23.020 & 2018-10-13T14:08:43.020 \\
 & EPIC-pn & X-ray Der. & \nodata & \nodata & \nodata & \nodata & \nodata & 2018-10-13T14:23:53.020 & 2018-10-13T14:03:33.020 \\
 & RGS & X-ray & 48.53 & 11.20 & 0.94 & \nodata & 2018-10-13T13:59:37.491 & 2018-10-13T14:48:09.491 & 2018-10-13T14:10:49.491 \\
28 & OM & UVW2 & 5.67 & 2.00 & 5.74 & $(1.9\pm0.1) \times 10^{31}$ & 2018-10-13T17:54:17.130 & 2018-10-13T17:59:57.130 & 2018-10-13T17:56:17.130 \\
 & EPIC-pn & X-ray & 195.67 & 28.17 & 8.36 & $(5.7\pm1.3) \times 10^{32}$ & 2018-10-13T17:51:23.020 & 2018-10-13T21:07:03.020 & 2018-10-13T18:19:33.020 \\
 & EPIC-pn & X-ray Der. & \nodata & \nodata & \nodata & \nodata & \nodata & 2018-10-13T18:28:43.020 & 2018-10-13T17:55:43.020 \\
29 & OM & UVW2 & 4.50 & 3.00 & 2.29 & $(7.3\pm0.5) \times 10^{30}$ & 2018-10-13T19:16:04.636 & 2018-10-13T19:20:34.636 & 2018-10-13T19:19:04.636 \\
30 & EPIC-pn & X-ray & 86.33 & 17.17 & 5.70 & $(1.9\pm0.4) \times 10^{32}$ & 2018-10-13T21:07:13.020 & 2018-10-13T22:33:33.020 & 2018-10-13T21:24:23.020 \\
 & EPIC-pn & X-ray Der. & \nodata & \nodata & \nodata & \nodata & \nodata & 2018-10-13T21:50:23.020 & 2018-10-13T21:39:03.020 \\
31 & OM & UVW2 & 2.33 & 0.67 & 3.22 & $(3.6\pm0.3) \times 10^{30}$ & 2018-10-13T23:20:26.268 & 2018-10-13T23:22:46.268 & 2018-10-13T23:21:06.268 \\
32 & EPIC-pn & X-ray & 100.67 & 54.50 & 6.04 & $(1.0\pm0.2) \times 10^{32}$ & 2018-10-13T23:16:53.020 & 2018-10-14T00:57:33.020 & 2018-10-14T00:11:23.020 \\
 & EPIC-pn & X-ray Der. & \nodata & \nodata & \nodata & \nodata & \nodata & 2018-10-14T00:14:23.020 & 2018-10-14T00:10:03.020 \\
33 & LCOGT & U & 17.45 & 6.08 & 0.05 & $(2.2\pm0.1) \times 10^{31}$ & 2018-10-13T23:53:16.000 & 2018-10-14T00:10:43.000 & 2018-10-13T23:59:21.000 \\
34 & EPIC-pn & X-ray & 180.17 & 83.83 & 10.34 & $(5.9\pm1.3) \times 10^{32}$ & 2018-10-14T12:21:16.970 & 2018-10-14T15:21:26.970 & 2018-10-14T13:45:06.970 \\
 & EPIC-pn & X-ray Der. & \nodata & \nodata & \nodata & \nodata & \nodata & 2018-10-14T13:55:06.970 & 2018-10-14T12:23:16.970 \\
 & RGS & X-ray & 214.80 & 97.20 & 0.67 & \nodata & 2018-10-14T12:15:11.942 & 2018-10-14T15:49:59.942 & 2018-10-14T13:52:23.942 \\
35 & EPIC-pn & X-ray & 35.83 & 8.83 & 21.00 & $(2.6\pm0.6) \times 10^{32}$ & 2018-10-15T01:12:06.970 & 2018-10-15T01:47:56.970 & 2018-10-15T01:20:56.970 \\
 & EPIC-pn & X-ray Der. & \nodata & \nodata & \nodata & \nodata & \nodata & 2018-10-15T01:26:16.970 & 2018-10-15T01:16:56.970 \\
 & RGS & X-ray & 42.60 & 12.60 & 0.91 & \nodata & 2018-10-15T01:12:11.942 & 2018-10-15T01:54:47.942 & 2018-10-15T01:24:47.942 \\
 & LCOGT & U & 29.73 & 6.22 & 0.67 & $(2.8\pm0.0) \times 10^{32}$ & 2018-10-15T01:12:14.000 & 2018-10-15T01:41:58.000 & 2018-10-15T01:18:27.000 \\
 & LCOGT & V & 11.53 & 4.15 & 0.05 & $(1.4\pm0.1) \times 10^{32}$ & 2018-10-15T01:14:19.067 & 2018-10-15T01:25:50.656 & 2018-10-15T01:18:28.089 \\
 & 2KCCD & V & 16.14 & 3.80 & 0.04 & $(1.6\pm0.1) \times 10^{32}$ & 2018-10-15T01:14:12.956 & 2018-10-15T01:30:21.269 & 2018-10-15T01:18:01.184 \\
 & CHIRON & H$\alpha$ & 48.60 & 9.67 & 0.57 & \nodata & 2018-10-15T01:10:11.100 & 2018-10-15T01:58:47.300 & 2018-10-15T01:19:51.400 \\
36 & EPIC-pn & X-ray & 35.83 & 13.17 & 10.17 & $(1.5\pm0.3) \times 10^{32}$ & 2018-10-15T03:14:26.970 & 2018-10-15T03:50:16.970 & 2018-10-15T03:27:36.970 \\
 & EPIC-pn & X-ray Der. & \nodata & \nodata & \nodata & \nodata & \nodata & 2018-10-15T03:21:26.970 & 2018-10-15T03:17:36.970 \\
 & RGS & X-ray & 49.80 & 12.60 & 0.80 & \nodata & 2018-10-15T03:14:35.942 & 2018-10-15T04:04:23.942 & 2018-10-15T03:27:11.942 \\
 & LCOGT & U & 8.35 & 2.27 & 0.21 & $(4.0\pm0.1) \times 10^{31}$ & 2018-10-15T03:15:24.000 & 2018-10-15T03:23:45.000 & 2018-10-15T03:17:40.000 \\
 & CHIRON & H$\alpha$ & 83.40 & 34.56 & 0.32 & \nodata & 2018-10-15T02:54:43.400 & 2018-10-15T04:18:07.500 & 2018-10-15T03:29:17.100 \\
37 & EPIC-pn & X-ray & 151.00 & 13.83 & 12.58 & $(5.8\pm1.3) \times 10^{32}$ & 2018-10-15T07:04:56.970 & 2018-10-15T09:35:56.970 & 2018-10-15T07:18:46.970 \\
 & EPIC-pn & X-ray Der. & \nodata & \nodata & \nodata & \nodata & \nodata & 2018-10-15T07:47:56.970 & 2018-10-15T07:13:06.970 \\
38 & EPIC-pn & X-ray & 21.33 & 10.33 & 5.90 & $(3.8\pm0.8) \times 10^{31}$ & 2018-10-15T10:04:56.970 & 2018-10-15T10:26:16.970 & 2018-10-15T10:15:16.970 \\
 & EPIC-pn & X-ray Der. & \nodata & \nodata & \nodata & \nodata & \nodata & 2018-10-15T10:19:06.970 & 2018-10-15T10:08:26.970 \\
 & LCOGT & U & 9.50 & 3.14 & 0.14 & $(2.3\pm0.1) \times 10^{31}$ & 2018-10-15T10:02:13.657 & 2018-10-15T10:11:43.485 & 2018-10-15T10:05:22.281 \\
39 & EPIC-pn & X-ray & 71.83 & 12.17 & 5.76 & $(8.4\pm1.8) \times 10^{31}$ & 2018-10-15T14:24:06.970 & 2018-10-15T15:35:56.970 & 2018-10-15T14:36:16.970 \\
 & EPIC-pn & X-ray Der. & \nodata & \nodata & \nodata & \nodata & \nodata & 2018-10-15T15:07:06.970 & 2018-10-15T14:35:46.970 \\
40 & EPIC-pn & X-ray & 28.67 & 14.33 & 7.27 & $(6.8\pm1.5) \times 10^{31}$ & 2018-10-15T19:19:16.970 & 2018-10-15T19:47:56.970 & 2018-10-15T19:33:36.970 \\
 & EPIC-pn & X-ray Der. & \nodata & \nodata & \nodata & \nodata & \nodata & 2018-10-15T19:33:26.970 & 2018-10-15T19:26:26.970 \\
 & RGS & X-ray & 28.20 & 10.80 & 0.51 & \nodata & 2018-10-15T19:19:23.942 & 2018-10-15T19:47:35.942 & 2018-10-15T19:30:11.942 \\
 & LCOGT & U & 16.49 & 1.56 & 0.16 & $(9.0\pm0.7) \times 10^{31}$ & 2018-10-15T19:24:12.845 & 2018-10-15T19:40:42.003 & 2018-10-15T19:25:46.278 \\
41 & EPIC-pn & X-ray & 35.83 & 12.17 & 4.07 & $(2.6\pm0.6) \times 10^{31}$ & 2018-10-15T20:31:16.970 & 2018-10-15T21:07:06.970 & 2018-10-15T20:43:26.970 \\
 & EPIC-pn & X-ray Der. & \nodata & \nodata & \nodata & \nodata & \nodata & 2018-10-15T20:45:26.970 & 2018-10-15T20:39:36.970 \\
42 & EPIC-pn & X-ray & 28.67 & 22.67 & 5.61 & $(6.2\pm1.3) \times 10^{31}$ & 2018-10-15T21:07:16.970 & 2018-10-15T21:35:56.970 & 2018-10-15T21:29:56.970 \\
 & EPIC-pn & X-ray Der. & \nodata & \nodata & \nodata & \nodata & \nodata & 2018-10-15T21:21:26.970 & 2018-10-15T21:16:26.970 \\
43 & OM & UVW2 & 1.50 & 0.67 & 1.84 & $(2.3\pm0.2) \times 10^{30}$ & 2018-10-15T22:02:43.412 & 2018-10-15T22:04:13.412 & 2018-10-15T22:03:23.412 \\
 & EPIC-pn & X-ray & 23.67 & 9.83 & 6.10 & $(3.9\pm0.8) \times 10^{31}$ & 2018-10-15T21:57:36.970 & 2018-10-15T22:21:16.970 & 2018-10-15T22:07:26.970 \\
 & EPIC-pn & X-ray Der. & \nodata & \nodata & \nodata & \nodata & \nodata & 2018-10-15T22:11:56.970 & 2018-10-15T22:03:36.970 \\
44 & EPIC-pn & X-ray & 21.50 & 10.17 & 6.33 & $(3.4\pm0.7) \times 10^{31}$ & 2018-10-16T01:33:36.970 & 2018-10-16T01:55:06.970 & 2018-10-16T01:43:46.970 \\
 & EPIC-pn & X-ray Der. & \nodata & \nodata & \nodata & \nodata & \nodata & 2018-10-16T01:47:56.970 & 2018-10-16T01:34:36.970 \\
 & CHIRON & H$\alpha$ & 35.63 & 2.15 & 0.18 & \nodata & 2018-10-16T01:28:47.600 & 2018-10-16T02:04:25.400 & 2018-10-16T01:30:56.700 \\
45 & LCOGT & U & 6.45 & 2.36 & 0.15 & $(2.6\pm0.1) \times 10^{31}$ & 2018-10-16T19:41:28.237 & 2018-10-16T19:47:55.014 & 2018-10-16T19:43:49.791 \\
46 & OM & UVW2 & 2.67 & 1.17 & 1.80 & $(3.6\pm0.4) \times 10^{30}$ & 2018-10-17T00:18:51.362 & 2018-10-17T00:21:31.362 & 2018-10-17T00:20:01.362 \\
 & EPIC-pn & X-ray & 37.33 & 31.17 & 6.01 & $(4.6\pm1.0) \times 10^{31}$ & 2018-10-17T00:20:09.980 & 2018-10-17T00:57:29.980 & 2018-10-17T00:51:19.980 \\
 & EPIC-pn & X-ray Der. & \nodata & \nodata & \nodata & \nodata & \nodata & 2018-10-17T00:43:09.980 & 2018-10-17T00:22:29.980 \\
47 & OM & UVW2 & 30.78 & 22.45 & 3.43 & $(4.0\pm2.6) \times 10^{31}$ & 2018-10-17T01:57:28.367 & 2018-10-17T02:28:15.276 & 2018-10-17T02:19:55.276 \\
 & EPIC-pn & X-ray & 129.50 & 44.00 & 8.67 & $(3.2\pm0.7) \times 10^{32}$ & 2018-10-17T01:26:29.980 & 2018-10-17T03:35:59.980 & 2018-10-17T02:10:29.980 \\
 & EPIC-pn & X-ray Der. & \nodata & \nodata & \nodata & \nodata & \nodata & 2018-10-17T02:38:19.980 & 2018-10-17T02:20:09.980 \\
 & RGS & X-ray & 129.15 & 59.85 & 0.92 & \nodata & 2018-10-17T01:26:34.575 & 2018-10-17T03:35:43.575 & 2018-10-17T02:26:25.575 \\
 & 2KCCD & V & 17.66 & 4.56 & 0.02 & $(1.4\pm0.1) \times 10^{32}$ & 2018-10-17T02:15:58.506 & 2018-10-17T02:33:37.951 & 2018-10-17T02:20:32.345 \\
 & CHIRON & H$\alpha$ & 128.67 & 60.41 & 0.30 & \nodata & 2018-10-17T01:26:11.500 & 2018-10-17T03:34:51.800 & 2018-10-17T02:26:36.400 \\
48 & OM & UVW2 & 1.17 & 0.67 & 3.21 & $(3.6\pm0.3) \times 10^{30}$ & 2018-10-17T04:07:23.271 & 2018-10-17T04:08:33.271 & 2018-10-17T04:08:03.271 \\
 & EPIC-pn & X-ray & 23.00 & 12.67 & 6.06 & $(2.2\pm0.5) \times 10^{31}$ & 2018-10-17T04:04:49.980 & 2018-10-17T04:27:49.980 & 2018-10-17T04:17:29.980 \\
 & EPIC-pn & X-ray Der. & \nodata & \nodata & \nodata & \nodata & \nodata & 2018-10-17T04:19:09.980 & 2018-10-17T04:09:29.980 \\
49 & EPIC-pn & X-ray & 28.67 & 6.83 & 6.73 & $(4.7\pm1.0) \times 10^{31}$ & 2018-10-17T05:45:39.980 & 2018-10-17T06:14:19.980 & 2018-10-17T05:52:29.980 \\
 & EPIC-pn & X-ray Der. & \nodata & \nodata & \nodata & \nodata & \nodata & 2018-10-17T05:59:59.980 & 2018-10-17T05:52:39.980 \\
50 & OM & UVW2 & 6.67 & 0.50 & 29.03 & $(4.0\pm0.1) \times 10^{31}$ & 2018-10-17T08:01:34.494 & 2018-10-17T08:08:14.494 & 2018-10-17T08:02:04.494 \\
 & EPIC-pn & X-ray & 35.83 & 8.00 & 10.35 & $(1.6\pm0.3) \times 10^{32}$ & 2018-10-17T08:02:29.980 & 2018-10-17T08:38:19.980 & 2018-10-17T08:10:29.980 \\
 & EPIC-pn & X-ray Der. & \nodata & \nodata & \nodata & \nodata & \nodata & 2018-10-17T08:16:39.980 & 2018-10-17T08:04:39.980 \\
51 & OM & UVW2 & 5.67 & 2.50 & 12.88 & $(3.2\pm0.1) \times 10^{31}$ & 2018-10-17T08:38:24.494 & 2018-10-17T08:44:04.494 & 2018-10-17T08:40:54.494 \\
 & EPIC-pn & X-ray & 28.67 & 9.00 & 13.13 & $(1.6\pm0.3) \times 10^{32}$ & 2018-10-17T08:38:29.980 & 2018-10-17T09:07:09.980 & 2018-10-17T08:47:29.980 \\
 & EPIC-pn & X-ray Der. & \nodata & \nodata & \nodata & \nodata & \nodata & 2018-10-17T08:45:29.980 & 2018-10-17T08:40:59.980 \\
 & RGS & X-ray & 28.35 & 10.80 & 0.75 & \nodata & 2018-10-17T08:38:34.575 & 2018-10-17T09:06:55.575 & 2018-10-17T08:49:22.575 \\
52 & OM & UVW2 & 5.33 & 0.17 & 2.76 & $(1.0\pm0.1) \times 10^{31}$ & 2018-10-17T14:47:19.184 & 2018-10-17T14:52:39.184 & 2018-10-17T14:47:29.184 \\
 & EPIC-pn & X-ray & 12.00 & 5.33 & 6.99 & $(4.0\pm0.9) \times 10^{31}$ & 2018-10-17T14:42:09.980 & 2018-10-17T14:54:09.980 & 2018-10-17T14:47:29.980 \\
 & EPIC-pn & X-ray Der. & \nodata & \nodata & \nodata & \nodata & \nodata & 2018-10-17T14:52:39.980 & 2018-10-17T14:45:09.980 \\
 & RGS & X-ray & 21.15 & 15.30 & 0.69 & \nodata & 2018-10-17T14:38:34.575 & 2018-10-17T14:59:43.575 & 2018-10-17T14:53:52.575 \\
53 & OM & UVW2 & 1.17 & 0.67 & 2.25 & $(2.9\pm0.3) \times 10^{30}$ & 2018-10-17T15:54:59.184 & 2018-10-17T15:56:09.184 & 2018-10-17T15:55:39.184 \\
54 & LCOGT & U & 44.10 & 4.55 & 0.06 & $(4.9\pm0.3) \times 10^{31}$ & 2018-10-18T01:39:37.000 & 2018-10-18T02:23:43.000 & 2018-10-18T01:44:10.000 \\
 & CHIRON & H$\alpha$ & 103.89 & 30.33 & 0.26 & \nodata & 2018-10-18T01:23:38.400 & 2018-10-18T03:07:32.100 & 2018-10-18T01:53:58.300 \\
55 & CHIRON & H$\alpha$ & 54.25 & 34.73 & 0.24 & \nodata & 2018-10-19T01:11:40.600 & 2018-10-19T02:05:55.900 & 2018-10-19T01:46:24.500 \\
56 & CHIRON & H$\alpha$ & 47.69 & 16.28 & 0.24 & \nodata & 2018-10-19T03:18:31.000 & 2018-10-19T04:06:12.600 & 2018-10-19T03:34:47.600 \\
57 & LCOGT & U & 31.22 & 19.87 & 0.06 & $(3.3\pm0.1) \times 10^{31}$ & 2018-10-21T00:37:12.000 & 2018-10-21T01:08:25.000 & 2018-10-21T00:57:04.000 \\
 & CHIRON & H$\alpha$ & 36.93 & 26.09 & 0.36 & \nodata & 2018-10-21T00:37:55.600 & 2018-10-21T01:14:51.700 & 2018-10-21T01:04:01.000 \\
58 & LCOGT & U & 14.12 & 3.02 & 0.14 & $(3.0\pm0.2) \times 10^{31}$ & 2018-10-21T03:07:20.000 & 2018-10-21T03:21:27.000 & 2018-10-21T03:10:21.000 \\
 & CHIRON & H$\alpha$ & 15.23 & 10.86 & 0.25 & \nodata & 2018-10-21T03:09:32.800 & 2018-10-21T03:24:46.500 & 2018-10-21T03:20:24.500 \\
59 & LCOGT & U & 30.42 & 11.35 & 0.06 & $(4.0\pm0.3) \times 10^{31}$ & 2018-10-22T01:25:40.000 & 2018-10-22T01:56:05.000 & 2018-10-22T01:37:01.000 \\
60 & LCOGT & U & 16.67 & 1.52 & 0.13 & $(4.6\pm0.2) \times 10^{31}$ & 2018-10-22T01:56:50.000 & 2018-10-22T02:13:30.000 & 2018-10-22T01:58:21.000 \\
61 & LCOGT & U & 12.87 & 7.58 & 0.18 & $(5.8\pm0.2) \times 10^{31}$ & 2018-10-22T02:28:16.000 & 2018-10-22T02:41:08.000 & 2018-10-22T02:35:51.000 \\
 & CHIRON & H$\alpha$ & 184.25 & 70.38 & 0.79 & \nodata & 2018-10-22T01:27:13.600 & 2018-10-22T04:31:28.800 & 2018-10-22T02:37:36.400 \\
62 & LCOGT & U & 29.03 & 4.55 & 0.08 & $(4.9\pm0.2) \times 10^{31}$ & 2018-10-22T23:57:44.000 & 2018-10-23T00:26:46.000 & 2018-10-23T00:02:17.000 \\
 & CHIRON & H$\alpha$ & 39.04 & 22.74 & 0.34 & \nodata & 2018-10-22T23:57:18.500 & 2018-10-23T00:36:21.200 & 2018-10-23T00:20:03.200 \\
63 & CHIRON & H$\alpha$ & 48.04 & 17.73 & 0.20 & \nodata & 2018-10-23T01:05:30.800 & 2018-10-23T01:53:33.400 & 2018-10-23T01:23:14.400 \\
64 & CHIRON & H$\alpha$ & 40.08 & 13.00 & 0.20 & \nodata & 2018-10-23T03:33:08.300 & 2018-10-23T04:13:13.300 & 2018-10-23T03:46:08.500 \\
65 & CHIRON & H$\alpha$ & 173.43 & 43.46 & 2.48 & \nodata & 2018-10-24T00:28:16.900 & 2018-10-24T03:21:42.900 & 2018-10-24T01:11:44.500 \\
66 & CHIRON & H$\alpha$ & 24.96 & 13.06 & 0.12 & \nodata & 2018-10-25T01:57:31.000 & 2018-10-25T02:22:28.400 & 2018-10-25T02:10:34.900 \\
67 & LCOGT & U & 29.82 & 11.33 & 0.07 & $(3.0\pm0.1) \times 10^{31}$ & 2018-10-26T00:56:19.000 & 2018-10-26T01:26:08.000 & 2018-10-26T01:07:39.000 \\
68 & LCOGT & U & 36.73 & 17.05 & 0.07 & $(6.9\pm0.2) \times 10^{31}$ & 2018-10-26T02:31:21.000 & 2018-10-26T03:08:05.000 & 2018-10-26T02:48:24.000 \\
69 & LCOGT & U & 9.49 & 1.58 & 0.12 & $(2.6\pm0.1) \times 10^{31}$ & 2018-10-26T12:25:42.500 & 2018-10-26T12:35:11.717 & 2018-10-26T12:27:17.331 \\
70 & LCOGT & U & 6.30 & 1.57 & 0.08 & $(9.0\pm0.8) \times 10^{30}$ & 2018-10-29T11:24:31.511 & 2018-10-29T11:30:49.366 & 2018-10-29T11:26:05.555 \\
71 & LCOGT & U & 8.65 & 4.72 & 0.07 & $(2.1\pm0.1) \times 10^{31}$ & 2018-10-29T12:23:17.353 & 2018-10-29T12:31:56.114 & 2018-10-29T12:28:00.679 \\
72 & LCOGT & U & 6.80 & 2.27 & 0.39 & $(9.2\pm0.2) \times 10^{31}$ & 2018-10-12T04:16:29.000 & 2018-10-12T04:23:17.000 & 2018-10-12T04:18:45.000 \\
73 & LCOGT & U & 17.52 & 2.28 & 2.86 & $(5.4\pm0.0) \times 10^{32}$ & 2018-10-12T04:23:17.000 & 2018-10-12T04:40:48.000 & 2018-10-12T04:25:34.000 \\
\enddata
\tablecomments{
(1) The units of Peak rate are counts s$^{-1}$ for \xmm{} OM, EPIC-pn, and RGS data, \AA{}  for CHIRON H$\alpha$ data, and relative counts for all other data. 
}
\end{deluxetable} 
\end{longrotatetable}

\clearpage

\section{Neupert Classification Visuals}
\label{appendix:neupert_figures}

\begin{figure}[!ht]
\centering
\includegraphics[width=0.8\linewidth]{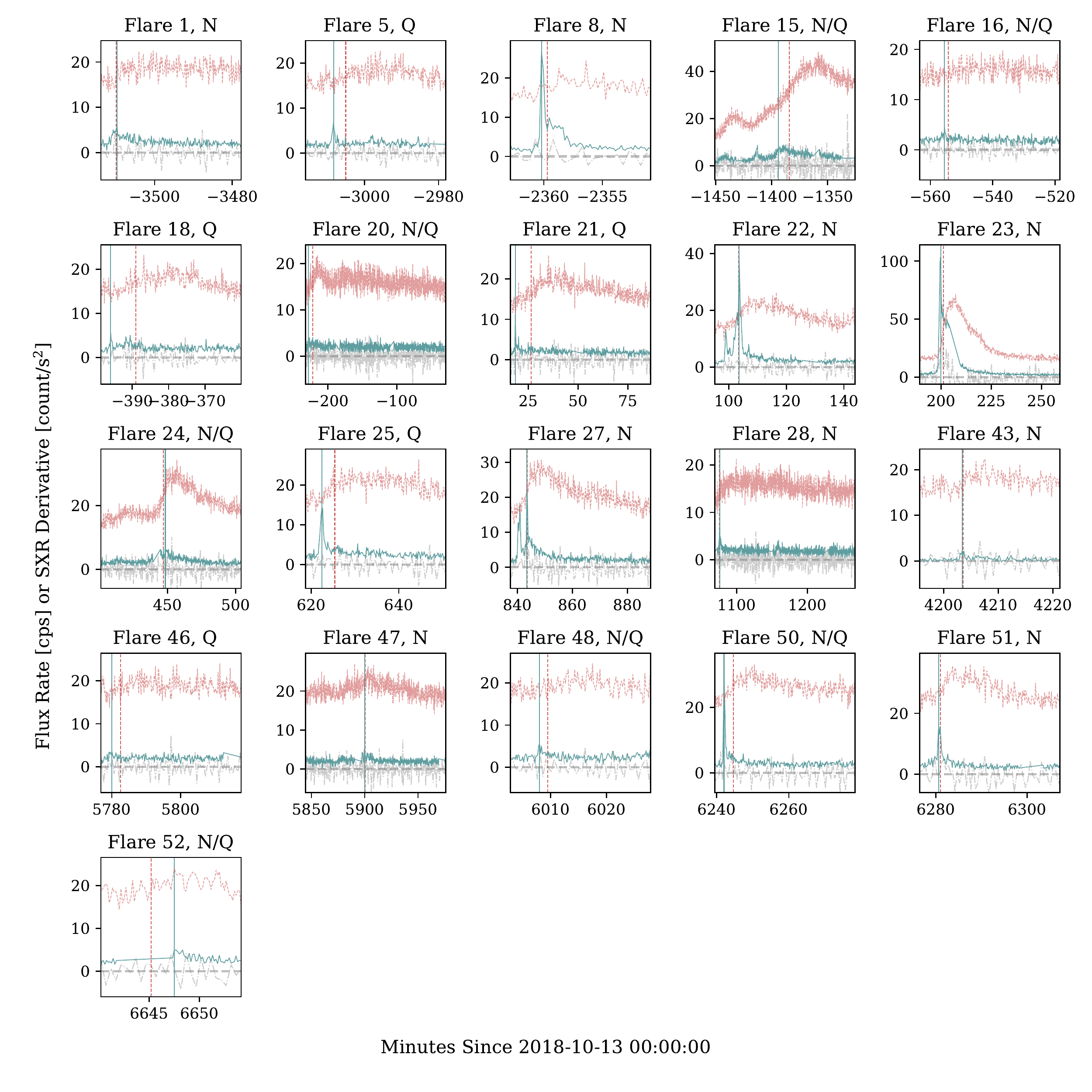}
\caption{See Figure \ref{fig:neuperttypes} for marker descriptions.
\label{fig:neupertfigall}
}
\end{figure}

\begin{figure}[!ht]
\centering
\includegraphics[width=0.8\linewidth]{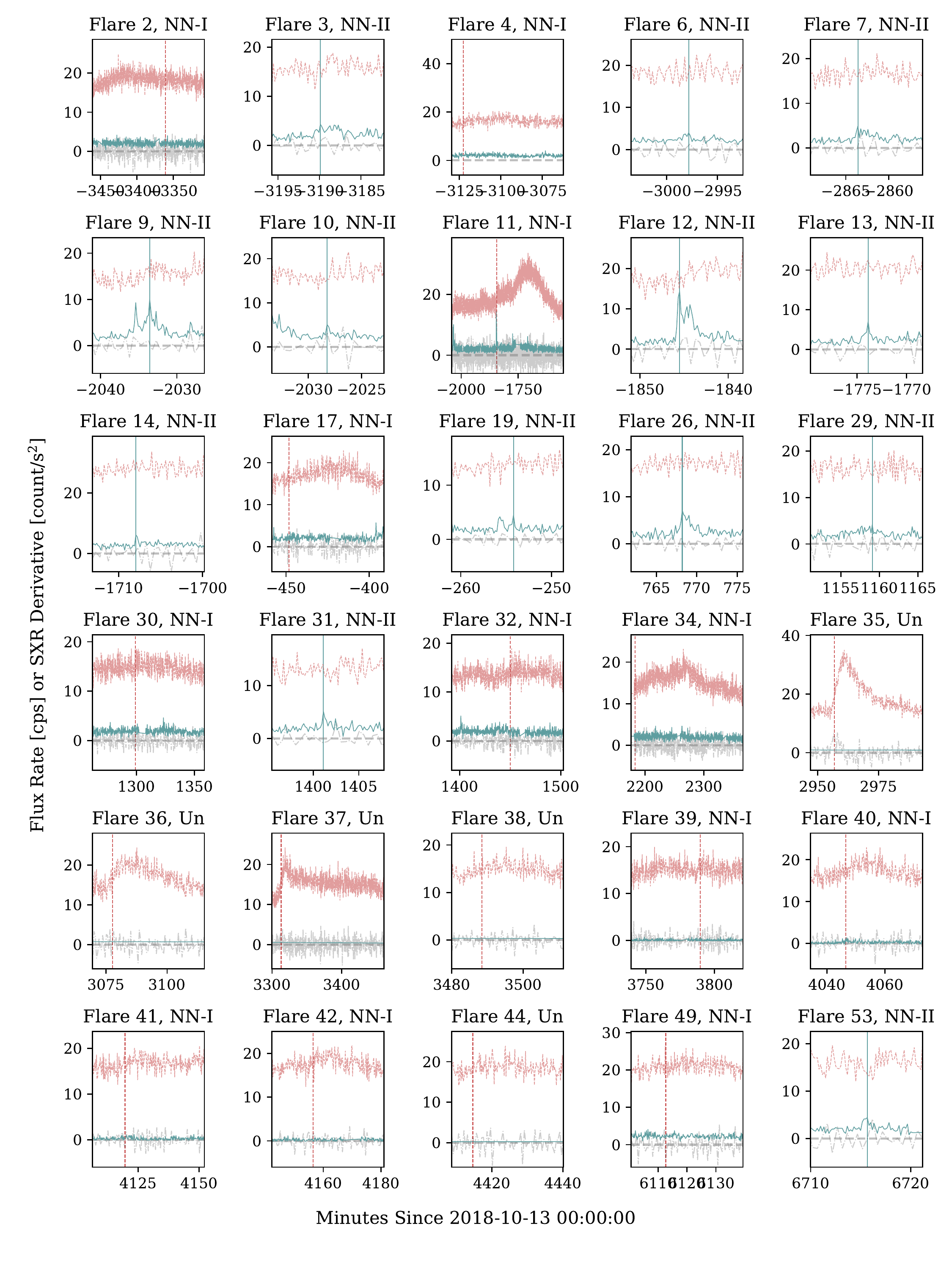}
\caption{See Figure \ref{fig:neuperttypes} for marker descriptions.
\label{fig:nonneupertfigall}
}
\end{figure}

We provide images similar to Figure \ref{fig:neuperttypes} for each \xmm{} flare. Figure \ref{fig:neupertfigall} shows Neupert and Quasi-Neupert flares, while Figure \ref{fig:nonneupertfigall} shows Non-Neupert flares of both types.

\clearpage

\section{Observation Logs} \label{app:obs}
We provide detailed observational logs for each observational dataset.
Note that Observation Windows are separated by at least 5,000 seconds on non-observation, unless stated otherwise in the table notes. Total durations here may be longer than in Table \ref{tbl:obs_summary}, as they include times with inclement weather or other issues. 

\begin{deluxetable}{cccc}[!ht]
\tabletypesize{\scriptsize}
\tablewidth{0pt}
\tablecaption{Observation Log - \xmm{} OM UVW2}
\tablehead{
\colhead{Observation Window} & \colhead{Start} & \colhead{End} & \colhead{Duration} \\
 & UTC & UTC & hrs}
\startdata 
1 & 2018-10-10T13:13:59.760 & 2018-10-12T01:42:19.177 & 36.47 \\
2 & 2018-10-12T13:06:33.808 & 2018-10-14T01:36:22.232 & 36.50 \\
3.1 & 2018-10-14T12:21:18.787 & 2018-10-14T18:48:06.449 & 6.45 \\
3.2 & 2018-10-15T10:35:09.418 & 2018-10-16T00:04:00.273 & 13.48 \\
4 & 2018-10-16T23:39:21.362 & 2018-10-17T18:17:33.004 & 18.64 \\
\enddata
\end{deluxetable} 

\begin{deluxetable}{cccc}[!ht]
\tabletypesize{\scriptsize}
\tablewidth{0pt}
\tablecaption{Observation Log - \xmm{} EPIC-pn X-ray}
\tablehead{
\colhead{Observation Window} & \colhead{Start} & \colhead{End} & \colhead{Duration} \\
 & UTC & UTC & hrs}
\startdata 
1 & 2018-10-10T13:13:57.030 & 2018-10-12T01:54:37.030 & 36.68 \\
2 & 2018-10-12T14:06:43.020 & 2018-10-14T02:33:43.020 & 36.45 \\
3 & 2018-10-14T12:21:16.970 & 2018-10-16T02:25:16.970 & 38.07 \\
4 & 2018-10-16T23:39:19.980 & 2018-10-17T18:29:59.980 & 18.84 \\
\enddata
\end{deluxetable} 

\begin{deluxetable}{cccc}[!ht]
\tabletypesize{\scriptsize}
\tablewidth{0pt}
\tablecaption{Observation Log - \xmm{} RGS X-ray}
\tablehead{
\colhead{Observation Window} & \colhead{Start} & \colhead{End} & \colhead{Duration} \\
 & UTC & UTC & hrs }
\startdata 
1 & 2018-10-10T13:07:48.868 & 2018-10-12T01:59:12.093 & 36.86 \\
2 & 2018-10-12T13:56:25.491 & 2018-10-14T02:38:26.807 & 36.70 \\
3 & 2018-10-14T12:15:11.942 & 2018-10-16T02:29:57.499 & 38.25 \\
4 & 2018-10-16T23:33:10.575 & 2018-10-17T18:34:40.232 & 19.02 \\
\enddata
\end{deluxetable}

\begin{longdeluxetable}{cccc}
\tabletypesize{\scriptsize}
\tablewidth{0pt}
\tablecaption{Observation Log - SMARTS V band}
\tablehead{
\colhead{Observation Window} & \colhead{Start} & \colhead{End} & \colhead{Duration} \\
 & UTC & UTC & hrs}
\startdata 
1 & 2018-10-10T02:40:42.667 & 2018-10-10T05:32:51.800 & 2.87 \\
2 & 2018-10-12T04:32:30.364 & 2018-10-12T05:35:17.543 & 1.05 \\
3 & 2018-10-12T23:55:49.511 & 2018-10-13T05:26:07.689 & 5.51 \\
4 & 2018-10-14T00:17:42.012 & 2018-10-14T02:27:56.587 & 2.17 \\
5 & 2018-10-14T05:06:33.806 & 2018-10-14T05:16:02.027 & 0.16 \\
6 & 2018-10-14T23:42:09.962 & 2018-10-15T05:16:13.908 & 5.57 \\
7 & 2018-10-15T23:43:58.108 & 2018-10-16T05:11:48.758 & 5.46 \\
8 & 2018-10-16T23:48:09.166 & 2018-10-17T05:04:31.086 & 5.27 \\
\enddata
\end{longdeluxetable} 

\begin{deluxetable}{cccc}[!ht]
\tabletypesize{\scriptsize}
\tablewidth{0pt}
\tablecaption{Observation Log - CHIRON H$\alpha$}
\tablehead{
\colhead{Observation Window} & \colhead{Start} & \colhead{End} & \colhead{Duration} \\
 & UTC & UTC & hrs}
\startdata 
 1 & 2018-10-10T01:15:17.900 & 2018-10-10T03:14:08.600 & 1.98 \\
 2 & 2018-10-12T04:30:48.900 & 2018-10-12T04:53:15.000 & 0.37 \\
 3 & 2018-10-12T23:37:17.900 & 2018-10-13T04:52:19.200 & 5.25 \\
 4 & 2018-10-14T00:31:47.800 & 2018-10-14T02:24:42.300 & 1.88 \\
 5 & 2018-10-14T23:49:19.900 & 2018-10-15T04:43:55.100 & 4.91 \\
 6 & 2018-10-15T23:43:54.700 & 2018-10-16T04:39:57.400 & 4.93 \\
 7 & 2018-10-16T23:48:17.000 & 2018-10-17T04:38:54.200 & 4.84 \\
 8 & 2018-10-18T00:09:43.300 & 2018-10-18T04:37:54.900 & 4.47 \\
 9 & 2018-10-18T23:56:59.300 & 2018-10-19T04:48:26.100 & 4.86 \\
10 & 2018-10-19T23:49:49.800 & 2018-10-20T04:51:57.500 & 5.04 \\
11 & 2018-10-20T23:57:55.900 & 2018-10-21T04:49:06.400 & 4.85 \\
12 & 2018-10-22T00:02:47.800 & 2018-10-22T04:43:22.200 & 4.68 \\
13 & 2018-10-22T23:51:56.100 & 2018-10-23T04:43:27.300 & 4.86 \\
14 & 2018-10-23T23:52:33.500 & 2018-10-24T04:33:04.800 & 4.68 \\
15 & 2018-10-25T00:00:38.300 & 2018-10-25T04:01:04.000 & 4.01 \\
\enddata
\end{deluxetable}

\begin{deluxetable}{cccc}[!ht]
\tabletypesize{\scriptsize}
\tablewidth{0pt}
\tablecaption{Observation Log - \swift{} UVOT W2 Band}
\tablehead{
\colhead{Observation Window} & \colhead{Start} & \colhead{End} & \colhead{Duration} \\
 & UTC & UTC & hrs }
\startdata 
1 & 2018-10-12T03:26:18.000 & 2018-10-12T03:53:48.000 & 0.46 \\
2 & 2018-10-12T04:50:18.000 & 2018-10-12T05:17:48.000 & 0.46 \\
3 & 2018-10-12T06:26:08.000 & 2018-10-12T06:53:48.000 & 0.46 \\
4 & 2018-10-12T08:01:08.000 & 2018-10-12T08:28:48.000 & 0.46 \\
5 & 2018-10-12T09:36:18.000 & 2018-10-12T10:04:48.000 & 0.47 \\
6 & 2018-10-14T03:03:13.000 & 2018-10-14T03:30:53.000 & 0.46 \\
7 & 2018-10-14T04:39:13.000 & 2018-10-14T05:06:53.000 & 0.46 \\
8 & 2018-10-14T06:18:43.000 & 2018-10-14T06:41:53.000 & 0.39 \\
9 & 2018-10-14T07:54:33.000 & 2018-10-14T08:17:53.000 & 0.39 \\
10 & 2018-10-14T09:31:53.000 & 2018-10-14T09:53:53.000 & 0.37 \\
\enddata
\tablecomments{A minimum separation of 3,000 seconds was used for this table.}
\end{deluxetable} 

\begin{deluxetable}{cccc}[!ht]
\tabletypesize{\scriptsize}
\tablewidth{0pt}
\tablecaption{Observation Log - \swift{} XRT X-ray}
\tablehead{
\colhead{Observation Window} & \colhead{Start} & \colhead{End} & \colhead{Duration} \\
 & UTC & UTC & hrs }
\startdata 
1 & 2018-10-12T03:26:20.000 & 2018-10-12T03:53:50.000 & 0.46 \\
2 & 2018-10-12T04:50:25.000 & 2018-10-12T05:17:50.000 & 0.46 \\
3 & 2018-10-12T06:26:25.000 & 2018-10-12T06:53:50.000 & 0.46 \\
4 & 2018-10-12T08:01:15.000 & 2018-10-12T08:28:50.000 & 0.46 \\
5 & 2018-10-12T09:36:40.000 & 2018-10-12T10:04:50.000 & 0.47 \\
6 & 2018-10-14T03:03:16.000 & 2018-10-14T03:30:51.000 & 0.46 \\
7 & 2018-10-14T04:39:26.000 & 2018-10-14T05:06:51.000 & 0.46 \\
8 & 2018-10-14T06:18:56.000 & 2018-10-14T06:41:51.000 & 0.38 \\
9 & 2018-10-14T07:54:41.000 & 2018-10-14T08:17:51.000 & 0.39 \\
10 & 2018-10-14T09:32:16.000 & 2018-10-14T09:53:51.000 & 0.36 \\
\enddata
\tablecomments{A minimum separation of 3,000 seconds was used for this table.}
\end{deluxetable}

\begin{deluxetable}{ccccc}
\tabletypesize{\scriptsize}
\tablewidth{0pt}
\tablecaption{Observation Log - LCOGT U band}
\tablehead{
\colhead{Observation Window} & \colhead{Start} & \colhead{End} & \colhead{Duration} & \colhead{MPC Code} \\
 & UTC & UTC & hrs & }
\startdata 
1 & 2018-10-10T17:40:11.000 & 2018-10-10T22:38:52.000 & 4.98 & K91 \\
2 & 2018-10-12T04:15:42.000 & 2018-10-12T04:40:46.000 & 0.42 & W85 \\
3 & 2018-10-12T17:43:00.000 & 2018-10-12T19:00:04.072 & 1.28 & K91--K93 \\
4 & 2018-10-12T23:46:56.000 & 2018-10-13T04:37:19.000 & 4.84 & W85 \\
5 & 2018-10-13T09:08:14.080 & 2018-10-13T09:43:55.716 & 0.59 & Q64 \\
6 & 2018-10-13T12:16:01.221 & 2018-10-13T13:47:54.839 & 1.53 & Q63 \\
7 & 2018-10-13T18:15:37.457 & 2018-10-13T18:55:42.289 & 0.67 & K93 \\
8 & 2018-10-13T23:47:10.000 & 2018-10-14T01:24:27.000 & 1.62 & W85 \\
9 & 2018-10-14T10:45:32.953 & 2018-10-14T13:03:16.208 & 2.30 & Q63 \\
10 & 2018-10-14T17:45:40.723 & 2018-10-14T20:01:27.405 & 2.26 & K93 \\
11 & 2018-10-14T23:46:50.000 & 2018-10-15T04:29:29.000 & 4.71 & W85 \\
12 & 2018-10-15T09:08:59.145 & 2018-10-15T13:41:52.258 & 4.55 & Q63 \\
13 & 2018-10-15T17:45:49.000 & 2018-10-15T20:57:38.729 & 3.20 & K91--K93 \\
14 & 2018-10-15T23:58:01.000 & 2018-10-16T04:25:31.000 & 4.46 & W87 \\
15 & 2018-10-16T09:10:10.842 & 2018-10-16T10:55:44.674 & 1.76 & Q63--Q64 \\
16 & 2018-10-16T12:15:32.307 & 2018-10-16T12:47:22.147 & 0.53 & Q64 \\
17 & 2018-10-16T17:45:34.514 & 2018-10-16T22:14:18.924 & 4.48 & K93 \\
18 & 2018-10-16T23:48:21.000 & 2018-10-17T04:22:27.000 & 4.57 & W85--W87 \\
19 & 2018-10-17T11:15:28.508 & 2018-10-17T12:38:51.859 & 1.39 & Q63 \\
20 & 2018-10-17T17:46:00.000 & 2018-10-17T22:12:27.000 & 4.44 & K91 \\
21 & 2018-10-17T23:49:12.000 & 2018-10-18T04:17:09.000 & 4.47 & W87 \\
22 & 2018-10-18T09:11:23.402 & 2018-10-18T13:28:15.164 & 4.28 & Q63 \\
23 & 2018-10-18T17:46:56.000 & 2018-10-18T22:07:41.000 & 4.35 & K91 \\
24 & 2018-10-18T23:52:55.000 & 2018-10-19T04:14:10.000 & 4.35 & W87 \\
25 & 2018-10-19T23:50:49.000 & 2018-10-20T03:59:50.000 & 4.15 & W87 \\
26 & 2018-10-20T23:51:35.000 & 2018-10-21T04:06:07.000 & 4.24 & W87--W85 \\
27 & 2018-10-21T23:52:26.000 & 2018-10-22T04:01:17.000 & 4.15 & W87 \\
28 & 2018-10-22T23:53:09.000 & 2018-10-23T03:57:44.000 & 4.08 & W85--W87 \\
29 & 2018-10-23T23:55:30.000 & 2018-10-24T03:55:35.000 & 4.00 & W87 \\
30 & 2018-10-24T23:54:48.000 & 2018-10-25T03:44:32.000 & 3.83 & W85--W87 \\
31 & 2018-10-26T00:00:48.000 & 2018-10-26T03:44:05.000 & 3.72 & W87--W85 \\
32 & 2018-10-26T09:18:27.644 & 2018-10-26T12:47:54.906 & 3.49 & Q64 \\
33 & 2018-10-27T00:15:49.000 & 2018-10-27T03:42:32.000 & 3.45 & W87 \\
34 & 2018-10-27T09:19:19.920 & 2018-10-27T09:45:29.010 & 0.44 & Q64 \\
35 & 2018-10-29T09:21:20.089 & 2018-10-29T12:35:53.349 & 3.24 & Q64 \\
\enddata
\tablecomments{Observation Windows here are determined by MPC code switches. Hyphenated MPC codes represent observations where fast changes between the two occurred. \\
MPC codes: \url{https://lco.global/observatory/sites/mpccodes/}
}
\end{deluxetable}

\begin{deluxetable}{ccccc}
\tabletypesize{\scriptsize}
\tablewidth{0pt}
\tablecaption{Observation Log - LCOGT V band}
\tablehead{
\colhead{Observation Window} & \colhead{Start} & \colhead{End} & \colhead{Duration} & \colhead{MPC Code} \\
 & UTC & UTC & hrs & }
\startdata 
1 & 2018-10-10T17:40:58.056 & 2018-10-10T22:29:40.256 & 4.81 & L09 \\
2 & 2018-10-11T20:15:26.104 & 2018-10-11T20:43:49.901 & 0.47 & Z17 \\
3 & 2018-10-12T17:41:26.188 & 2018-10-12T19:00:30.838 & 1.32 & L09 \\
4 & 2018-10-12T20:15:14.446 & 2018-10-12T20:43:43.565 & 0.47 & Z17 \\
5 & 2018-10-12T23:45:20.277 & 2018-10-13T04:14:26.415 & 4.49 & W79--W89 \\
6 & 2018-10-13T09:07:21.487 & 2018-10-13T10:04:22.760 & 0.95 & Q58 \\
7 & 2018-10-13T12:45:19.897 & 2018-10-13T13:48:02.765 & 1.05 & Q58 \\
8 & 2018-10-13T18:15:11.532 & 2018-10-13T18:55:26.561 & 0.67 & L09 \\
9 & 2018-10-13T20:15:18.121 & 2018-10-13T20:44:09.402 & 0.48 & Z17 \\
10 & 2018-10-13T23:45:32.916 & 2018-10-14T02:15:57.115 & 2.51 & W79--W89 \\
11 & 2018-10-14T10:45:17.467 & 2018-10-14T11:13:25.150 & 0.47 & Q58 \\
12 & 2018-10-14T13:15:18.881 & 2018-10-14T13:43:47.126 & 0.47 & Q58 \\
13 & 2018-10-14T20:16:27.059 & 2018-10-14T20:45:35.146 & 0.49 & Z17 \\
14 & 2018-10-14T23:46:17.290 & 2018-10-15T04:12:48.908 & 4.44 & W79--W89 \\
15 & 2018-10-15T04:49:55.571 & 2018-10-15T05:35:45.104 & 0.76 & T04--T03 \\
16 & 2018-10-15T09:08:38.998 & 2018-10-15T13:39:34.949 & 4.52 & Q59--Q58 \\
17 & 2018-10-15T17:45:30.897 & 2018-10-15T21:00:09.517 & 3.24 & L09 \\
18 & 2018-10-15T23:47:04.258 & 2018-10-16T03:33:45.227 & 3.78 & W79--W89 \\
19 & 2018-10-16T09:09:31.799 & 2018-10-16T12:47:22.397 & 3.63 & Q58--Q59 \\
20 & 2018-10-16T17:45:21.743 & 2018-10-16T21:43:55.774 & 3.98 & L09 \\
21 & 2018-10-16T23:47:49.052 & 2018-10-17T03:33:25.768 & 3.76 & W79--W89 \\
22 & 2018-10-17T04:48:24.104 & 2018-10-17T07:29:54.560 & 2.69 & T04--T03 \\
23 & 2018-10-17T11:15:17.099 & 2018-10-17T12:39:13.651 & 1.40 & Q58 \\
24 & 2018-10-17T17:45:28.091 & 2018-10-17T21:43:48.517 & 3.97 & L09--Z17 \\
25 & 2018-10-17T23:48:37.268 & 2018-10-18T03:46:44.435 & 3.97 & W79--W89 \\
26 & 2018-10-18T05:15:30.544 & 2018-10-18T07:26:09.755 & 2.18 & T03 \\
27 & 2018-10-18T09:11:13.301 & 2018-10-18T11:43:22.343 & 2.54 & Q59--Q58 \\
28 & 2018-10-18T12:15:19.076 & 2018-10-18T13:27:40.087 & 1.21 & Q59 \\
29 & 2018-10-18T17:46:21.786 & 2018-10-18T21:43:51.576 & 3.96 & L09 \\
30 & 2018-10-18T23:49:24.075 & 2018-10-20T03:33:32.772 & 27.74 & W89--W79 \\
31 & 2018-10-20T23:51:03.345 & 2018-10-21T03:44:24.467 & 3.89 & W89 \\
32 & 2018-10-21T23:52:11.034 & 2018-10-22T03:47:11.722 & 3.92 & W89--W79 \\
33 & 2018-10-22T23:52:42.075 & 2018-10-25T03:33:44.589 & 51.68 & W89--W79 \\
34 & 2018-10-26T00:00:18.789 & 2018-10-26T03:29:23.243 & 3.48 & W89--W79 \\
35 & 2018-10-26T06:00:18.472 & 2018-10-26T06:53:46.685 & 0.89 & T03 \\
36 & 2018-10-26T09:18:10.923 & 2018-10-26T12:56:00.095 & 3.63 & Q58--Q59 \\
37 & 2018-10-26T17:53:56.020 & 2018-10-26T21:14:11.359 & 3.34 & L09 \\
38 & 2018-10-27T00:00:19.949 & 2018-10-27T03:03:13.800 & 3.05 & W89--W79 \\
39 & 2018-10-27T09:19:07.294 & 2018-10-27T10:43:09.378 & 1.40 & Q58 \\
40 & 2018-10-27T10:45:15.196 & 2018-10-27T12:54:04.931 & 2.15 & Q59--Q58 \\
41 & 2018-10-27T18:45:14.346 & 2018-10-27T21:14:07.773 & 2.48 & L09 \\
42 & 2018-10-29T09:20:57.973 & 2018-10-29T12:35:00.467 & 3.23 & Q58--Q59 \\
\enddata
\tablecomments{Observation Windows here are determined by MPC code switches. Hyphenated MPC codes represent observations where fast changes between the two occurred. \\
MPC codes: \url{https://lco.global/observatory/sites/mpccodes/}
}
\end{deluxetable}

\clearpage
\bibliography{bibl}

\end{document}